\documentclass[ppnp,12pt]{elsarticle}
\usepackage[utf8]{inputenc}
\usepackage{amssymb}
\usepackage{amsmath}
\usepackage{hyperref}

\usepackage{aas_macros}
\usepackage{sidecap}
\usepackage{wrapfig}
\usepackage{dsfont}
\usepackage{color}

\newcommand{\lr}{\left( }
\newcommand{\rr}{\right) }

\newcommand{\be}{\begin{equation}}
\newcommand{\ee}{\end{equation}}
\newcommand{\bea}{\begin{align}}
\newcommand{\eea}{\end{align}}

\newcommand{\veck}{\mathbf{k}}
\newcommand{\vecj}{\mathbf{j}}
\newcommand{\vecv}{\mathbf{v}}
\newcommand{\vecp}{\mathbf{p}}
\newcommand{\vecx}{\mathbf{x}}
\newcommand{\vecb}{\mathbf{B}}
\newcommand{\vece}{\mathbf{E}}
\newcommand{\ompe}{\omega_\mathrm{pe}}
\newcommand{\ompi}{\omega_\mathrm{pi}}
\newcommand{\omce}{\Omega_\mathrm{ce}}
\newcommand{\omci}{\Omega_\mathrm{ci}}
\newcommand{\ompj}{\omega_{\mathrm{p},j}}
\newcommand{\omcj}{\Omega_{\mathrm{c},j}}
\newcommand{\omi}{\omega_\mathrm{\cal I}}
\newcommand{\oml}{\Omega_\mathrm{L}}
\newcommand{\ms}{M_\mathrm{s}}
\newcommand{\ma}{M_\mathrm{A}}
\newcommand{\mi}{m_\mathrm{i}}
\newcommand{\me}{m_\mathrm{e}}
\newcommand{\thbn}{\Theta_\mathbf{Bn}}
\newcommand{\thbncrit}{\Theta_{\mathbf{Bn},\mathrm{crit}}}
\newcommand{\lse}{\lambda_\mathrm{s,e}}
\newcommand{\vsh}{v_\mathrm{sh}}

\newcommand{\jn}{\textcolor{black}}
\newcommand{\mpo}{\textcolor{black}}
\newcommand{\mh}{\textcolor{black}}
\newcommand{\parder}[2]{\frac{\partial {#1}}{\partial {#2}}}

\journal{Progress in Particle and Nuclear Physics}

\begin{document}

\begin{frontmatter}

\title{PIC Simulation Methods for Cosmic Radiation and Plasma Instabilities}

\author[a1,a2]{M.\ Pohl}
\address[a1]{Institut f\"ur Physik und Astronomie, Universit\"at Potsdam,
 D-14476 Potsdam, Germany}
\address[a2]{DESY, D-15738 Zeuthen, Germany}
\author[a3]{M.\ Hoshino}
\address[a3]{Department of Earth and Planetary Science, University of Tokyo, 7-3-1 Hongo, Bunkyo-ku, Tokyo 113-0033, Japan}
\author[a4]{J.\ Niemiec}
\address[a4]{Institute of Nuclear Physics Polish Academy of Sciences, PL-31342 Krakow, Poland}

\begin{abstract}
{Particle acceleration in collisionless plasma systems is a central question in astroplasma and astroparticle physics. The structure of the acceleration regions, electron-ion energy equilibration, preacceleration of particles at shocks to permit further energization by diffusive shock acceleration,}
require knowledge of the distribution function of particles besides the structure and dynamic of electromagnetic fields, and hence a kinetic description is desirable. Particle-in-cell simulations offer an appropriate, if computationally expensive method of essentially conducting numerical experiments that explore kinetic phenomena {in} collisionless {plasma}. We review recent results of PIC simulations of {astrophysical plasma systems}, particle acceleration, and the instabilities that shape them. 

\end{abstract}


\end{frontmatter}


\newpage
\tableofcontents
\newpage


\section{Introduction}
\label{sec:1}
\subsection{Kinetic plasma physics}
One of the main challenges in plasma astrophysics is the wide range of scales. Cosmic objects such as the remnants of supernova explosions are a few light-years in size. Their structure and evolution are shaped by nonthermal particles whose acceleration and transport is governed by processes that operate on scales down to $10^{8}$~cm, or a few light-milliseconds, ten orders of magnitude smaller than the object that is influenced. There is no single technique that permits a simultaneous study of the processes on all scales, and specific methods are employed for smaller ranges of scales. Here we review Particle-in-Cell simulations, henceforth referred to as PIC simulations, that are designed to describe kinetic processes at the low end of scales. Originally conceived more than 50 years ago \citep{da62}, both algorithm development and advances in computer hardware have enabled PIC simulations to mature and to become an instrument with which today we conduct highly detailed computer experiments of processes like magnetic reconnection and particle acceleration at collisionless shocks \citep{he81,da83,bl05}. This review is intended to give an overview of the current status of research and recent results.

Kinetic processes are important in collisionless systems, which are so called on account of the low frequency of two-body collisions, through which particles can exchange energy and momentum. In an ionized medium, the relevant two-body collisions would be Coulomb scattering. The mean free path, $l_\mathrm{c}$, and the interaction rate, $\nu_\mathrm{ee}$, in a medium of density $n_e$ and temperature $T$ are\mpo{
\be
l_\mathrm{c} \simeq \frac{1}{4\pi}\left(\frac{v}{c}\right)^4\,\frac{1}{n_e\,\sigma_T\,\ln\Lambda}\qquad\qquad
\nu_\mathrm{ee} =\frac{v}{l_c}\ ,
\label{int1}
\ee}
where $\sigma_T$ is the Thomson cross-section, $k_B$ the Boltzmann constant, $m_e$ the electron mass, $c$ the speed of light, \mpo{$\ln\Lambda$ is the Coulomb logarithm,} and
\mpo{$v$ is the typical speed of the electrons.}
If Coulomb scattering, or neutral-neutral scattering of atoms, were the fastest interaction process, the distribution function of particles would relax to a Maxwellian. The microscopical interactions would have provided a local equilibrium, that we can describe with macroscopic parameters such as density, pressure, or temperature, whose variation with time and location can be followed treating the ensemble of particles as a fluid. \mpo{The characteristic speed of electrons can then be replaced with the electron thermal velocity, $v_{th,e}=({3k_BT/m_e})^{1/2}$.} Hydrodynamics (HD) and magnetohydrodynamics (MHD) are based on this notion. In collisionless systems they are therefore valid only on very large spatial and temporal scales. On smaller scales the particle ensemble has typically not relaxed to an equilibrium state, and so one does not know the distribution function. Instead, knowing the distribution function is one of the major challenges.

In space particles interact collectively, and the medium is referred to as a plasma, as opposed to an ionized gas. Systematic perturbations in the position and movement of charged particles lead to oscillating electromagnetic fields that can be understood as superposition of waves. The electromagnetic waves interact with charged particles and thus modify their distribution function. Many types of waves exist, the simplest of which are periodic collective displacements of electrons around their average positions with the frequency
\be
\ompe = \sqrt{\frac{4\pi\,e^2\,n_e}{m_e}}\simeq (5.6\cdot 10^4\ {\rm Hz})\,
\sqrt{\frac{n_e}{1\ {\rm cm^{-3}}}}\ ,
\label{int2}
\ee
where $e$ denotes the electric charge.
The so-called electron plasma waves, \mpo{or Langmuir waves, are longitudinal,} electrostatic waves that are excited if \mpo{the partial derivatives of the electron distribution function satisfy certain conditions at a velocity that is in resonance with the wave, $\veck\vecv=\omega$. In a 1D description the
distribution function of electrons must increase with momentum, i.e., be inverted.} In this case, the electrons see a constant electron field and can change their energy. Likewise, electron plasma waves are damped, if the distribution function falls off with momentum at the velocity resonance. This process is known as \emph{Landau damping}.

Electrons can also collectively shield other charges. If we added a test charge, $Q$, its potential would displace electrons as far as in a statistical equilibrium the Boltzmann statistic would permit. 
At a distance $r$ from the test charge
the potential is then truncated,
$$
\phi\simeq \frac{Q}{r}\,\exp\left(-\frac{r}{\lambda_D}\right)
$$
with
\be
\lambda_\mathrm{D}=\sqrt{\frac{k_B\,T_e}{4\pi\,n_e\,e^2}}=\frac{v_{th,e}}{\sqrt{3}\,\ompe}\simeq (6.9\ {\rm cm})\,
\sqrt{\frac{T_e/(1\ {\rm K})}{n_e/(1\ {\rm cm^{-3}})}}\ 
\label{int3}
\ee
This so-called Debye shielding is the result of a collective action of the electrons. The \emph{Debye length}, $\lambda_\mathrm{D}$, is one of the fundamental length scales of plasma physics. Another one is the \emph{skin depth}, the ratio of the speed of light and the plasma frequency, which for electrons is
\be
\lse= \frac{c}{\ompe}\simeq
(5.3\cdot 10^5\ \mathrm{cm})\ \lr\frac{n_e}{1\ {\rm cm^{-3}}}\rr^{-1/2}\ .
\label{int4}
\ee
\mpo{The name derives from a certain similarity to the skin effect in a conductor, as it is the attenuation length of an electromagnetic wave in plasma with $\omega\ll \ompe$.}
In resolving these length scales, PIC simulations represent plasma computer experiments that do not rely on assumptions on the microscopic behavior of the plasma.

Debye shielding can occur only, if the system size, $L$, is much larger than the Debye length. Collective action requires a large number of particles in a Debye sphere, otherwise there were no electron
to shield a test charge. The number of electrons in a Debye sphere, \mpo{$N_\mathrm{D}$}, is similar to the ratio of electron plasma frequency and the Coulomb collision rate,\mpo{
\be
N_\mathrm{D}=\frac{4\pi\,n_e\,\lambda_\mathrm{D}^3}{3}\simeq\frac{\ln\Lambda}{6}\,\frac{\ompe}{\nu_\mathrm{ee}}\ , 
\label{int5}
\ee}
and so a large number of electrons in a Debye sphere implies that collective interactions, here exemplified by electron plasma waves are faster than two-body collisions. \mpo{Note that for a classical definition of the minimal impact parameter the argument of the Coulomb logarithm is essentially the left-hand side of equation~\ref{int5}.} In this case the distribution function must be determined by solving the Vlasov equation, also known as the collisionless Boltzmann equation. PIC simulations do this by solving the equations of motion of a large number of computational particles representing electrons and ions, that move in time-dependent electromagnetic fields that they evolve on account of their charge and current density.

Plasma waves are driven by instabilities that thrive on particular feature of the distribution function of the electrons or ions, which may be inversion, anisotropy, or their carrying a current. \mpo{There is a whole zoo of such waves that can be sorted by the angle between the wave vector and the fluctuating electric field, and by the ranking of the frequency and wavenumber as well as the the skin depth and the Larmor radius for electrons and ions. \ref{appendixB} and \ref{appendixC} give an overview over some of the relevant wave types and what drives them.} 

The plasma waves then feed back on the particle ensemble, typically reducing the trait that led to their growth. The interaction can be described as a scattering process. It is obvious that an electrostatic waves can change both the direction and the modulus of a particle's momentum. A subluminal transverse electromagnetic wave, such as an Alfv\'en wave, carries only magnetic fluctuations in its frame of motion, and so it would elastically scatter a particle in this particular frame of reference. In any other frame of reference, this may involve a change in energy. 

Scattering on plasma waves is one of the central processes that shape the transport of plasma particles. Frequent kicks from particle-wave interactions change the trajectories of the particles from the initial ballistic motion (or helical path in the presence of a large-scale magnetic field) to a quasi-random walk. On large scales a diffusion-advection ansatz may be a useful way of describing the propagation of particles, but the obvious difficulty lies in finding the appropriate diffusion coefficient in \mpo{position} and momentum space. PIC simulation follow individual particles in their self-generated electromagnetic environment, and so they may help to understand the statistical properties of wave-particle interactions. They do this in a time-dependent fashion and can capture the non-linear feedback of the scattering processes. Wave-particle interactions are then described in a much more self-consistent way than is possible analytically with, e.g., linear growth rates of waves and a Fokker-Planck treatment of their impact on the particles.

\subsection{Acceleration processes}
Besides undergoing scattering particles may systematically gain energy in certain situations which requires the presence of electric field. A simple motional electric field, $\vece=\jn{-(1/c)}\vecv\times \vecb$, is not sufficient, because that field would disappear in the frame moving with velocity $\vecv$. If instead magnetic fluctuations moved with a range of velocities, either stochastically as turbulence or systematically as at shocks or in shear flows, there would be no frame in which the electric field disappeared, and particle acceleration would result. There are also situations in which an electric field can exist that is at least partially parallel to the magnetic field and hence is not entirely motional, \mpo{for example in magnetic reconnection. In shock-drift acceleration particles drift along the motional electric field.} The efficacy in particle acceleration of all these processes can be investigated with PIC simulations. In the following we give a brief introduction into each of them.

\subsubsection{Magnetic reconnection}
\mpo{In many plasma settings in the universe, the magnetic field structure contains a neutral sheet where the magnetic-field polarity changes direction, and during a change in magnetic field topology, magnetic reconnection is widely known as being important to rapidly release magnetic field energy \citep{1993noma.book.....B}. In the course of reconnection, the reconnected magnetic field line exerts a Lorentz force, and the bulk plasma can be accelerated up to the Alfven speed, $v_A=c \sqrt{\sigma/(1+\sigma)}$, where $\sigma=B^2/(4 \pi \rho c^2)$ is the so-called magnetization parameter.  Associated with the bipolar Alfvenic jets, not only hot plasma but also non-thermal particles can be rapidly generated on the order of the Alfven transit time $\lambda/v_A$, where $\lambda$ is the thickness of the neutral sheet.}
The process is well observed on the surface of the sun, and it is believed to play an important role near black holes, pulsars, and in stars. It is also seen in simulations of shocks \citep{2015Sci...347..974M,2015ApJ...814..137Z}.

Turbulence is a decisive agent in determining in what fraction of the volume reconnection operates and what its speed is \citep{1999ApJ...517..700L,2008PhRvL.100w5001L}. With the advent of 3D simulations it became clear that magnetic reconnection can be triggered by small initial perturbations and then feed on self-produced turbulence to eventually occupy a large volume and proceed quickly \citep{2011EL.....9365001L,2012NPGeo..19..251L,2018JPlPh..84a7102B}.

Converging, oppositely oriented magnetic-field structures convert to magnetic islands. Whereas simulations in 2D initially suggested that coherent electric field in the reconnection region might be most instrumental in accelerating particles, the contraction of the magnetic islands turned out to be important \citep{2006Natur.443..553D}. Trapped particles bouncing off the magnetic walls of the islands gain energy with each bounce. In addition, the islands move, and so particles residing outside may collide with them and stochastically gain energy.
\subsubsection{Stochastic acceleration}
The concept of stochastic acceleration was first introduced by \citet{1949PhRv...75.1169F} as potential source process for galactic cosmic rays. Particles elastically bounce off magnetic structures in their rest frame. Once these structure move with speed $V$ in arbitrary direction, particles can gain or loose energy, depending on the angle between the velocities of the particle and the scatterer, $\theta$. The relative change in energy of a particle moving at speed $v$ is
\be
\frac{\Delta E}{E}\simeq 2\,\frac{V^2}{c^2} + \frac{2\,v\,V\,\cos\theta}{c^2}\ .
\label{int6}
\ee
The first term is quadratic in the usually small parameter $V/c$, so it is very small itself. The second
term is linear in $V/c$, i.e. not so small, but it can change sign depending in the
direction of the incoming particle. There is a weak preference for frontal collisions
compared with overtaking collisions, which means that correctly averaging over angle the
second term does not yield zero, but a small number scaling also quadratically in $V/c$. In total, the process can be described as diffusion in momentum space, which corresponds to a continuous energy gain superimposed on stochastic redistribution in energy space. 

The rate of stochastic acceleration is governed by the scattering frequency with the structures that provides acceleration. These may be not the structures (or plasma waves)  that are the best scatterers and hence control the spatial transport of particles \citep{2008ApJ...673..942Y,2012PhPl...19j2901S}. Equation~\ref{int6} suggests that stochastic acceleration is relevant only where the phase velocity of plasma waves is high or generally as a secondary process that modifies the spectra of particles that were accelerated by some other means \citep[e.g.][]{2015A&A...574A..43P}. Even the latter may require a very large power for sustained operation \citep{2017A&A...597A.117D}, and so stochastic acceleration may be an efficient damping process for turbulence.

\subsubsection{Diffusive shock acceleration}
\mpo{In a nutshell, shocks represent converging flows and hence locations of significant heating. In space most of the shocks are in fact completely non-collisional, and processes at the smallest plasma scales dominate their physics. Collisionless shocks are not sharp jumps, but have a finite width. \mpo{Strictly parallel nonrelativistic shocks can also deviate from the MHD picture, in which the magnetic field is irrelevant for the shock profile \cite{2018JPlPh..84f9004B}.} The processes through which the incoming plasma flow is decelerated and practically isotropized are collective interactions with self-excited plasma waves, whose nature and operation depends on the orientation of the large-scale magnetic field \cite{2019PhPl...26c2106Z}. It turns out that for a moderate angle between the shock normal and the large-scale magnetic field, $\thbn$, the shock structure is similar to that strictly parallel shocks, and hence they are classified as \emph{quasi-parallel shocks}. Likewise, shocks with large $\thbn$ can be subsumed as \emph{quasi-perpendicular shocks}. }

\mpo{A second, independent distinction is provided by the magnetic-field orientation angle beyond which a particle traveling along the downstream large-scale magnetic field can no longer return to the shock. In that case the shock is referred to as superluminal, \mpo{and in the normal shock frame, in which the shock is stationary and the flow antiparallel to the shock normal,} the shock speed must obey
\be
v_\mathrm{sh}\, \tan\thbn \ge c\ .
\label{eq:dHT}
\ee
Otherwise, the shock is subluminal,
and particles can in principle return to the shock. If Eq.~\ref{eq:dHT} applies, one can describe the shock in the so-called de Hoffman-Teller frame, in which the flow of the upstream plasma is parallel to the magnetic field, and hence the motional electric field is absent. In a frame of reference, in which the upstream flow is along the shock normal, $\mathbf{V}_\mathrm{u} \parallel \mathbf{n}$, the de Hofmann-Teller frame moves with velocity
\be
\mathbf{V}_\mathrm{HT}=\frac{\mathbf{n}\times (\mathbf{V}_\mathrm{u}\times \mathbf{B}_\mathrm{u})}
{\mathbf{n}\cdot \mathbf{B}_\mathrm{u}}= - V_\mathrm{u}\,\tan \thbn \,\mathbf{e}_\mathrm{z},
\label{dht}
\ee
where $\mathbf{e}_\mathrm{z}$ denotes the direction of the perpendicular component of the magnetic field.
A slightly different definition of superluminal/subluminal shocks arises from the distinction whether or not a particle can in principle move ahead of the shock in the upstream region, which requires $v_\mathrm{sh} < c\, \cos\thbn$ \mpo{in the upstream frame}.}

Particles crossing the shock are elastically scattered on either side in the local rest frame, and there is a systematic energy gain. At a non-relativistic shock we can expect the particle distribution function to be isotropic, and the typical energy gain per cycle is
\be
\frac{\Delta E}{E}\simeq \frac{4}{3}\,\frac{\Delta V}{c}=
\frac{4}{3}\,\frac{\kappa -1}{\kappa}\,\frac{V_\mathrm{sh}}{c}\ ,
\label{int7}
\ee
where $\Delta V$ is the difference in flow speed between the upstream and the downstream region, that can be expressed in term of the shock speed, $V_\mathrm{sh}$, and the compression ratio, $\kappa$. As the escape probability toward the far-downstream region also scales with the shock speed, the equilibrium particle spectrum \mpo{only} depends on the compression ratio and corresponds to a power law with index $s=2$ ($E^{-s}$) for a compression ratio $\kappa=4$, \mpo{which one observes strong shocks in a mono-atomic gas in the absence of significant cosmic-ray pressure. The latter condition is known as the test-particle limit.}

If the accelerated particles carry a significant fraction of the energy and momentum flux at the shock, they will modify it, and the particle spectrum is expected to deviate from the test particle solution \citep{1987PhR...154....1B}. Typically, at very high energies the spectrum is harder than the test-particle solution ($s<2$) \citep[e.g.][]{2014ApJ...789..137B}.

The acceleration rate scales \mpo{inversely} with the spatial-diffusion coefficient which together with the shock speed determines how far from the shock a particle propagates before it is turned around and moves back to the shock. Scattering turbulence must be continuously and efficiently built in the upstream region, otherwise particles would escape to the far-upstream region and that acceleration process would terminate.

Much of the above also applies to relativistic shocks. An exception is that the distribution function of energetic particles near a relativistic shock can not be assumed to be isotropic. In fact, in the upstream region scattering by an angle $\theta_\mathrm{s}\approx 1/\Gamma_\mathrm{sh}$ is sufficient to provide transport back to a shock moving with Lorentz factor $\Gamma_\mathrm{sh}$, and so beyond the first half-cycle the energy gain is only $\Delta E \approx E$ \citep{2012SSRv..173..309B}. A second exception is that the magnetic field in the immediate downstream region tends to be highly oblique, and diffusive return to the shock becomes inefficient, unless large-angle scattering is invoked. Particle spectra are then typically rather soft \citep{2004APh....22..323E,2006ApJ...650.1020N}.

\subsubsection{Shock drift acceleration}
If the large-scale magnetic field has a component perpendicular to the shock normal, it would be compressed at the shock. Whereas for parallel shocks the large-scale magnetic field is dynamically irrelevant, at perpendicular shocks it is amplified by the shock compression ratio, $\kappa$. The structure of oblique shocks can be likened to that parallel and perpendicular shocks, and so one denotes them as quasi-parallel or quasi-perpendicular.

At quasi-perpendicular shocks, the reflection of particles off the compressed magnetic field can cause them to drift along the shock surface and be accelerated by the motional electric field. This coherent process should operate, if the particles do not efficiently collide with turbulence or waves, and it is referred to as \textit{shock-drift acceleration}. In the de Hofmann-Teller frame, the acceleration can be understood as mirror reflection off the magnetic-field gradient and subsequent transformation back to the normal frame of reference, without considering electric fields \cite{1989JGR....9415367K}. 

The energy gain is proportional to the kinetic energy in the transverse motion in the de Hofmann-Teller frame, $(m/2)\, \mathbf{V}_\mathrm{HT}^2$, and can be large for quasi-perpendicular shocks. For nearly perpendicular shocks the de Hofmann-Teller frame does not exist, because it would formally have a superluminal speed. Efficient shock-drift acceleration is thus expected for only a narrow range of parameters. In addition, established features of a collisionless shock, such as a cross-shock electric field or a magnetic overshoot, also have an impact on the number of accelerated particles and the maximum energy that can be achieved \cite{2001PASA...18..361B}.

\subsection{Scope of the review}
There is a wide variety of astrophysical environments in which kinetic plasma processes are an important agent that modifies the distribution function of particles, determines their transport properties, and may provide particle acceleration. The basic processes in these environments may be similar, but the physical parameters certainly are not. In some systems the characteristic particle speeds are close to the speed of light, in others they are much lower than that. In some objects the magnetic field carries the lion's share of the energy density, in others its amplitude is modest. This review is organized by object class, which translates to a certain regime of parameter values. Following a discussion of the technical aspects of PIC simulations, we shall turn our attention to relativistic systems, subdivided into strongly magnetized environments such as pulsar-wind nebulae (PWN) and objects harboring weak magnetic fields, for example the jets of Active Galactic Nuclei (AGN) or Gamma-Ray Bursts (GRBs). We shall then cover recent simulation results pertaining to nonrelativistic systems. There are three categories that we discuss individually. High-velocity outflows as found in Supernova Remnants (SNR) can drive shocks with very large Mach numbers. Among the low-velocity environments we further distinguish environments with large thermal energy density compared to that of magnetic field, e.g., shocks in clusters of galaxies. The ratio of thermal to magnetic pressure is called the plasma $\beta$, 
\be
\beta=\frac{8\pi\,n_e\,k_\mathrm{B}T}{B^2}\ .
\label{int9}
\ee
An example of low-velocity, low-$\beta$ environments is a planetary bow shock. An overview of the main plasma instabilities is found in the appendix.

\vfill
\newpage
\section{Method}
\subsection{Kinetic description of collisionless plasma}

\jn{The fully microscopic description of plasmas requires knowledge of positions and momenta of all plasma particles in function of time. For $N$ particles of a given type $l$ (e.g., $l=e$ for electrons or $l=i$ for ions)
one can define the probability distribution function in phase-space
\be
N_l(\vecx,\vecp,t)=\sum_{i=1}^{N}\delta(\vecx-\vecx_i(t))\delta(\vecp-\vecp_i(t)),
\ee
where particle positions $\vecx_i(t)$ and momenta $\vecp_i(t)$ evolve according to the particle equations of motion that use microscopic interaction forces. This is the so-called Klimontovich function, whose time evolution is obtained by applying the Liouville's theorem for conservation of phase-space. The resulting Klimontovich equation together with Maxwell's equations provide fundamental description of plasmas that accounts for a discrete nature of the particles \citep[e.g.,][]{2018PhR...783....1K,2016PhRvE..93c3203Y}. It allows rigorous treatment of binary collisions or spontaneously-emitted fluctuations, e.g., within the weak turbulence theory. The {\it statistical} properties of a plasma system are fully determined through particle distribution function
\be
F_l(\vecx_1,\dots,\vecx_N,\vecp_1,\dots,\vecp_N,t),
\label{distFN}
\ee
that determines the probability of finding at time $t$ particle positions and momenta in the phase-space element $d\vecx_1,\dots,d\vecx_N,d\vecp_1,\dots,d\vecp_N$ around point $\vecx_1,\dots,\vecx_N,\vecp_1,\dots,\vecp_N$. 
For numerous applications simpler reduced distribution functions are often used, such as one-particle distribution function, $f(\vecx, \vecp, t)$, that is obtained from equation~\ref{distFN} by integration over positions and momenta of all but one particle and multiplying by the particle number. The function $f$ represents the particle density in the
six-dimensional (6D) phase-space at point $(\vecx, \vecp)$ and time $t$, and is the ensemble-averaged Klimontovich function: 
\be
f_l(\vecx, \vecp, t)=\langle N_l(\vecx,\vecp,t)\rangle.
\ee
Time evolution of the distribution function $f$ is then obtained from the Klimontovich equation, and is known as the Vlasov equation
\citep[for detailed introduction see, e.g.,][]{1973ppp..book.....K}.}
%
For electromagnetic forces the \jn{relativistic} Vlasov equation takes the form:
\begin{equation}
 \parder{f_l}{t}+\vecv\cdot \parder{f_l}{\vecx} + \jn{{q_l} [\vece(\vecx, t) + \frac{1}{c}\vecv \times \vecb(\vecx, t)] \cdot \parder{f_l}{\vecp}} = 0, 
\label{vlasov}  
\end{equation}
where 
the electric charge $q_e = -e$ and \mpo{$q_i = Z_i\,e$}, \jn{and relativistic particle momentum $\vecp=\gamma m_l \vecv$, where $m_l$ is the particle rest mass and $\gamma = (1-(v/c)^2)^{-1/2}$ is the Lorentz factor. The third term in equation~\ref{vlasov} contains the Lorentz force:
\be
\mathbf{F}=q(\vece + \frac{1}{c}\vecv \times \vecb) .
\label{Lorentz}
\ee}
The electromagnetic fields
are generated self-consistently through long-range collective interactions between plasma particles as well as external charges and currents. 
Because of this inherent nonlinear coupling between the electromagnetic fields and particles, the Vlasov description must be complemented with Maxwell equations. Defining the charge and current densities:
\be
\rho (\vecx,t) = \sum_l q_l \jn{\int{f_l(\vecx,\vecp,t) \ d^3 p}} \qquad
\vecj (\vecx,t) = \sum_l q_l \jn{\int{f_l(\vecx,\vecp,t) \ \vecv \ d^3 p}}\ ,
\label{sources}
\ee
we have:
\begin{equation}
  \mathbf{\nabla} \cdot\vece =4\pi\rho \ , 
  \mathbf{\nabla} \cdot\vecb =0 \ ,
  \mathbf{\nabla} \times \vece =-\frac{1}{c}\parder{\vecb}{t} \ ,
  \mathbf{\nabla} \times \vecb =\frac{1}{c}\parder{\vece}{t}+\frac{4\pi}{c}\vecj \ .  
\label{Maxwell}  
\end{equation}
The set of equations \ref{vlasov}-\ref{Maxwell} describes the full dynamics of the collisionless plasma, \jn{i.e., it can be used to study plasma behavior on times scales much shorter than characteristic time scales of binary collisions.}

The derivation of exact analytical solutions of the Vlasov equation is impossible for an arbitrary distribution function. In the limit of 
weakly turbulent plasmas, linear theory can provide dispersion relations, $\omega(\veck)$, that define normal plasma modes and their growth or damping.
\jn{Nonlinear evolution of these modes can also be studied with quasilinear theory or the weak turbulence theory \citep[e.g.,][]{2016PhRvE..93c3203Y}.}
Although indispensable in plasma physics studies, such solutions cannot be evolved to \jn{strongly} nonlinear stages \jn{and applied to very complex plasma systems}, for which one needs to adhere to numerical methods. A~direct numerical integration of the Vlasov equation requires representation of the distribution function on a discrete mesh of phase-space. Though such methods have been proposed and used successfully, their application is usually limited to problems of restricted dimensionality. Multidimensional 6D problems are computationally very  expensive, in particular if high resolution in velocity space is needed in long-time simulations. However, the Vlasov equation can be solved more efficiently with so-called particle methods that approximate the plasma by a finite number of computational particles.

The core idea of particle methods stems from the fact that a solution, $f$, of the Vlasov equation satisfies 
$\frac{d}{dt} f(x (t), \jn{p(t)}, t) = 0$ on a trajectory in phase-space given by the solutions of the ordinary differential equations system:
\be
 \frac{d \vecx}{dt}= \vecv(t) \qquad
 \jn{\frac{d \vecp}{dt}}= \mathbf{F}(\vecx(t),t)/m \,,
\label{traj}  
\ee
which are called the characteristics of the Vlasov equation. These are \jn{the well-known relativistic} particle equations of motion with the Lorentz force, $\mathbf{F}$ \jn{(eq.~\ref{Lorentz})}. 
Thus, for a given initial condition, $f(\vecx,\vecp,t=0)$, for particle positions and velocities, the equations of motions (eq.~\ref{traj}) yield an ensemble of characteristic curves that represents a surface in phase-space that is a solution of the Vlasov equation \mpo{\citep[e.g.][]{Bernstein68}}.
As the particle techniques solve particle equations of motions, they represent a solution of the Vlasov equation with the method of characteristics.
 
The particle-in-cell technique is a particle method in which plasma is represented by an ensemble of macro-particles. Each macro-particle corresponds to many particles of the real plasma. Thus the charge and mass of a macro-particle are numerically different from $e$ and $m_e$ or $m_p$, but the equations of motion are the same as for the real particles, as only the charge-to-mass ratio, 
$q/m$, enters the equations. The forces between particles are not calculated directly because that is not feasible even if modern Pflop/s supercomputers are used. Instead, macro-particles interact through electric and magnetic fields that are defined on a computational mesh of the physical space. Electromagnetic fields are thus discretized in space but particles can have arbitrary positions on the grid. To calculate forces acting on particles the field values are interpolated from the grid points to the position of the particles.
This significantly reduces the arithmetic operation count that now grows linearly with the number of simulated particles, compared to a quadratic dependence in the direct method. All quantities are discretized in time.

As \mpo{demonstrated in equation~\ref{int5},} the collisionless plasma is characterized by a large number of particles in a Debye sphere, $N_{\rm D}\gg 1$. In this limit the Coulomb collision rate, \mpo{
\be
\frac{\nu_{ee}}{\ompe}\sim \frac{\ln\Lambda}{6\,N_{\rm D}}\,,
\label{coll}
\ee}
goes to zero, and particles interact through long-range collective forces. For typical parameters of, e.g., SNR shocks,  $N_{\rm D}$ may be as large as $N_{\rm D}\sim 10^{14}$. To facilitate computer simulations for such systems the PIC method offers a way of modeling the conditions of collisionless plasma with $N_{\rm D}\sim 10$ or even smaller \jn{(albeit with some drawbacks -- see section~\ref{method_impl}). The method aims at damping of the short-range forces causing the particle collisions.}
The use of a spatial grid for the fields can already be considered as an effective elimination of forces occurring at sub-grid scales. The other feature is the application of the finite-size particle description, that is fundamental for the PIC technique.    
Here the electric charge of a macro-particle is spread in a volume of finite size. At a large distance, $r$, the electrostatic force between two such particle \emph{clouds} assumes the same asymptotic form of $F \propto q^2/r^2$ as the force acting between two point charges. However, the short-range force responsible for collision effects can effectively disappear at particle distances smaller than the particle radius, if the size of the charge cloud is comparable to or larger than the Debye length. The collision rate is very low in this case and the dominant particle interactions are the collective ones.
Therefore, the finite-size particle approach describes the collisionless plasma. Being a method of the Vlasov equation solution, the PIC technique can be then considered a first-principle (\emph{ab-initio}) model of collisionless plasma.

\subsection{Implementation of the PIC simulation method \label{method_impl}}
\begin{figure}
\begin{center}
\includegraphics[width=12cm]{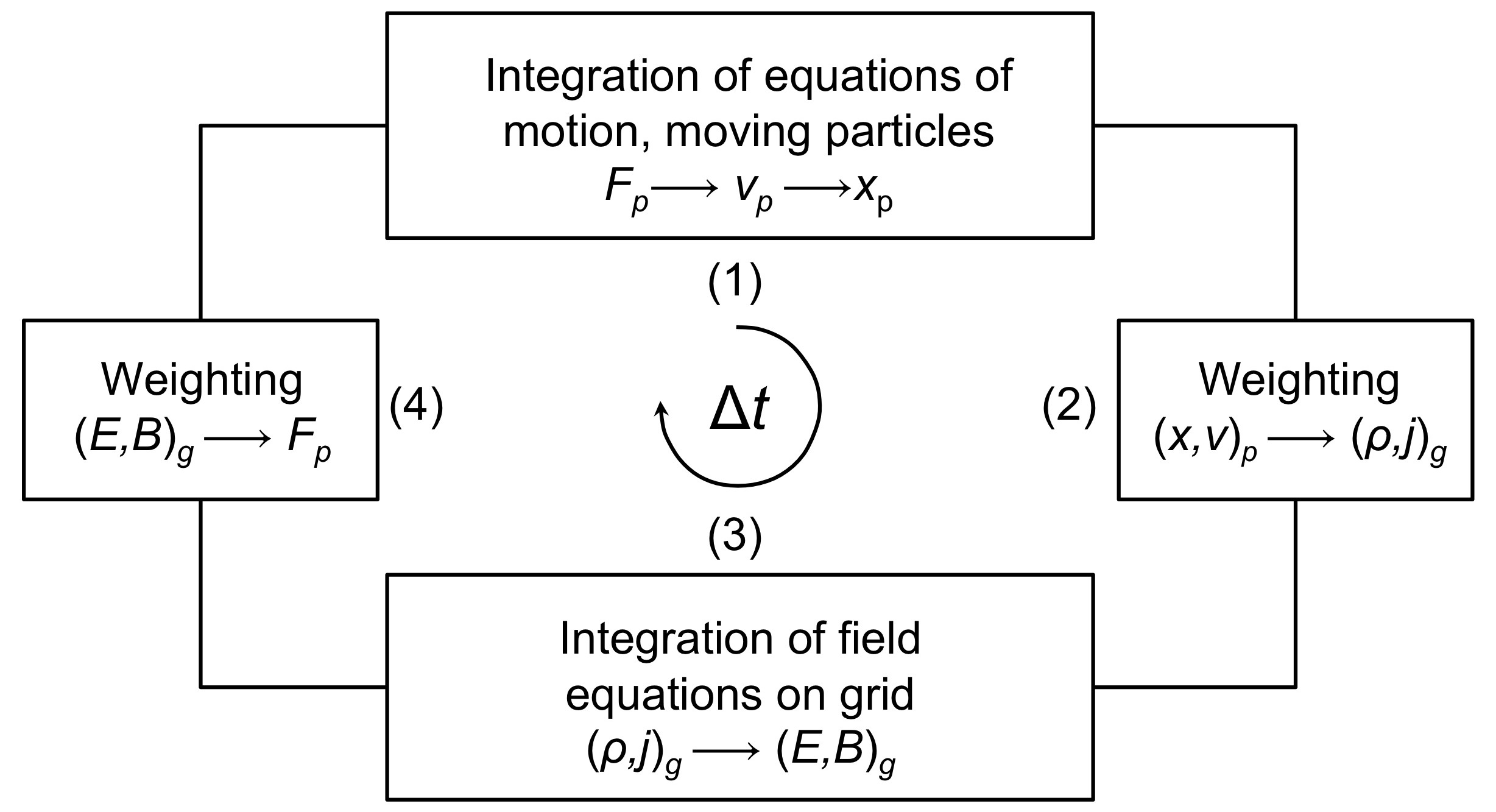}
\end{center}
\caption{A computational cycle of a PIC code. Particles are numbered by index $p$, and the grid index is $g$. At each time step the cycle goes through four stages. First, particles are advanced to new positions by integration of their equations of motion (eq.~\ref{traj}) with the Lorentz force. In the second stage electric charges and/or currents are accumulated on (weighed to) a spatial grid from particle data. This corresponds to a discretized form of eq.~\ref{sources}. Next, field equations (eq.~\ref{Maxwell}) are integrated on the grid. In the forth stage, new forces acting on particles are calculated by interpolation of the new magnetic and electric fields to particle positions. The next computational cycle begins. Taken  with  permission from \citet{Bohdan2017}.
}
\label{fig:loop}
\end{figure}

The main stages of the PIC code computational cycle that solves the system of equations \ref{Maxwell}-\ref{traj} are presented in Figure~\ref{fig:loop}. 
Most of the presently used codes solve the full set of Maxwell's equations, and thus are termed ``electromagnetic". Approximations to Maxwell's equations, e.g., electrostatic codes that integrate only the Poisson's equation and do not contain light waves, have been widely used in the past to reduce the computational cost. The need for such models has been alleviated in the era of peta-flop computing. Also, most astrophysics applications require simulating the full electromagnetic response of the system. The most widely used codes nowadays for astrophysical plasma computing solve Maxwell's equations formulated for $\vece$ and $\vecb$ fields, as in the flow-chart of Figure~\ref{fig:loop}. However, implementation of the Maxwell's equations solver through the vector and scalar potentials, $\mathbf{A}$~and~$\Phi$, is also possible \cite[e.g.][]{2019CoPhC.235...16O}. Here we discuss the $\vece$, $\vecb$ codes only. 
A relationship between variables defining fields and particles is given by the procedures of charge and/or current deposition from particle positions to grid points and interpolation of forces to the positions of particles. These procedures depend on the applied method of \emph{weighting} the charge/current contributions of a given particle to adjacent grid points, through which particles attain a certain shape that is seen by the grid.

\begin{figure}[ht!]
\begin{center}
\includegraphics[width=10cm]{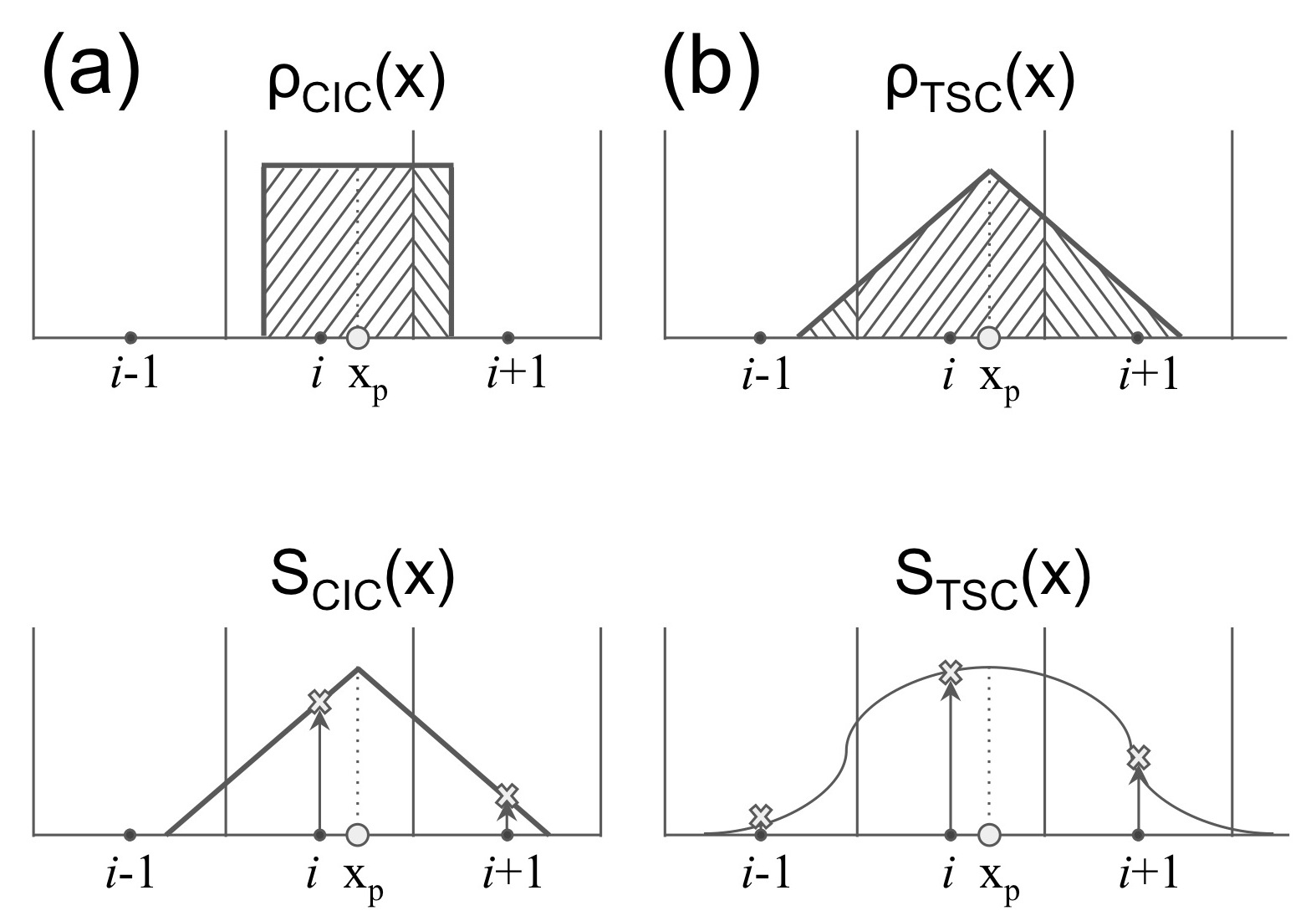}
\end{center}
\caption{Distribution of the electric charge for a single macro-particle and corresponding shape factors in a 1D grid for (a) the cloud-in-cell (CIC) and (b) the triangular-shape-cloud (TSC) approximations. The shape factors determine the fractions of particle charge deposited at given grid points \jn{that are illustrated with different shadings in the upper panels}.
}
\label{fig:shape}
\end{figure}

Two shapes that are often used for finite-size macro-particles are illustrated for a 1D grid in Figure~\ref{fig:shape}. The most commonly used Cartesian grid is considered. 
In the cloud-in-cell (CIC, Fig.~\ref{fig:shape}a) approximation a linear interpolation scheme is used.
On 1D grid with uniform spacing, $\Delta x$, the electric charge $q$ of a particle at location $x_{\rm p}$ makes contributions to two nearest grid points at $x_{\rm i}$ and $x_{\rm i+1}$ with weights linearly dependent on the particle position relative to the grid points: 
\be
q_{\rm i} = q\frac{x_{\rm i+1} -x_{\rm p}}{\Delta x}, \qquad
q_{\rm i+1} = q\frac{x_{\rm p}-x_{\rm i}}{\Delta x} \,.
\label{cic}  
\ee
The linear weighting scheme involves 4 points in 2D (area weighting) and 8 points (volume weighting) in 3D. Thus in 2D (3D) particles seem to have the shape of a square (cube) of the size of the grid cell, $\Delta x$. Such finite-size particles do not rotate and can freely pass through each other. The weighting method in the triangular-shape-cloud (TSC, Fig.~\ref{fig:shape}b) approximation uses the 3 nearest grid points in 1D, so that particles attain the shape of a triangle with a base equal to $2\Delta x$. A smoother particle shape provides for a lower level of the numerical noise compared with the linear interpolation scheme. In the TSC model on a 2D grid the particle charge is distributed to 9 grid points, and in 3D to 27 grid points.
The total charge deposited to a given grid point, $q_{\rm i}$, by all particles with charges $q_j$ can be obtained by calculating the sum $q_{\rm i}=\sum_j q_j S(x_j)$, with the shape factor (effective particle shape, assignment function) $S(x)$ that, e.g., for the linear weighting of equation~\ref{cic} is $S(x)= 1-|x-x_{\rm p}|/\Delta x$ for $|x-x_{\rm p}|/\Delta x\leq 1$ and $S(x)=0$ otherwise. 
Assuming the shape of macro-particles in concordance to the geometry of the computational grid (i.e., squares in 2D instead of circles) thus greatly simplifies calculations.
Similar procedure can be applied for the current density (see below).

An important requirement for a PIC code is that the same interpolation scheme is used to compute forces acting on particles as is applied for the charge deposition to the grid. In this way momentum conservation is ensured -- forces between two particles are equal and have opposite direction, and particles do not interact with themselves. If the weighting schemes at the second and fourth phases of the computational cycle (Fig.~\ref{fig:loop}) are different, the so-called self-force is not zero and may lead to unphysical particle acceleration.

\begin{figure}
\begin{center}
\includegraphics[width=13.5cm]{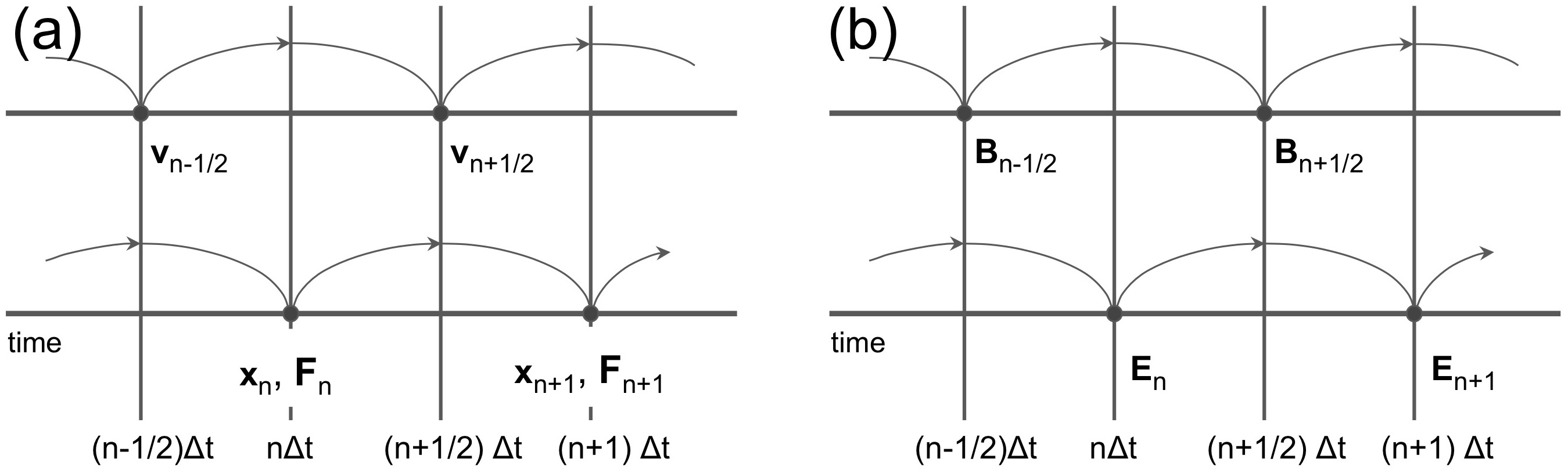}
\end{center}
\caption{Leapfrog scheme for time-integration of particle and field quantities in a single cycle of an electromagnetic PIC simulation code. Panel (a) shows time-centering of $\vecv$ for advancing $\vecx$, and of force $\mathbf{F}$ for the $\vecv$ pusher. The scheme for electromagnetic fields, $\vece$~and $\vecb$, is shown in panel (b). Full-integer time steps, $n\Delta t$, and half-integer time steps, $(n+1/2)\Delta t$, are indicated.}
\label{fig:leapfrog}
\end{figure} 

\begin{figure}[t!]
\begin{center}
\includegraphics[width=7.5cm]{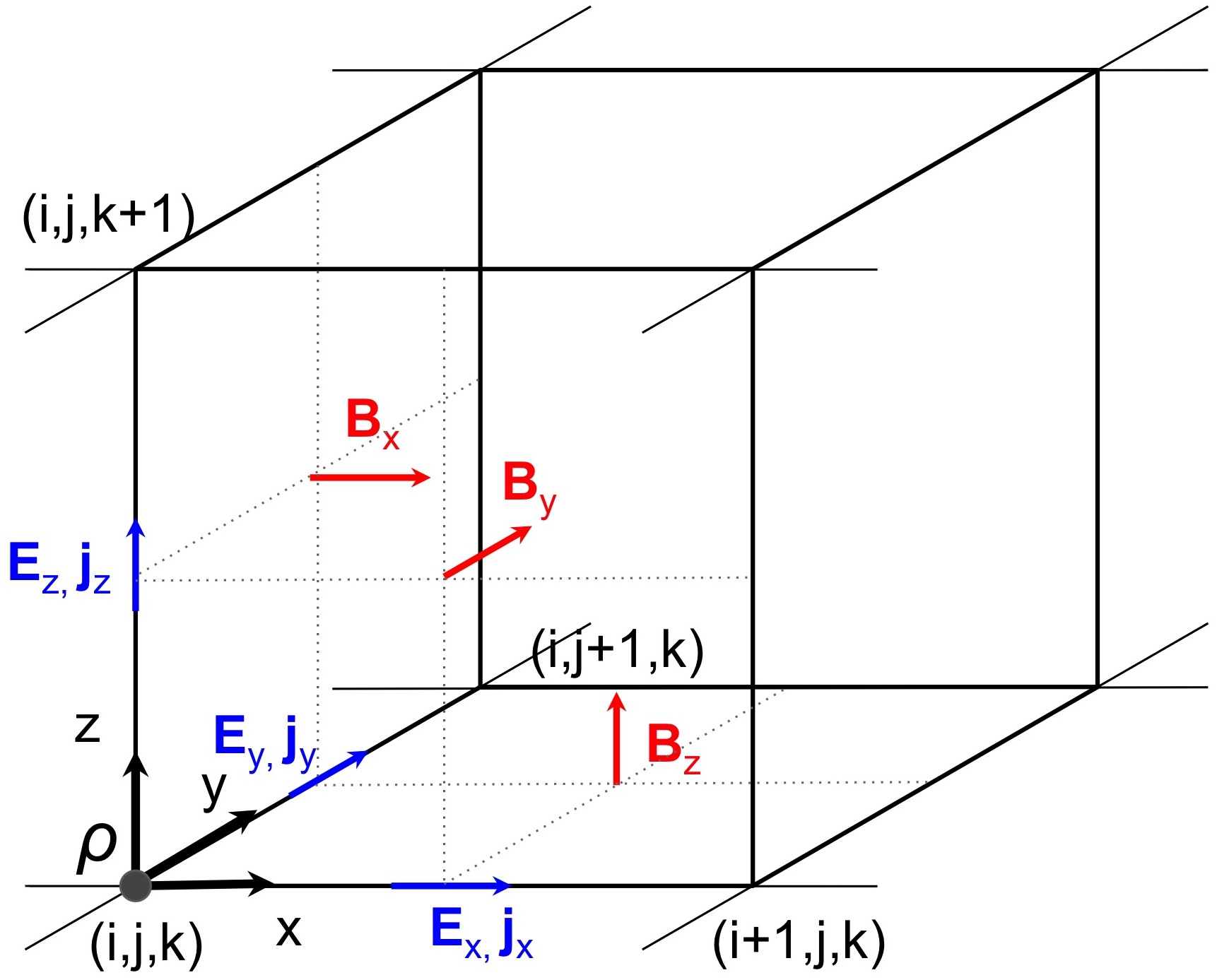}
\end{center}
\caption{Electric and magnetic field components and the electric charge and currents defined on the Yee lattice in 3D. A single cubic grid voxel of size $\Delta x = \Delta y = \Delta z =1$ originating at the grid point $(i,j,k)$ is illustrated. 
}
\label{fig:yee}
\end{figure}

The differential equations \ref{Maxwell} and \ref{traj} in a PIC code are solved by employing discretization in time and finite-difference methods. Any algorithm for integration of these differential equations should fulfill the following four major criteria: convergence, accuracy, stability, and efficiency. A consistent method means that a numerical solution of a differential equation converges to the exact solution in the limit of time-step $\Delta t\rightarrow 0$ and grid spacing $\Delta x\rightarrow 0$. The method should also reflect the features of the original equations, e.g., symmetry in time. An efficient algorithm should be fast, i.e., use the lowest operation count possible per time-step, and minimize RAM access (a number of time-steps back in time a quantity must be stored in the computer memory to push it to a new time-step). A method that \jn{in times of early developments was achieving} 
the best balance between accuracy, stability, and efficacy, and \jn{has been} the most widely used in PIC modeling is the finite-difference time-domain (FDTD) technique \jn{\cite{1973ppp..book.....K}}. 
This simple method uses centered-difference approximations to the space and time partial derivatives. This means that time integration proceeds in a leapfrog scheme, as illustrated in Figure~\ref{fig:leapfrog}, and spatial derivatives operate on a staggered mesh that is usually based on the Yee lattice (Fig.~\ref{fig:yee}). The FDTD method achieves second-order accuracy in space and time that is sufficient in most applications.

The leapfrog scheme for particle equations of motion (eq.~\ref{traj}) in the simplest nonrelativistic limit ($\gamma \rightarrow 1$), takes a form:
\begin{equation}
  \frac{\vecx^{ \,n+1}-\vecx^{ \,n}}{\Delta t} = \vecv^{ \,n+1/2}, \qquad
 m \ \frac{\vecv^{ \,n+1/2}-\vecv^{ \,n - 1/2}}{\Delta t} = \mathbf{F}^{\,n}   \,  
    \label{frog-coord}
\end{equation}
where $n$ indexes the time step.
For the Lorentz Force 
\begin{equation}
\mathbf{F}^{ \,n}=q(\vece^{\,n}+1/2\jn{c}(\vecv^{ \,n+1/2} + \vecv^{ \,n-1/2})\times\vecb^{ \,n})
\end{equation}
the velocity pusher has an \emph{implicit} form because the velocity at the new time, $\vecv^{ \,n+1/2}$, appears on both sides of the equation. However, simple \emph{explicit} formulation of the particle pusher,
that uses velocity values only at previous time, 
is possible \cite{Boris1970,2008PhPl...15e6701V} and widely used. A comparison of various integrators is found in \citet{2018ApJS..235...21R}.

Most of the PIC codes solve only the time-dependent Maxwell equations (eq.~\ref{Maxwell}). Writing the time derivative in the most popular explicit form, one advances fields in a single time-step $\Delta t$ as (compare Fig.~\ref{fig:leapfrog}):
\begin{eqnarray}
\vece^{\,n+1} &=& \vece^{\,n}+ c\Delta t (\mathbf{\nabla}\times\vecb)^{\,n+1/2}-\jn{4\pi}\mathbf{j}^{\,n+1/2}, \nonumber \\
\vecb^{\,n+1/2} &=& \vecb^{\,n-1/2}- c\Delta t (\mathbf{\nabla}\times\vece)^{\,n} \, .
\end{eqnarray}
An implicit method for solving Maxwell's equations can also be used to suppress aliasing errors at short wavelengths,  
\begin{eqnarray}
\vece^{\,n+1} &=& \vece^{\,n}+ c\Delta t (\alpha \mathbf{\nabla}\times\vecb^{\,n+1}+(1-\alpha)\mathbf{\nabla}\times\vecb^{\,n})-\jn{4\pi}\mathbf{j}^{\,n+1/2}, \nonumber \\
\vecb^{\,n+1} &=& \vecb^{\,n}+ c\Delta t (\alpha \mathbf{\nabla}\times\vece^{\,n+1}+(1-\alpha)\mathbf{\nabla}\times\vece^{\,n}),
\end{eqnarray}
where $\alpha=1/2$ applies to the central difference scheme in time, and $\alpha=1$ corresponds to the backward difference scheme. The parameter $\alpha$ is usually made slightly larger than $1/2$.  

The spatial derivatives for both methods are calculated on the Yee lattice, in which the electric field and electric current density are defined at mid-cell edges, and the magnetic field at mid-cell surfaces. This ensures that the change of $\vecb$ flux through a cell surface equals the negative circulation of $\vece$ around that surface, and the change of $\vece$ flux through a cell surface equals the circulation of $\vecb$ around that surface minus the current through it. Decentering also maintains $\mathbf{\nabla}\cdot\vecb=0$ to machine precision. 
However, Poisson's equation requires special care. This is because electric currents are defined at different grid points than charges and so they are interpolated to the grid with a different shape function. In consequence, the continuity equation 
\be
\partial \rho/\partial t + \mathbf{\nabla}\cdot\mathbf{j}=0
\ee
may be not satisfied. A solution to this problem lies in methods of rigorous charge conservation \cite{2001CoPhC.135..144E,2003CoPhC.156...73U,1992CoPhC..69..306V} that are widely used \jn{for current assignment schemes} in modern codes. Alternatively, one may add a correction $\delta\vece$ to the electric field computed from Ampere's law to ensure that  $\mathbf{\nabla} \cdot\vece =\jn{4\pi}\rho$ is maintained.

Simple and accurate explicit FDTD algorithms have, however, constraints as to the choice of maximum $\Delta t$ and $\Delta x$.
A numerical analysis of the vacuum light waves shows that their dispersion relation $\omega^2=c^2k^2$ is modified on the grid to
$\Omega^2=c^2K^2$, with $\Omega=2\sin(\omega\Delta t/2)/\Delta t$ and $K=2\sin(k\Delta x/2)/\Delta x$. This relation gives real (stable) solutions for $\omega$ only if the Courant-Friedrichs-Lewy (CFL) \cite{cfl} condition is met: $c\sqrt{D}< \Delta x/\Delta t$, where $D$ is the number of spatial dimensions. If this is the case, no phase or magnitude errors in $\vece$ and $\vecb$ fields are present, and errors in $\omega$ and direction of $k$ are of second-order.
The accuracy of the explicit methods thus requires the time steps much smaller than any characteristic frequencies in the system.

Discussion of the stability of the FDTD schemes is closely related to the issue of the numerical noise and its filtering. 
Due to the loss of displacement invariance for discretized physical quantities, nonphysical modes, so-called aliases, appear in the system and couple to the physical modes producing nonphysical instabilities, numerical noise, and spurious forces.
Aliases cannot be distinguished from physical modes. Therefore numerical techniques must be used to eliminate them. In PIC simulations the most straightforward way for reducing the aliasing effect is to use a large number of particles per cell, $N_{\rm ppc}$. This is because the mean amplitude of short-scale fluctuations decreases as $N_{\rm ppc}^{-1/2}$. Substantial noise damping can also be achieved by using higher-order particle shape functions, $S(x)$, that effectively serve as a low-pass filter. For example, increasing the order of the shape function from CIC to TSC can result in an order of magnitude lower noise amplitude with the same $N_{\rm ppc}$. The amount of the noise reduction needed depends on the system under investigation, and should be such so that the spurious forces no longer dominate the physical forces on the particles. In practice, the limited computational resources at one's disposal do not allow for a large $N_{\rm ppc}$ in a simulation. Therefore, additional noise damping is usually applied through digital filtering of currents or charge distributions in PIC codes in the configuration space, or direct filtering of Fourier spectra of physical variables in the Fourier-based codes. 

The analysis of the plasma properties performed for a chosen numerical model and including the effects of aliases also delivers other constraints for the maximum time and grid spacing.  For oscillations at the plasma frequency, $\ompe$, aliases lead to unstable fluctuations in cold plasma if $\ompe\Delta t > 2$. The instability threshold is reduced to $\ompe\Delta t > 1.62$ for thermal plasma. However, accurate solutions still require considerably smaller $\Delta t$ that should fulfill $\ompe\Delta t \ll 1$. In a similar way aliases impose restrictions on the value of the grid spacing, $\Delta x$. The controlling quantity is then the Debye length, $\lambda_{\rm D}$. The impact of aliases is negligible if $\lambda_{\rm D} \geq 0.3 \Delta x$ for linear weighting and the threshold can be even lower for higher-order particle shapes. If this condition is not fulfilled, nonphysical fluctuations of the electric field heat the plasma until 
$\lambda_{\rm D} \sim \Delta x$, that might result in a higher noise level than acceptable.

\begin{SCfigure}[0.84]
\includegraphics[width=7.cm]{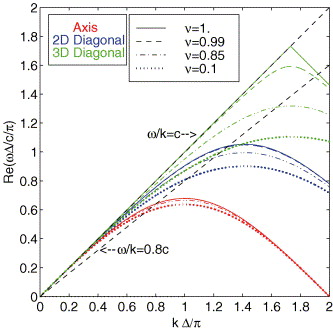}
\caption{FDTD numerical dispersion relation for light waves propagating on the Yee lattice. Propagation along a grid axis, a~2D grid diagonal, and a~3D grid diagonal is shown for different time-steps, $\Delta t=\nu\Delta t_{\rm max}$ (where $\Delta t_{\rm max}=\Delta x/c/\sqrt{3}$ is given by the CFL condition) and compared to the physical dispersion relation, $\omega/k=c$. The particle beam mode with $\omega=v_bk$ and $v_b=0.8c$ is superluminal at high frequencies, leading to nonphysical Cherenkov radiation. Taken with permission from~\citet{2004JCoPh.201..665G}.

}
\label{cherenkov}
\end{SCfigure}

By far the most important limitations of the FDTD schemes come from the nonphysical Cherenkov instability, as illustrated in Figure~\ref{cherenkov}. The instability arises at short scales when
relativistic particles travel faster than the numerical speed of light waves. High-frequency radiation quickly nonlinearly couples through wave-particle resonances to physical frequencies and disrupts the simulation. The effect is severe for plasma flows with relativistic speed against the grid. Several solutions have been proposed to mitigate this instability, such as strong smoothing of currents, damping the electromagnetic fields using dedicated filtering methods, e.g., the Friedman filter \cite{2004JCoPh.201..665G,2011JCoPh.230.5908V}, application of the spectral solvers \cite{2014JCoPh.266..124Y,2015CoPhC.192...32Y} or higher-order computational schemes with tunable coefficients for integration of the Maxwell's equations 
\cite{2004JCoPh.201..665G,2011JCoPh.230.5908V}, or some combination of these solvers \cite{2017CoPhC.214....6L}. It was also found that for some numerical models the Cherenkov instability growth is greatly inhibited for a carefully chosen CFL number \cite{2013JCoPh.248...33G,2015PASJ...67...64I,2011JCoPh.230.5908V}. 
Although these methods can effectively damp the instability, it can still remain an issue in simulations that cover large temporal scales over which relativistic particles drift on the grid. Such a situation is typically met in shock simulations. Here, the alleviation is often done through limiting the particle beam travel time on the grid by adopting a particle injection scheme in which the injection layer moves away from the shock interaction region \cite[e.g.][]{2012ApJ...759.73N,2009ApJ...698.1523S}. Another method is to reduce the beam drift velocity against the grid through a change of the reference frame \cite{2016PhRvE..94e3305L,2007PhRvL..98m0405V}. 

As mentioned above, the solutions of the Maxwell's equations, instead of the FDTD method, can be obtained in Fourier space employing fast Fourier transforms between the coordinate and Fourier space. Such solutions are quite accurate and allow one to easily deal with the filtering of the numerical effects. However, the application of the Fourier methods is difficult in systems with non-periodic boundary conditions and decomposed for efficient parallel computing.


An alleviation of the constraint for the maximum time-step is offered by \emph{implicit} methods of time integration. They can be applied to the field equations alone, as well as to the coupled system of particle equations of motion and the Maxwell's equations. In the latter fully-implicit approach, the resulting set of nonlinear coupled equations can be solved via nonlinear iteration with Newton-Krylov solvers \cite{Kelley2003,2004JCoPh.193..357K,2011JCoPh.230.7037M}, or -- after linearization of particle-field coupling -- 
with the methods of linear algebra, extrapolation, or iteration. Successful implementations of the latter in the form of the direct implicit method 
\cite[e.g.,][]{1983JCoPh..51..107L,2010JCoPh.229.4781D} or the implicit moment method \cite[e.g.,][]{1982JCoPh..46..271B,1985mts..conf.....B} are known.
Although, due to computationally challenging algorithms, in the past PIC codes based on fully-implicit methods have not been as popular as the explicit codes, the advances in computing hardware and numerical methods brought a re-birth of the implicit PIC codes. 
Also novel efficient solutions are proposed that allow for exact energy conservation and eliminate many of the constraints of the explicit PIC codes on the resolution of the spatial and temporal scales \cite{2017JCoPh.334..349L}. Such codes are promising for the treatment of multiple-scale problems, in which the focus is on the macroscopic or ion-scale processes, and the electron-scale physics does not need to be well resolved \cite{2017JPlPh..83b7005L}.  

Modern PIC experiments in astrophysics, space physics, and plasma physics use billions of macroparticles to simulate plasma systems on ever growing macroscopic scales. 2D simulations are the standard now, and the first large-scale 3D experiments have recently become feasible. PIC simulations pose serious computational demands and therefore must be run on high-performance computing systems that exploit massively parallel approach and scale up to $10^5-10^6$ CPU cores.
The most recent code developments thus focused on hardware-specific code optimization and parallelization. Parallelism is inherent in PIC codes because particle calculations and most field solvers are local. Particle advance dominates the computational cost of a simulation, and so parallel models based on grid decomposition into spatial domains with the Message Passing Interface (MPI) framework are widely used. To further enhance the efficiency, a shared memory parallelism is often utilized through hybrid MPI and OpenMP codes. Significant developments have also been made in implementing PIC algorithms in the accelerator hardware, such as graphics processing units (GPUs).
Some challenges still remain, the most important of which is the load imbalance that seriously deteriorates scalability. Solutions are sought through the implementation of adaptive grids \cite{2013PPCF...55l4011F}, dynamic assignment of domain patches to the MPI processes \cite{2016JCoPh.318..305G}, or via merging and splitting macro-particles \cite[e.g.,][]{2015CoPhC.191...65V}. Another computational issue is the visualization of particle and field data, whose volume in large-scale experiments can be in excess of Petabytes. 

A noteworthy PIC codes development concerns extensions of the standard PIC method. Inclusion of radiative cooling enables calculations of synthetic time-dependent photon emission spectra that allow to connect simulations with astronomical observations \cite{2013ApJ...770..147C,2013PhPl...20f2904H,2005PhDT.........2H,2010ApJ...724.1283R}. These codes also take into account the radiation reaction force on simulated particles \cite[e.g.,][]{2012PhRvE..86c6401C,2009PhRvL.103g5002J}. Binary Coulomb collisions are also featured by many codes \cite{2016JCoPh.318..305G,2013PhPl...20f2904H}, as well as ionisation \cite{2013JCoPh.236..220C,2018PhLA..382.3412D} and quantum electrodynamics effects, such as Compton scattering \cite{2013PhPl...20f2904H} or the creation of electron-positron pair cascades \cite{2017PhRvE..95b3210G,2011PhRvL.106c5001N,2010MNRAS.408.2092T}.

Though modern state-of-the-art supercomputers operate at maximum performances well exceeding 1 petaflop, performing simulations that include spatial and temporal scales larger than several proton gyroradii or gyrotimes is computationally challenging, as codes must usually at the same time resolve small electron scales. A common practice to overcome this difficulty is to simulate in 2D and use a reduced proton-to-electron mass ratio. \mpo{Also, many published results are based on simulations with low grid resolution, grid size, or a small number of particles per cell. Care must be exercised in choosing the setup, because the ranking of physical processes may depend on the dimensionality or the mass ratio. \jn{Some effects may also not be properly resolved with too low resolution and weakly growing or small-amplitude wave modes require substantial noise damping and long simulation times, thus significant computing resources.} Repeating a simulation with different reduced mass ratios at least allows for extrapolation to the behaviour at $\mi/\me=1842$. In the following sections we shall discuss this aspect for a few examples.} Most 2D models apply a 2.5D approach, in which all three components of particle velocity and electromagnetic fields are followed. Size-restricted 3D test simulations are typically performed to validate the results of the 2D simulations. However, a full scrutiny of the 3D physics must await the coming era of exascale computing.     

\jn{In addition to challenges with dealing with the large data volumes produced in PIC experiments, diagnostics and interpretation of the phenomena observed in simulations is often complex since all processes act simultaneously and usually are let to evolve to nonlinear stages. Therefore a common practice is to verify early-time results against the predictions of the linear theory. This allows one to identify the dominant wave modes, although in most cases it is not possible to calculate spectra for each eigenmode separately. Early nonlinear evolution of the system can be compared with expectations of the quasilinear theory. In this respect semi-analytic methods, such as numerical solutions of equations of weak turbulence theory can also be useful. It has been recently demonstrated that both weak turbulence theory and PIC simulations are in good agreement with each other in describing processes involving weak wave growth and low wave energy density when compared to the particle thermal energy density \citep{2019ApJ...871...74L}.} 

\subsection{PIC-MHD-hybrid}
An advantage of the PIC method is its ability to describe collisionless plasma from first principles. In most implementations the requirements of the stability and accuracy of computations pose a need of resolving the electron scales down to the smallest scales given by the Debye length and the electron plasma frequency. However, space physics and astrophysics systems typically represent multiple-scale problems, in which physical processes operate not only on the vast ranges of scales from micro to macro, but also most often the microphysical, or in general kinetic, processes considerably influence the macro state of an object. Due to computational constraints PIC models can deliver a system description up to several thousand proton plasma times or several hundred proton gyrotimes, but may at the same time not capture the spatial scale of the proton gyroradius. Simplifications to the kinetic modeling of plasmas are therefore needed for the description of much larger spatial and temporal scales. Relevant questions also appear as to the limits of validity of different approaches.     

Large-scale phenomena in magnetized plasmas can be described in the MHD (or fluid) approach. MHD equations couple moments of the Boltzmann equation with the Maxwell's equations (without the displacement current) and the plasma equation of state. They assume that local thermodynamic equilibrium is provided through short-range binary collisions between particles, and thus the particle distributions are Maxwellians. Only small departures from the Maxwellian distributions are allowed to  model transport phenomena, such as viscosity or thermal conductivity. 
\jn{Since MHD is a reduced description of plasma and underlying single-fluid equations use in addition many simplifying assumptions,}
it is impossible to clearly determine the limits of applicability of the fluid plasma description. Nevertheless, if particle collisions cannot be neglected, the fluid approach will hold for system sizes, $L$, much larger than the Coulomb scattering mean free path, $L\gg l_{\rm c}$. The MHD plasma is also quasi-neutral, and the response of electrons and ions is equal (charge-separation effects are not included). This leads to constraints that $L\gg\lambda_{\rm D}$ and time-scales $T\gg\ompe^{-1}$. The magnetic field can also keep the plasma particles in a fluid element of the size of the Larmor radius, $r_{\rm L}$. Therefore, for $L\gg r_{\rm L}$, a fluid  model can be used even for collisionless plasma. On the other hand, on scales smaller than $\sim r_{\rm L}$ the MHD description is inadequate, as microphysical plasma processes that depend on the $\vecv$ phase-space variable are not taken into account. This applies also to phenomena in plasma in equilibrium. For non-Maxwellian distribution functions the MHD approach may not recover all physical processes correctly. Also purely kinetic effects, such as Landau damping, the Weibel instability, particle acceleration, particle-wave interactions, etc., require fully kinetic description. 

A compromise between PIC and MHD models are so-called hybrid models, in which a fully kinetic description is retained for ions, whereas the electrons are described as a fluid. Electron scales are thus not resolved; in fact the electrons are considered to be
magnetized, following the magnetic-field lines. The relevant spatial scales are the ion gyroradius and the ion skindepth, and the times scale is the inverse ion gyrofrequency. The evolution of the physical system is thus possible for much longer times than possible with the PIC method. In the simplest implementation, the electron mass is neglected in a generalized Ohm's law, plasma is assumed to be quasi-neutral, and the electron pressure is scalar. Extended hybrid models may include resistivity effects, electron
inertia, or electron pressure effects \cite{Lipatov2002}. All of them neglect the displacement current in Amp\'ere’s law, i.e.,    
$\mathbf{\nabla} \times \vecb =\jn{(4\pi/c)}\vecj$, thus eliminating the propagation of light waves, which is consistent with neglecting high-frequency oscillations due to electrons. Waves with frequencies around and below the ion cyclotron frequency are adequately described.

If the physical scales of interest are even much larger than the ion inertial length, a coupled MHD-PIC approach can be formulated. A typical scenario for such a model is high-energy cosmic-ray feedback on the background plasma upstream of nonrelativistic SNR shocks. The model treats cosmic rays in a kinetic PIC way, while the ions and electrons of the thermal plasma are described through MHD equations \cite{2000MNRAS.314...65L,2015ApJ...809...55B,2018ApJ...859...13M}. Another approach, that goes the opposite way, is to supply initial and boundary conditions from MHD simulations to the PIC code. Then, localized regions of a global MHD plasma simulation can be followed with PIC to study the microphysics of the processes of interest \cite{2014JCoPh.268..236D,2013PhPl...20f2904H}. Such a model was used to study active regions in the solar corona \cite{2013ApJ...771...93B} and planetary magnetospheres \cite[e.g.,][]{2017JGRA..12210318C,2018JGRA..123.3742M}. The most recent development toward a coupled MHD-PIC method is to apply so-called polymorphic computational particles. The latter can be either kinetic or fluid, and can change their morph when necessary \cite{2018FrP.....6..100M}. 

\jn{Notwithstanding the efforts to combine various plasma simulation methods to reach into macrophysical scales, another}
novel possibility that was mentioned above is to avoid the coupling between different approaches and study multi-scale problems within a PIC model that is capable of providing kinetic electron information but does not need to resolve all electron scales. Such semi-implicit PIC method has recently been proposed in \cite{2017JPlPh..83b7005L} and applied to several test cases. 

\subsection{Establishing a shock}
In PIC simulations, one can initiate collisionless \mpo{shocks} in a number of ways, among them the injection method \citep{burgess}, the flow-flow method \citep{1992JGR....9714801O}, the relaxation method \citep{leroy1981,leroy1982}, and the magnetic-piston method \citep{1992PhFlB...4.3533L}. The injection method uses a plasma beam that is reflected off a conducting wall. \mpo{This is a computationally efficient method, but the reflecting wall corresponds to an ideal contact discontinuity that should be extended on the plasma scale. It is quite conceivable that there is an initial unphysical reflection of particles and/or electromagnetic waves, arising from, e.g., imperfect balancing of $\mathbf{\nabla}\times (\vecv\times\vecb)$ in simulations of oblique or perpendicular shocks. }   
In the flow-flow method two counterstreaming plasma beams are continuously injected at the sides of the computational box  that collide and eventually form a system of two shocks separated by a discontinuity. \mpo{This method is computationally expensive, but can involve two different shocks in one simulation. The discontinuity is kinetically modelled, and care must be exercised to distinguished particles reflected off the discontinuity from those interacting with the shock.} 
The relaxation method uses a simulation box filled with plasma that is separated by a discontinuity into two uniform plasma slabs that are supposed to initially satisfy the shock jump conditions \citep[see also][]{2006EP&S...58E..41U}. \mpo{One difficulty lies in knowing the distribution function of particles prior to conducting the simulation. The properties and extent of the foreshock is particularly critical.} The magnetic piston method applies an electromagnetic-field transient that in the plasma develops into a shock \citep[for a more detailed account of the shock excitation methods see, e.g.,][]{2003LNP...615...54L}. \mpo{This method can realistically describes situations in which Poynting flux injects a lot of energy, e.g. in laser-plasma interactions, but it is unclear how much relaxation time is needed to develop a shock in a statistical steady state.}

\vfill
\newpage
\section{Relativistic magnetized outflows: Pulsar Wind Nebulae}
\subsection{Nonthermal particle acceleration and $\sigma$ problems}
It is widely accepted that relativistic plasma flows emanate from central objects in many astrophysical settings, and pulsars are known to have a~relativistic, magnetized outflow, whose energy is provided by the spin-down power of the central neutron star with a strong surface magnetic field.  The Crab pulsar and its surrounding nebula is one of the nearest-at-hand examples of a relativistic plasma outflow in collisionless plasma system, and due to the interaction of the outflow plasma with its surrounding medium the outflow kinetic energy is released.  In fact, a broadband spectrum extending from the radio band to the X-ray and gamma-ray band is observed from the Crab pulsar wind nebula, and the energy spectrum of the radio band is well approximated by a power-law $F_{\nu} \propto \nu^{-p}$ with spectral index of $p=0.0 \sim 0.3$, suggesting synchrotron radiation.

High energy particles can be generated during the supernova explosion stage, but the synchrotron cooling time of those electrons that emit X-rays and gamma rays is less than the age of the Crab nebula.  Therefore, it is believed that non-thermal high energy particles should be continuously accelerated during the expansion of the nebula by, for example, shock waves and magnetic reconnection, etc., and as the result synchrotron radiation are observed.  Yet the energy transfer mechanism from the central object to the non-thermal particle and the synchrotron radiation remains to be solved.

Another important issue is the so-called ``$\sigma$'' problem.  While the pulsar magnetosphere inside the light-cylinder is commonly believed to be occupied by strongly magnetized outflow plasma due to a strong magnetic field of the neutron star, whose magnetic field magnitude reaches up to almost $10^{12}$ Gauss, observations of pulsar wind nebulae indicate weakly magnetized plasma with non-thermal particles.  Therefore, the efficient energy transfer of the magnetic energy into particle kinetic energy must happen somewhere between the pulsar magnetosphere and the nebula.  The parameter $\sigma$ is a measure to show the relation between the bulk flow energy and the magnetic field energy, defined by,
\begin{equation}
  \sigma = \frac{B^2}{4 \pi N m c^2 \gamma},
  \label{eq:sigma}
\end{equation}
where $B$, $N$, and $\gamma$ are the magnitude of the magnetic field, the plasma density, and the Lorentz factor of the wind speed, respectively.  \mh{Note that $\sigma$ is a Lorentz invariant}. In the pulsar wind $\sigma$ is thought to be large, of the order of $10^6$, but in the pulsar wind nebula $\sigma$ is regarded to be small, around $10^{-3}$ \cite{Arons79,Kennel84}.

It might be simply argued that the termination shock formed at the boundary between the pulsar wind and the nebula, that is located about $10^{17}$ cm from the neutron star for the Crab nebula, can provide the energy conversion from the bulk outflow energy into the non-thermal particles.  However, a serious problem is that efficient particle acceleration by shock waves requires a low-$\sigma$ flow upstream in a weakly magnetized plasma.  In fact, it is argued that a pulsar wind shock with $\sigma = 10^{-3}$ is needed to explain the observed synchrotron radiation in the Crab nebula \cite{Kennel84}.  \mh{As the Alfvenic Mach number is given by $M_{\rm A}=\sqrt{(1+\sigma)/\sigma}$ for a relativistic shock}, a shock with a large $\sigma$, i.e., a low Mach number, is not favorable for synchrotron emission.

In addition, another controversial issue is the geometry of the shock front and the magnetic field.  In the pulsar wind shock, the magnetic field vector would be perpendicular to the normal of the shock front in analogy to the solar wind in Heliosphere, and this topology is called a perpendicular shock.  The controversial issue is whether or not diffusive shock acceleration can happen at a perpendicular relativistic shock. While multiple crossing of a non-thermal particle across the shock front back and forth is required for diffusive shock acceleration, the crossing process will be strongly suppressed for a perpendicular shock, because the cross-field diffusion is in general weak.  However, it is more or less widely accepted that some sort of the shock acceleration is operating at the termination shock, and that the shock plays an important role in generating non-thermal particles downstream by dissipating the bulk-flow energy upstream.  

\subsection{Striped wind with magnetic reconnection}
A promising process for leading to a low-$\sigma$ wind would be magnetic dissipation/magnetic reconnection in a current layer, where the magnetic field polarity is switched and the electric current is concentrated.  In a pulsar wind with magnetic field lines stretched by the outflow, the current layer can be formed in the equator, but for an obliquely rotating neutron star, where the magnetic moment of the neutron star is not parallel to the pulsar rotation axis, the current layer appears periodically in a finite equatorial zone with a cone angle of $2 \theta$, where $\theta$ is the angle between the rotation axis and the magnetic moment of the neutron star \cite{Michel71, Usov94}.

Before arguing magnetic energy dissipation in a pulsar wind, it would be better to pay attention to the radial evolution of the current layer. As the pulsar wind propagates outward, the amplitude of the magnetic field decays as $1/r$, whereas the column density of the charged particle confined in the current layer decreases as $1/r^2$.   If no magnetic-field dissipation occurs, and if the thickness of the current sheet does not change, beyond some distance the charge particles cannot support the electric current maintained by the surrounding magnetic field, namely the so-called charge starvation will happen.  Therefore we would expect that the current sheet begins to bring additional plasma from the surrounding magnetized plasma through the process of magnetic reconnection/magnetic-field annihilation.  

Another issue in an expanding pulsar wind that we need to pay attention to is the time scale required for the nonlinear evolution of reconnection.  Roughly speaking, the time scale of reconnection is not shorter than $L/c$, where $L$ and $c$ are the thickness of the current sheet and the speed of light (or the Alfven speed in a relativistic regime), respectively. With increasing distance, $r$, from the central neutron star, 
\mh{the current sheet layer will expand with scale dependence $L \sim r (L_0/r_0)$, where $\theta_0=L_0/r_0$ is the opening angle of the beginning of the current sheet.  On the other hand, the elapse time of the wind in the proper frame (i.e., in the plasma frame) is $r/(\gamma c)$.  Therefore, the condition required for the evolution of reconnection should read $\theta_0 \gamma < 1$,}
and this condition is not obviously satisfied in a pulsar wind because $\gamma \gg 1$. Moreover, if reconnection happens during the expansion, the current sheet is heated and work is done. Then the wind can be accelerated and the wind Lorentz factor $\gamma$ becomes large.  This process dilates the elapse time given the above condition \cite{Lyubarsky01,Kirk03}.  However, the nonlinear evolution of magnetic reconnection/magnetic-field annihilation depends on the energy dissipation process coupled with the macroscopic expanding current structure and the microscopic magnetic diffusion process, and many unresolved issues remain to be solved.  The puzzle of the energy conversion of a Poynting-flux dominated wind to a kinetic-energy-flux wind may be a ubiquitous problem for any spherical relativistic flow that contains a toroidal magnetic field.

\begin{figure}[tb]
\begin{center}
\includegraphics[width=14cm]{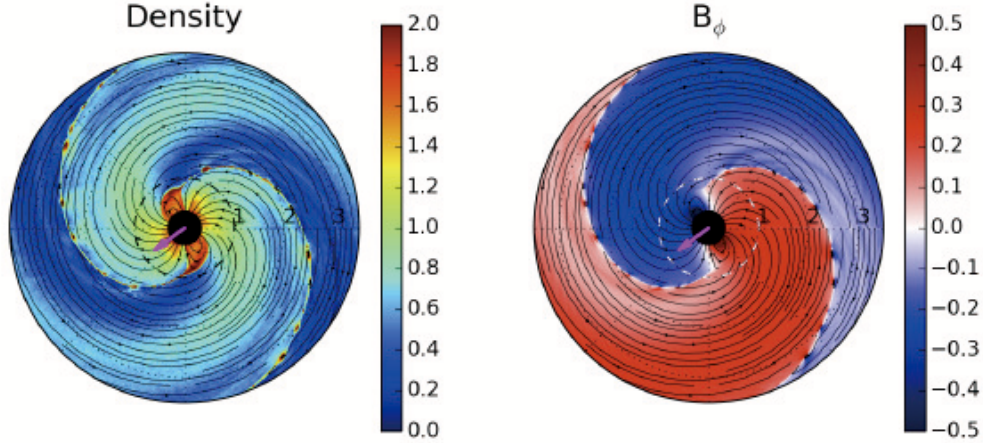}
\end{center}
\caption{Striped wind structure obtained by PIC simulation.  Left is the plasma multiplicity $n/n_{GJ}$ in logarithmic color scale, and right is the toroidal magnetic field $(r/r_*) (B_{\phi}/B_*)$. Reproduced with permission from \citet{Cerutti17}. \label{fig1}}
\end{figure}

\mh{There are some significant advances in our understanding of the pulsar magnetosphere over the last decade by using the particle-in-cell (PIC) simulations \cite[e.g.][]{Chen14,Belyaev15,Cerutti17,Philippov18,Kalapotharakos18}.  The global dynamics of the pulsar magnetosphere has been addressed by paying attention to the current and charge distribution, the role of pair plasma production, dissipation processes, electromagnetic emission and so on.
In order to resolve the $\sigma$ problem of the global current-sheet dissipation in a striped pulsar wind structure,}
\citet{Cerutti17} investigated a striped wind in the equator of an inclined split-monopole by a two-dimensional PIC code.  As initial condition, magnetic field lines are assumed to be purely radial out of a split-monopole, and they start to co-rotate with the central neutron star.  As shown in Figure \ref{fig1}, shortly after commencing co-rotation the magnetic field lines are wound and form two nested Parker Archimedean spirals, and the magnetic-field structure collapses into a closed magnetosphere inside the light cylinder and an open magnetosphere extended beyond the light cylinder. In this standard pulsar magnetospheric structure, sporadic formation of current-sheet tearing close to the light cylinder has been observed, hence magnetic reconnection occurs actively and a chain of plasmoids propagating outward is formed.  Then the reconnection started around the light cylinder continues to develop with increasing distance from the pulsar, $r$, and the broadening of the current sheet thickness was found to be proportional to $r$ in the PIC simulation.  In addition, they found the radial dependence of the drift velocity carrying the electric current is weak.  From the Ampere's law, they found that the distance at which the magnetic energy completely dissipated to a low-$\sigma$ wind can be given by,
\begin{equation}
 r_\mathrm{diss}= \pi \Gamma_\mathrm{lc} \kappa_\mathrm{lc} R_\mathrm{lc},
\end{equation}
where $\Gamma_\mathrm{lc}$ and $\kappa_\mathrm{lc}$ are the bulk Lorentz factor and the plasma multiplicity at the light cylinder. For the Crab nebula case, the multiplicity, $\kappa_\mathrm{lc}$, is believed to be of the order of $10^3 \sim 10^4$ \cite{Timokhin13}, and the bulk Lorentz factor around $\Gamma_\mathrm{lc} \le 10^2$.  Then the dissipation distance, $r_\mathrm{diss}/R_\mathrm{lc}$, can be estimated as $10^5 \sim 10^6$, but the distance of the termination shock is known to be $r_{term}/R_\mathrm{lc} \approx 10^9$.  Therefore, the dissipation distance, $r_\mathrm{diss}$, becomes much less than the radius of the termination shock, and this nonlinear simulation suggests that the stripes could dissipate far before the wind reaches the termination shock in the Crab nebula.  The reason why the $\sigma$ problem does not appear is probably that relativistic magnetic reconnection switches on near the light cylinder, and that the reconnection could effectively dissipate the stripes in the inner magnetosphere.

\subsection{Particle acceleration by magnetosonic shocks \label{msshocks}}
It is widely believed that diffusive shock Fermi acceleration can produce nonthermal power-law energy spectra with $dN/dE \propto E^{-s}$ with the ubiquitous slope $s = 2$ determined only by the shock compression ratio, by particle scattering back and forth across the shock front \cite{2006ApJ...645L.129L}. The outflows from pulsars, however, are believed to be highly relativistic with a bulk Lorentz factor $\Gamma \gg 1$, and these shocks are expected to have a large magnetic field component perpendicular to the shock normal direction in the shock front frame. The standard diffusive Fermi shock acceleration may not necessarily work in such a shock geometry, which is so-called perpendicular shock, because the particle cannot be easily diffused across the magnetic-field lines \cite{2019PhRvL.123c5101L,2019PhRvE.100c3209L,2019PhRvE.100c3210L}.  

Instead of the diffusive shock acceleration model, it has been argued that the wave-particle interaction of collective plasma phenomena can lead to efficient particle acceleration. \citet{Hoshino92} proposed that the synchrotron resonance process can efficiently accelerate electrons and positrons near the shock front region, if heavy ions have contaminated the pulsar wind with its dominant constituent of pair plasma \citep{Amato06}. As a part of the incoming particles can be ubiquitously reflected off the shock front back into the upstream region in the form of gyromotion, and the gyrational energy around the magnetic field can be released by the synchrotron maser instability \cite{Zheleznyakov72}.  Not only the leptonic component of positrons and electrons but also the hadronic component of heavy ions drive the emission of X-mode electromagnetic waves by the synchrotron maser instability \cite{Hoshino91}, and the X-mode waves generated by the heavy ion population can be absorbed by both positrons and electrons.  In one-dimensional PIC simulations, the time scale of the synchrotron maser process is of order of tens of the gyro-period \citep{Hoshino92,Amato06}, which is quite short compared to the time scale of the conventional diffusive shock acceleration.  They also found that the pair plasma generates hard energy spectra downstream of the termination shock, if the number density of the heavy ion contamination is larger than the mass ratio of heavy ion to electron/positron pairs. However, the contamination of heavy ions may be still an open question.

\subsection{Particle acceleration in striped wind}
As an alternative model of the nonthermal particle acceleration in the pulsar wind nebula, \citet{Nagata08} and \cite{Sironi11} discussed the interaction of the striped wind and the magnetosonic shock front.  The interaction of the fast mode shock and the tangential discontinuity may lead to additional particle acceleration and/or magnetic energy dissipation: for example, the interaction of shock and discontinuity may generate another magnetosonic wave propagating downstream.  The collision of the plasma sheet (i.e., tangential discontinuity) with the shock front may drive reconnection, which in turn leads to a rapid magnetic energy dissipation.

\citet{Nagata08} studied with a one-dimensional PIC simulation the interaction of the magnetosonic shock and the plasma sheet for a striped wind with relatively low $\sigma=0.1$. They found that after the interaction of the tangential discontinuity both a large amplitude magnetosonic wave and a tangential discontinuity are formed in the downstream region, and high energy particles can be accelerated in association with those structures, if the thickness of the tangential discontinuity in the upstream region is larger than the electron inertial length.  In addition, the separation distance between the striped wind/tangential discontinuity is a key agent, and the separation distance in the Crab pulsar wind would be estimated as the light cylinder of $r_\mathrm{sep} = c/\Omega \approx 1600$ km, while the typical gyro-radius $r_g$ in the downstream plasma may be estimated as $r_g = c/(eB/\gamma m c) \approx 10^8$ km, where $\gamma \approx 10^6$ is the wind bulk Lorentz factor. In this parameter regime with $r_g \gg r_\mathrm{sep}$, the particles originally trapped inside the discontinuity in the upstream region cannot be trapped downstream and could travel over the striped structures.  Therefore, the magnetic energy could be quickly dissipated by the finite Larmor radius effect.

\begin{figure}[tb]
\begin{center}
\includegraphics[width=14cm]{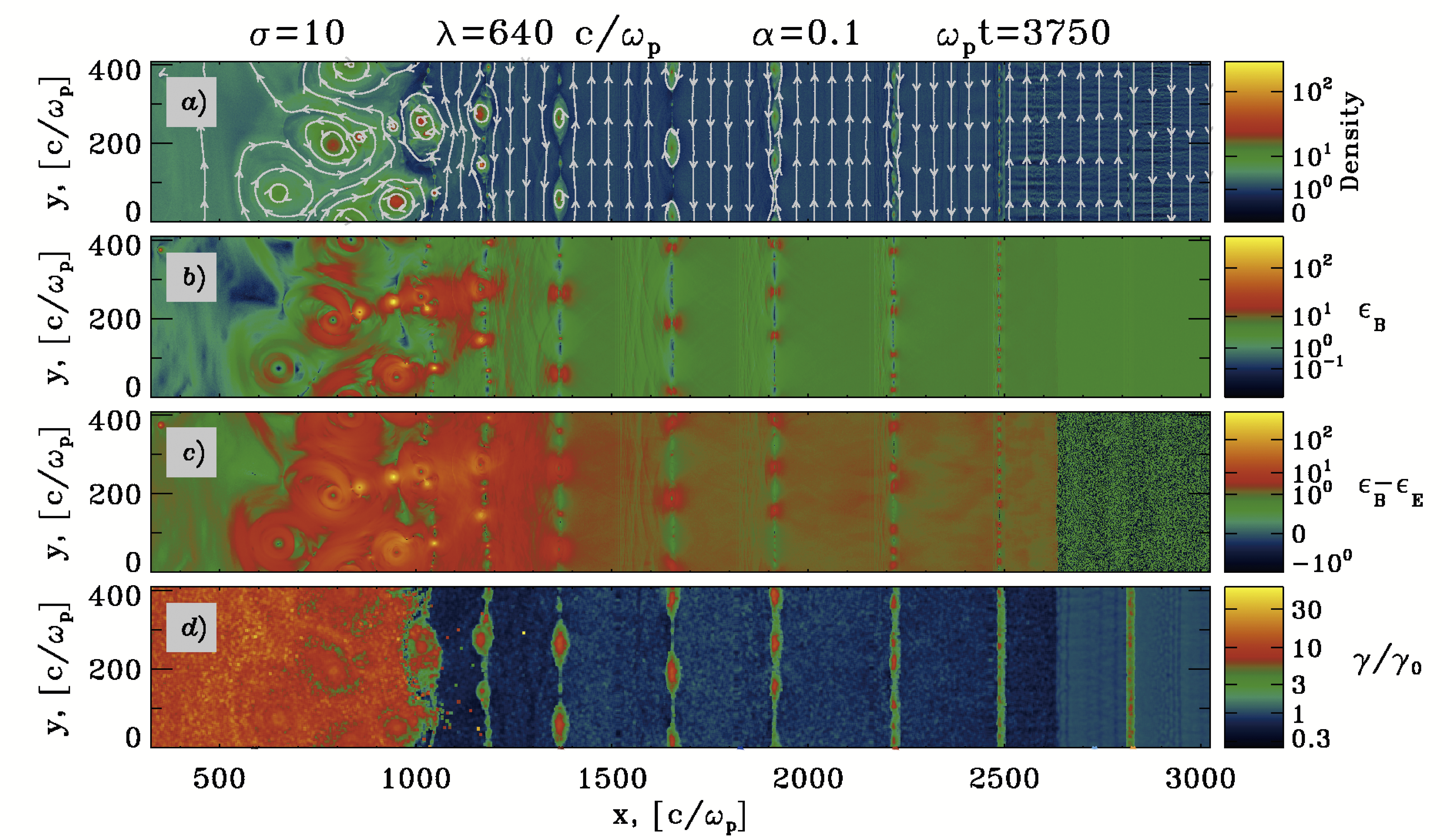}
\end{center}
\caption{Striped wind structure around the termination shock. A relativistic strip wind that was injected from the right-hand boundary interacts with the termination shock around $x=2600$, and then magnetic reconnection is initiated.  From the top, the panels show the plasma density with magnetic field lines (white lines), the magnetic energy density, the difference of the magnetic energy density and the electric energy density, and the mean kinetic energy per particle in units of Lorentz factor. Reproduced with permission from \citet{Sironi11} (\textcopyright  AAS). \label{fig2}}
\end{figure}

\citet{Sironi11} investigated the interaction of the striped wind and the shock using two- and three-dimensional PIC simulations for relatively large $\sigma =10 \sim 100$, and studied whether or not the Poyinting flux-dominated striped wind can dissipate the magnetic field energy.  As shown in Figure \ref{fig2}, they found that although the shock compression is weak due to a large $\sigma$ wind, magnetic reconnection can be driven when the striped wind just attached to the weak fast mode shock front at $x/(c/\omega_p) = 2600$, and then in association with nonlinear evolution of magnetic reconnection,
the formation of magnetic islands and their interaction of coalescence leads to a rapid energy dissipation and particle acceleration in the downstream region.  If the separation distance, $r_\mathrm{sep}$, is large enough, the nonthermal energy spectrum becomes broader, and the power-law energy tail shows a hard spectrum with its spectral index around $-1.5$, probably reflecting the mechanism of relativistic reconnection acceleration \cite{Zenitani01}.

\subsection{Particle acceleration in magnetic reconnection} 
We have already mentioned that magnetic reconnection plays an important role in various phenomena, but in this section, we quickly review the basic process of reconnection by focusing on particle acceleration \cite{Hoshino12b,Blandford17}. The reconnected magnetic field line exerts a Lorentz force, and the bulk plasma can be accelerated up to the Alfven speed \mpo{that can be written as function of the magnetization parameter, $\sigma$, as $v_A=c \sqrt{\sigma/(1+\sigma)}$ (cf. Eq.~\ref{eq:sigma}). Hot plasma and} also non-thermal particles can be rapidly generated on the order of the Alfven transit time $\lambda/v_A$, where $\lambda$ is the thickness of the neutral sheet.

Astrophysical applications of reconnection are accretion disks, pulsar winds, AGN jets, magnetars and so on, and the observed non-thermal signatures from these objects are often believed to be at least partially attributed to magnetic reconnection specially in the relativistic regime where the Alfven speed is close to a speed of light.  Among different numerical approaches of simulation, the process of particle acceleration can be treated from first principle by means of kinetic particle-in-cell (PIC) simulations.  Many PIC simulation studies have been carried out so far: earlier studies of relativistic reconnection in two-dimensional space reported that strong non-thermal particles can be quickly generated in and around the magnetic diffusion region, the so-called X-type neutral point, and they lead to the formation of a hard power-law spectrum with $N(\varepsilon) \propto \varepsilon^{-p}$ where $p=1 \sim 2$ \cite{Zenitani01,2009PhRvL.103g5002J}. Decaying turbulence involving reconnection appears to generically lead to power-law spectra of particles, although a fair fraction of the energization happens by stochastic interactions with turbulence \cite{2018PhRvL.121y5101C}. For the $\sigma>>1$ regime, it is shown that the power-law index becomes $p=1$, and most of the released energy is converted into non-thermal particles \cite{Guo14,Sironi14}. In a pair-proton plasma, one finds that the post-reconnection energy is shared roughly equally between magnetic fields, pairs, and protons \cite{2019ApJ...880...37P}. For very high $\beta$, the particle spectrum transitions to a Maxwellian \cite{2018ApJ...862...80B}.

While magnetic reconnection can happen in two-dimensional space where the anti-parallel magnetic fields are included in the simulation plane, the drift-kink instability, which is known as another important magnetic energy dissipation process, can be excited in the out-of-plane direction, i.e., in the plane including the electric current direction \cite{Zenitani05}.  The linear growth rate of the drift-kink instability (DKI) for the anti-parallel magnetic field case is larger than that of reconnection for a thin current sheet, namely the thickness current sheet is less than a gyro-radius of the thermal plasma. However, the growth rate of DKI is slower than that of reconnection for a thick current sheet \cite{Zenitani07}.  During the nonlinear time evolution of DKI, the thickness of the current sheet gets broader as time goes by.  In fact, the time evolution in three-dimensional simulations with a thin current sheet shows the drift-kink unstable current structure in the early stage, and then the reconnection structure with a magnetic island/plasmoid in the later phase after the thickening of the current sheet.  As another aspect, DKI can be easily suppressed by imposing a finite magnetic field parallel to the electric current.  Therefore, in a realistic three-dimensional system with a magnetized plasma with a large $\sigma$, magnetic reconnection is known to produce intense non-thermal particles \cite{Sironi14}.  In addition, the interaction of charged particles with many magnetic islands generated in a large reconnection system can generate efficiently non-thermal particles through the Fermi process \cite{Hoshino12}.

The maximum attainable energy of a charge particle is, in general, limited by the size of the acceleration region, $L$, and given by $\varepsilon_{max} < e E L$, where $e$ is the charge of particle and $E$ is the electric field, which can be replaced by the motional electric field defined by $v B/c \sim B$ in a relativistic environment. \citet{2018MNRAS.481.5687P} find that the energy of particles continues to grow, albeit slowly, and the decisive field strength is that in the plasmoids, not the average field that enters the $\sigma$ parameter. However, radiation losses during the particle acceleration may decrease the above simple limit. In the case of synchrotron radiation loss, the balance between the electric acceleration rate and the synchrotron radiation loss rate give the maximum attainable energy as $\varepsilon_{sync} = (9mc^2/4 \alpha_F)(E/B) < 160$~MeV, where $\alpha_F$ is the fine structure constant and $mc^2$ is the rest mass energy of electron.  During energy acquisition by relativistic reconnection, the effect of radiation loss in PIC simulations has been investigated by including the radiation reaction term of Lorentz-Abraham-Dirac form in the Lorentz equation.  Due to radiative cooling, the gas pressure in the plasma sheet is reduced, and then in order to maintain the pressure balance between the plasma sheet and the magnetic field dominated inflow region, fast reconnection can be initiated \cite{2009PhRvL.103g5002J}.  It is also shown that $\varepsilon_{sync} > 160$~MeV can be generated by preferential acceleration in a weak magnetic field region \cite{2013ApJ...770..147C}. 
A high radiation intensity above $1$~MeV can lead to a significant opacity for pair production which adds particles and hence reduces the effective magnetization of the plasma, leading to a lower high-energy cut-off in the emission spectrum \cite{2019ApJ...877...53H}. 

A promising area of magnetic-reconnection research are spatially coupled MHD and PIC simulations which reproduce the dynamics of fully kinetic simulations that Hall-MHD does not capture \cite{2018PhPl...25h2904M}.

\vfill
\newpage

\section{Weakly magnetized relativistic systems}
\subsection{Relativistic shocks}
Nonthermal emission observed from astrophysical sources with relativistic outflows, such as gamma-ray bursts (GRBs), jets of Active Galactic Nuclei (AGN) and microquasars, and Pulsar Wind Nebulae, is frequently modeled as synchrotron or inverse Compton emission of electrons. These electrons are supposed to be accelerated in a first-order Fermi or DSA process at relativistic collisionless shocks, although some studies suggest that magnetic reconnection may be involved as well \cite[e.g.][]{2016MNRAS.462.3325P}.  

At relativistic shocks the bulk flow speed is comparable to the speed of the particles. In effect, the distributions of accelerated particles are highly anisotropic at the shock.
In contrast to the case of nonrelativistic shocks, the DSA process at relativistic shocks is therefore very sensitive to the background conditions of the upstream plasma, such as the Lorentz factor of the bulk flow, the strength of the magnetic field and its orientation with respect to the shock normal, and the structure of electromagnetic turbulence responsible for particle scattering. This has been first demonstrated in the test-particle limit through semi-analytic calculations \cite{1987ApJ...315..425K,1988MNRAS.235..997H,2000ApJ...542..235K,2005PhRvL..94k1102K} and Monte Carlo simulations \cite{2004ApJ...610..851N,2006ApJ...641..984N,2006ApJ...650.1020N,2003ApJ...589L..73L,2004APh....22..323E}.
For quasi-perpendicular superluminal shocks -- the most typical configuration for ultrarelativistic shocks --
these studies showed that particle acceleration 
is very inefficient, and the particle spectra are typically very steep, unless a highly turbulent magnetic field exist near the shock, that is equivalent to very low plasma magnetization \cite{2002A&A...394.1141O,2006ApJ...645L.129L,2006ApJ...650.1020N}. Only under these conditions highly relativistic shocks with $\Gamma_\mathrm{sh}\gg 1$ produce broad-range power-law energy spectra with the ``asymptotic'' index $s=2.2-2.3$, that are compatible with the spectra of synchrotron-radiating electrons derived from modeling of GRB afterglows \cite{1998PhRvL..80.3911B,1999MNRAS.305L...6G,2001MNRAS.328..393A,2003ApJ...589L..73L,2004APh....22..323E,2006ApJ...650.1020N}.     
However, the test-particle treatment assumes {\it ad-hoc} turbulence structures, and an in-depth understanding of relativistic shocks and particle acceleration at them was only recently made possible with large-scale multi-dimensional PIC simulations.

\begin{figure}[tb]
\begin{center}
\includegraphics[width=0.48\textwidth]{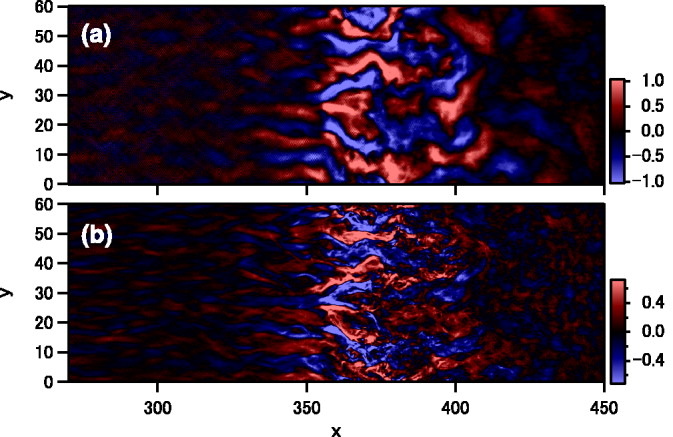} 
\includegraphics[width=0.48\textwidth]{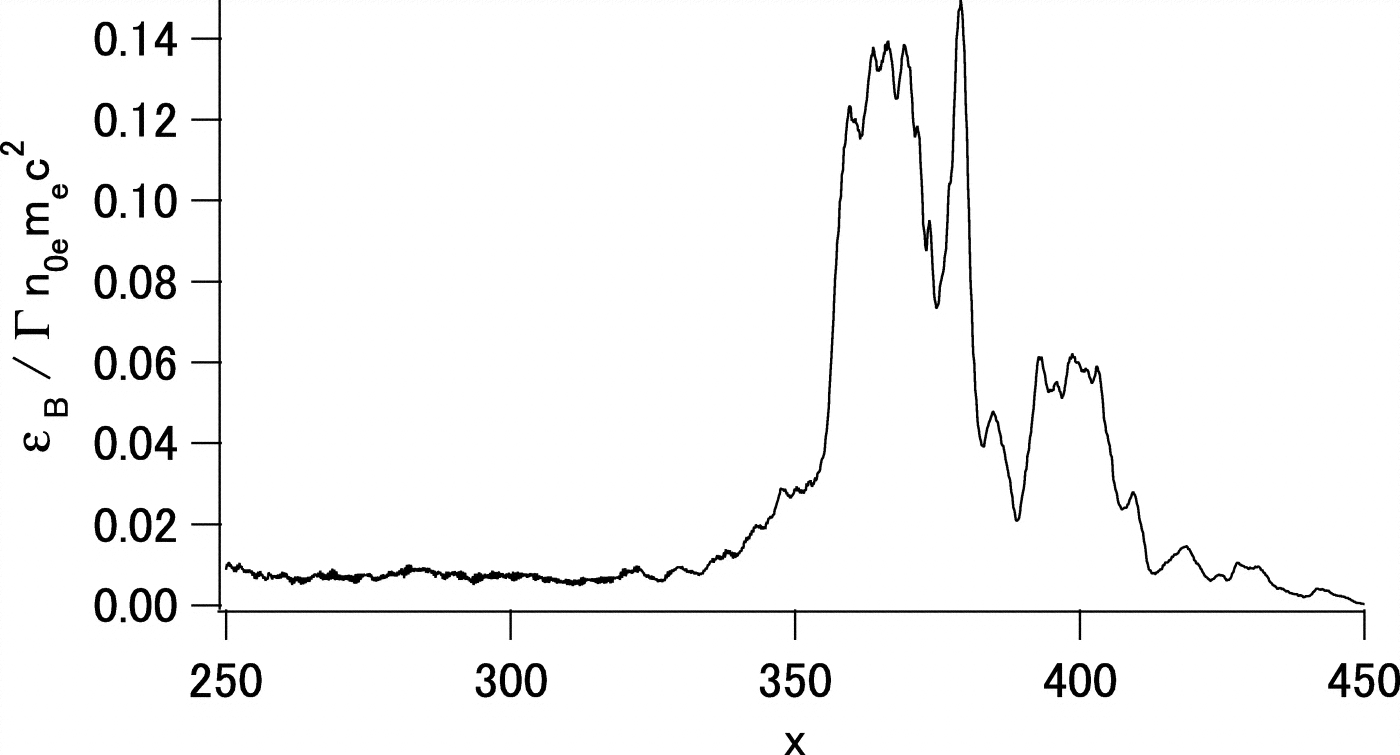}
\end{center}
\caption{The structure of a Weibel-mediated relativistic shock in a 2D PIC simulation of electron-positron plasma. The left panel shows the out-of-plane 
component of the magnetic field, $B_z$ (a), and the associated component of the electric current density, $J_x$, in the direction of the upstream plasma flow (from left to right). The right panel presents the transversely averaged profile of the magnetic energy density normalized to the upstream bulk kinetic energy density. The shock transition extends roughly from $x=350$ to $x=400$ in units of upstream plasma skin depth. 
Reproduced with permission from \citet{2007ApJ...668..974K} (\textcopyright  AAS). \label{Kato_filaments}}
\end{figure}

\mpo{Below a well-defined critical density, collisions do not mediate the shock, and the shock formation builds on plasma instabilities \cite{2018JPlPh..84c9011B}.} Collisionless shocks and the associated electromagnetic fields and accelerated particles self-consistently evolve, driven by the anisotropy of the particle distribution function that is inherent to relativistic shocks. Depending on physical conditions, many competing instabilities can develop at astrophysical shocks \cite{2009ApJ...699..990B}.
It was first recognized by Medvedev \& Loeb \cite{1999ApJ...526..697M}, that in the ultrarelativistic regime the Weibel \mpo{or filamentation instability} has the highest linear growth rate\footnote{\mpo{Technically, the filamentation instability is driven by a beam of particles and the Weibel instability by temperature anisotropy. Both have wavevectors perpendicular to the streaming direction or the high-$T$ axis. It is commonplace to label both as Weibel modes \cite{2005PhRvE..72a6403B}.}}.  
As predicted analytically, \mpo{shown in dedicated studies \cite{2014PhPl...21g2301B}} ,and observed in PIC simulations \cite{2003ApJ...596L.121S,2005ApJ...618..822J}, the instability generates micro-scale, mostly magnetic turbulence, with a characteristic coherence scale of the order of the relativistic skin depth. The fields are organized into filaments associated with electric-current filaments formed by particles of opposite charges and elongated in the direction of the plasma flow. The fields are thus mostly transverse. The instability saturates when the magnetic field becomes strong enough to deflect particles in the filaments. The bulk plasma flow can then be isotropized and slowed-down, providing a means for dissipation of relativistic flows through shocks. 

Weibel-mediated formation of relativistic shock has been demonstrated in numerous PIC simulations for both electron-positron and electron-ion plasma \cite{2004ApJ...608L..13F,2007ApJ...668..974K,2003ApJ...595..555N,2005ApJ...622..927N,2005AIPC..801..345S,2008ApJ...673L..39S,2008ApJ...674..378C,2009ApJ...698L..10N,2011ApJ...739L..42H}. The instability is most robust in unmagnetized plasma, $\sigma=0$ \cite{2005ApJ...623L..89H}.
Figure~\ref{Kato_filaments} (left) illustrates the structure of the shock transition, that corresponds to the peak in the magnetic-field energy (Fig.~\ref{Kato_filaments} (right)). The compression of the plasma is in line with the hydrodynamic jump conditions.
After saturation, current filaments begin to interact with each other and coalesce to form larger-scale quasi-regular structures downstream of the shock. The magnetic field strength can reach a level moderately below equipartition (at $\epsilon_B\sim 10\%-15\%$) with the upstream flow energy at the shock. However, the magnetic field decays downstream of the shock, where the filaments are destroyed \cite{2008ApJ...674..378C}. Although in PIC simulations the early instability growth results from the initial anisotropy, usually set up through counterstreaming plasma beams, later the shock is a self-sustained structure, in which the free energy for the Weibel instability is provided by the incoming flow of the upstream plasma and the beam of hot particles reflected from the shock back into the upstream region. 
The upstream electron/positron plasma is also heated on the way towards the shock. This is due to small-scale transverse electric fields in the shock precursor generated between the filaments and associated with space-charge effects \cite{2006PhRvL..96j5002T}. Electrons (positrons) oscillating in the electromagnetic fields of the growing and merging current filaments can be efficiently heated. In electron-ion plasma this mechanism was described by \citet{2004ApJ...617L.107H}, and the electrons were shown to be energized close to equipartition with the ions before they reach the downstream region \cite{2008ApJ...673L..39S}. When they enter the shock, the relativistic mass of the electrons is thus comparable to that of the ions, which renders similar the physics of ultrarelativistic electron-positron and electron-ion shocks~\cite{2011ApJ...726...75S,2013ApJ...771...54S}.

\begin{SCfigure}[1.2]
\includegraphics[width=7.5cm]{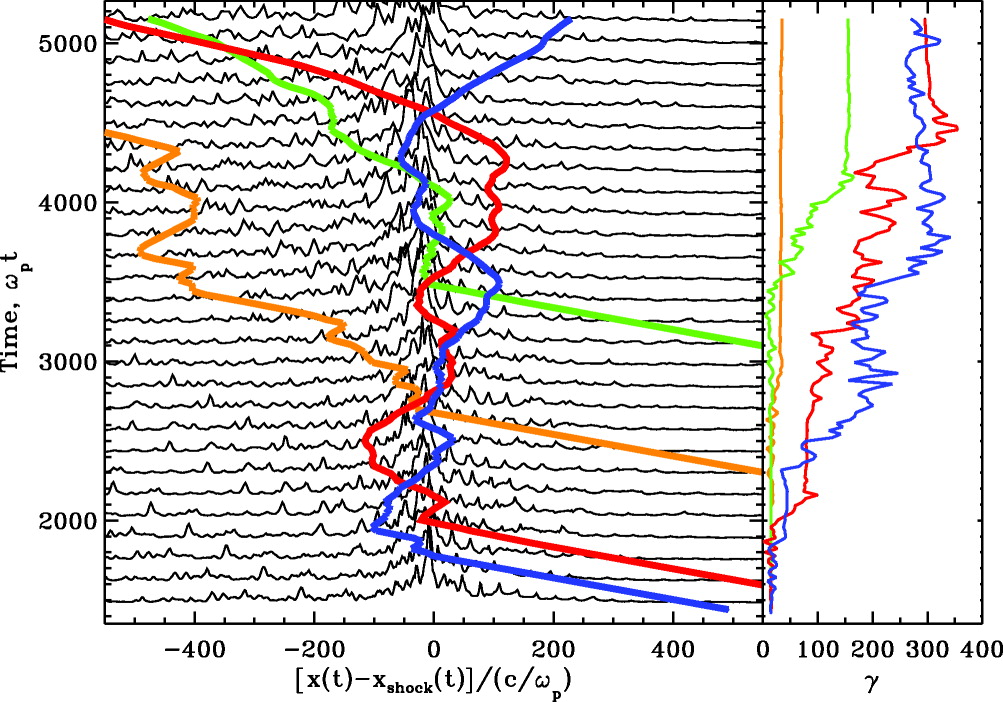}
\caption{Illustration of Fermi-like particle acceleration at a ultrarelativistic shock in pair plasma. Left panel: Horizontal position as a function of time for four representative particles (colored lines) overplotted on transversely averaged profiles of magnetic energy (gray lines). Right panel: the energy of the four particles (shown with corresponding color lines), likewise as function of time. All horizontal positions are shifted by $x_{\mathrm{shock}\,}( t)$  to align them with the shock location. All quantities are measured in the downstream frame. Reproduced with permission from \citet{2008ApJ...682L...5S} (\textcopyright AAS).
}
\label{Fermi-like}
\end{SCfigure} 

Ultrarelativistic unmagnetized shocks have been demonstrated to be efficient particle accelerators. As first shown by \citet{2008ApJ...682L...5S} for pair shocks and \citet{2009ApJ...695L.189M} and \citet{2011ApJ...739L..42H} for electron-ion shocks,
accelerated particles gain their energies through multiple reflections between the upstream and the downstream region of the shock (see Fig.~\ref{Fermi-like}), where they scatter on micro-scale turbulence. At each reflection, the energy gain is $\Delta E\propto E$, as expected for a first-order Fermi process. The final particle energy is acquired in the evolving shock structures after only a few collisions, and the scattering time between collisions increases as the particles are accelerated \cite{2009ApJ...695L.189M,2011ApJ...739L..42H}. The process is thus termed a Fermi-like acceleration.
The resulting particle energy spectra downstream of the shock have power-law tails with spectral indices between $s=2.3$ and $s=2.6$. The tail contains $\sim 1\%-3\%$ of the particles and carries about$10\%$ of the plasma energy. The maximum Lorentz factor grows in time as $\gamma_{\rm max}\propto t^{1/2}$, slower than expected in the Bohm limit ($\gamma_{\rm max}\propto t$), which reflects the nature of the small-scale turbulence seeded by the Weibel instability \cite{2013ApJ...771...54S}. The acceleration process thus operates in a manner that has been earlier described in analytical and Monte Carlo simulations \cite[see also above]{2001MNRAS.328..393A,2006ApJ...645L.129L,2006ApJ...650.1020N}.

The PIC results permit a qualitative estimate of the maximum energy to which particles can be accelerated at the external shocks of GRBs. Because proton acceleration is too slow, their maximum energy of $10^{16}$~eV is too low to explain ultra-high-energy cosmic-rays (UHECRs) reaching energies beyond $10^{20}$ eV \cite{2013ApJ...771...54S}. On the other hand, electrons accelerated at these shocks can account for the observed MeV-GeV emission of the early GRB afterglows, if they produce synchrotron radiation. 

A new development is the implementation of radiation techniques into PIC codes (see Section~\ref{method_impl}), whioch allow the calculation of photon emission spectra from first principles. As unmagnetized relativistic shocks are mediated by the Weibel instability that generates strong small-scale magnetic turbulence, it has been proposed that electron emission at shocks would result from the so-called ``jitter'' radiation, rather than the synchrotron process \cite{2000ApJ...540..704M,2006ApJ...637..869M,2006ApJ...638..348F}.
Photon spectra obtained from PIC simulations of the Weibel instability show agreement with the ``jitter'' emission in the initial linear phase of the instability, taking place in the foot of the shock precursor \cite{2010ApJ...722L.114F,2011ApJ...737...55M}. On the other hand, closer to the shock, where the magnetic filaments merge and grow in size, the radiation spectrum resembles ordinary synchrotron radiation. Spectra consistent with synchrotron emission have been calculated for the close downstream region of the shock in PIC simulations that self-consistently followed the shock formation process \cite{2009ApJ...707L..92S}.
However, significant contributions to the emission should be expected from the extended regions downstream of the shock. This includes the far downstream region, in which the magnetic field decays and so radiation may again enter the ``jitter'' regime \cite{2011ApJ...737...55M}. As the statistical structure of the magnetic field far behind the shock is still not known (see below), the exact nature of the photon emission from unmagnetized shocks remains to be resolved. 

The level of sub-equipartition magnetic field, $\epsilon_B\simeq 0.1$, the considerable fraction of energy transferred to the electrons downstream, and the power-law particle spectra in an extended energy range, all support the canonical picture of GRB afterglows as being manifestations of ultrarelativistic shocks propagating in a very weakly magnetized medium \cite[see, e.g.,][]{2004RvMP...76.1143P}. However, as noted, the magnetic field produced in the shock is small-scale, of the order of plasma skin depth, $\lambda_{\rm p}$, and quickly decays downstream of the shock. As the radiation modelling of GRB observations implies near-equipartition fiels across a macroscopic region of $\sim 10^{10}\lambda_{\rm p}$, the self-consistent generation of such fields still needs to be demonstrated. 
\citet{2009ApJ...693L.127K} showed that particles undergoing continued acceleration at the shock drive magnetic-field growth on progressively larger scales and in a growing region around the shock, demonstrating at the same time nonlinear feedback of the accelerated particles on the shock structure. Their simulations did not converge to a steady state, implying that the process may proceed further, in line with unlimited acceleration in unmagnetized plasma, $\sigma=0$ \cite[see above,][]{2013ApJ...771...54S}. A caveat is that, if weak magnetic fields are present in the pre-shock medium, particle energy growth saturates at $\gamma_{\rm max}\propto \sigma^{-1/4}$, because the self-generated turbulence is confined to increasingly smaller region around the shock, and there is no magnetic turbulence that can scatter highest-energy particles \cite{2013ApJ...771...54S}. Consequently, magnetic-field generation by accelerated particles may not work in these conditions.   

The properties of shock formation and efficient particle acceleration processes as described above hold for shocks in electron-positron plasma with upstream magnetization $\sigma \lesssim 10^{-3}$, and in electron-ion plasma for $\sigma \lesssim 3\times 10^{-5}$, regardless of shock obliquity, $\thbn$ \cite{2011ApJ...726...75S,2013ApJ...771...54S}. However, \mpo{for larger magnetizations synchrotron maser modes are induced that speed up the shock formation through magnetic reflection \cite{2016NJPh...18j5002S}.} The shock structure and particle acceleration strongly depend on $\thbn$ \cite{2009ApJ...698.1523S,2011ApJ...726...75S}. As shown by \citet{2009ApJ...698.1523S} for electron-positron plasma,
the shock transition in parallel shocks can be mediated by the Weibel filamentation instability up to a moderate plasma magnetization, $\sigma \lesssim 1$.
This instability also mediates the formation of oblique subluminal shocks, although with increasing shock obliquity the process of magnetic reflection of the incoming particles off the shock-compressed magnetic field \cite{1988PhFl...31..839A} becomes important and dominates for $\thbn$ approaching $\thbncrit$.
In the regime of superluminal shocks ($\thbn\geq \thbncrit$) particle acceleration was found to be inefficient, because such shocks cannot self-consistently produce large-amplitude turbulence to allow significant cross-field diffusion, as noticed already in Monte-Carlo studies \cite{2002A&A...394.1141O,2006ApJ...650.1020N}. Acceleration is possible at subluminal shocks with $\thbn\lesssim 34^{\rm o}/\gamma_0$, where $\gamma_0$ is the Lorentz factor of upstream flow, and its efficiency increases from $\sim 4\%$ of the flow energy at quasi-parallel shocks ($\thbn\approx 0$) to more than $10\%$ for obliquity angles close to $\thbncrit$.
At the same time, the high-energy spectral tail becomes harder, in agreement with semi-analytic and Monte Carlo simulations \cite[see, e.g.,][]{2004ApJ...610..851N,2004APh....22..323E}.
The increase in the acceleration efficiency is associated with a~growing importance of SDA that operates much faster than DSA. The~Fermi-like process energizes particles at quasi-parallel shocks, in which upstream scattering is provided by oblique waves that are self-consistently driven by shock-reflected particles. 
The above picture holds also for shocks propagating in electron-ion plasma \cite{2011ApJ...726...75S}. A notable difference in this case is that at subluminal shocks ions are more efficiently accelerated than electrons. The nonthermal tail in the ion spectra contains approximately $30\%$ of ion energy and its slope is $s\simeq 2.1$. Nonthermal electrons carry about $ 10\%$ of energy and their power-law spectra are steep with slope $s\simeq 3.5$. The lower acceleration efficiency reflects the stronger binding of electrons to the magnetic-field lines, as electrons despite their heating in the shock precursor enter the shock with lower energy than ions. 

Radiation from blazar jets and the prompt emission of GRBs are usually attributed to electrons accelerated to form nonthermal spectra at relativistic internal shocks. Mildly relativistic shocks are also thought to be hosted in the hot spots of AGN jets. As these shocks are most probably magnetized and quasi-perpendicular (superluminal), the expected inefficiency of particle acceleration remains to be in tension with those models.

PIC simulation results confirm the finding of earlier Monte-Carlo studies that the DSA process should not be considered as the main mechanisms for production of the observed radiating electrons \cite{2006ApJ...650.1020N}. Conversely, if the emission models require particle acceleration through the Fermi process at relativistic shocks, they should involve only nearly parallel and/or weakly magnetized shock conditions.

As discussed in Section~\ref{msshocks}, the synchrotron maser instability is a significant dissipation mechanism at relativistic superluminal magnetized shocks. In the context of particle acceleration at AGN and GRBs, it has been proposed by \citet{2008ApJ...672..940H} that electrons and ions can be quickly accelerated through wakefields that are generated when precursor waves propagate into the upstream electron-ion plasma \cite{2006ApJ...652.1297L}. In fact, the wakefield acceleration (WFA) has been proposed as a model of UHECR production in GRBs \cite{2002PhRvL..89p1101C}. Recent high-resolution PIC simulations of ultrarelativistic pair shocks by Iwamoto et al. \cite{2017ApJ...840...52I,2018ApJ...858...93I} showed that the precursor wave emission is inherent in realistic magnetized shocks and persists even at lower magnetizations, in which the Weibel instability dominates. The wave amplitude is sufficiently large \cite{2008ApJ...682L.113K} for WFA to operate at highly-relativistic electron-ion shocks.  
 
\subsection{Shear flows}
Active Galactic Nuclei and Gamma-ray bursts feature collimated relativistic outflows that are likely embedded in a slower wind \citep[e.g.][]{2012SSRv..173..309B}. Unavoidably there will be shear between the fast jet and the slower sheath around it that may trigger instabilities and also provide particle acceleration \citep{2006ApJ...652.1044R}.

\citet{2012ApJ...746L..14A} presented the first 3D simulation of the kinetic Kelvin-Helmholtz (kKH) instability in unmagnetized shear flows. The initial evolution involves the formation of electron current filaments. Eventually a strong large-scale DC magnetic field is generated, that is a few tens of electron skinlengths thick and extends of the entire length of the shear surface.  This kinetic feature arises from the leakage of electrons of one flow into the other across the shear layer, which ions cannot undergo on account of their inertia, and so it would not be captured in MHD simulations \citep[e.g.][]{2009MNRAS.393.1141M,2011ApJS..193....6B}. For electron-positron plasma the structure is similar to that produced by the Weibel instability, consisting of alternating currents and magnetic fields \citep{2014ApJ...793...60N}. The ratio of the energy density of magnetic field to that of particles matches the values inferred from radiation modelling of GRB afterglows, $\epsilon_\mathrm{MF}/\epsilon_\mathrm{P} \approx 3\cdot 10^{-3}$ \citep{2002ApJ...571..779P}. In the simulations of \citet{2013PhRvL.111a5005G} the lifetime of the DC magnetic field at the shear interface reached a few hundred ion inertial times, which is likewise commensurate with that required in radiation modelling of GRBs. The structural modification imposed by the kKH instability is of essential importance for the injection of particles into shear acceleration, as one also observes particle energization at the shear boundary \citep{2013ApJ...766L..19L}. Comparing in their simulations various compositions of the plasma (electron-positron, electron-ion, and mixtures of the two), \citet{2013ApJ...779L..27L} find that only for a hybrid composition one finds the power-law particle spectra that would lead to the observed power-law radiation spectra. As the energy source for the generation of magnetic field and the acceleration of particles is the kinetic energy of the shear flow, the operation of the kKH instability will invariably lead to a deceleration of the flow \citep{2016PPCF...58a4025A}.

The kKH instability is not the only instability that operates at the shear layer. \citet{2015PhRvE..92b1101A} observe a new kinetic mode in their simulations that they refer to as the Mushroom instability on account of the mushroom-like structures that emerge in the electron density. Linear theory verifies that this mode can grow faster than the kKH instability, even if the velocity shear is initially smeared over a length scale $L_\mathrm{s}$. The growth rate of the Mushroom instability decreases with increasing $L_\mathrm{s}$, but for relativistic flows remains higher than that of the kKH instability for the same $L_\mathrm{s}$.

\begin{figure}[tb]
\begin{center}
\includegraphics[width=\textwidth]{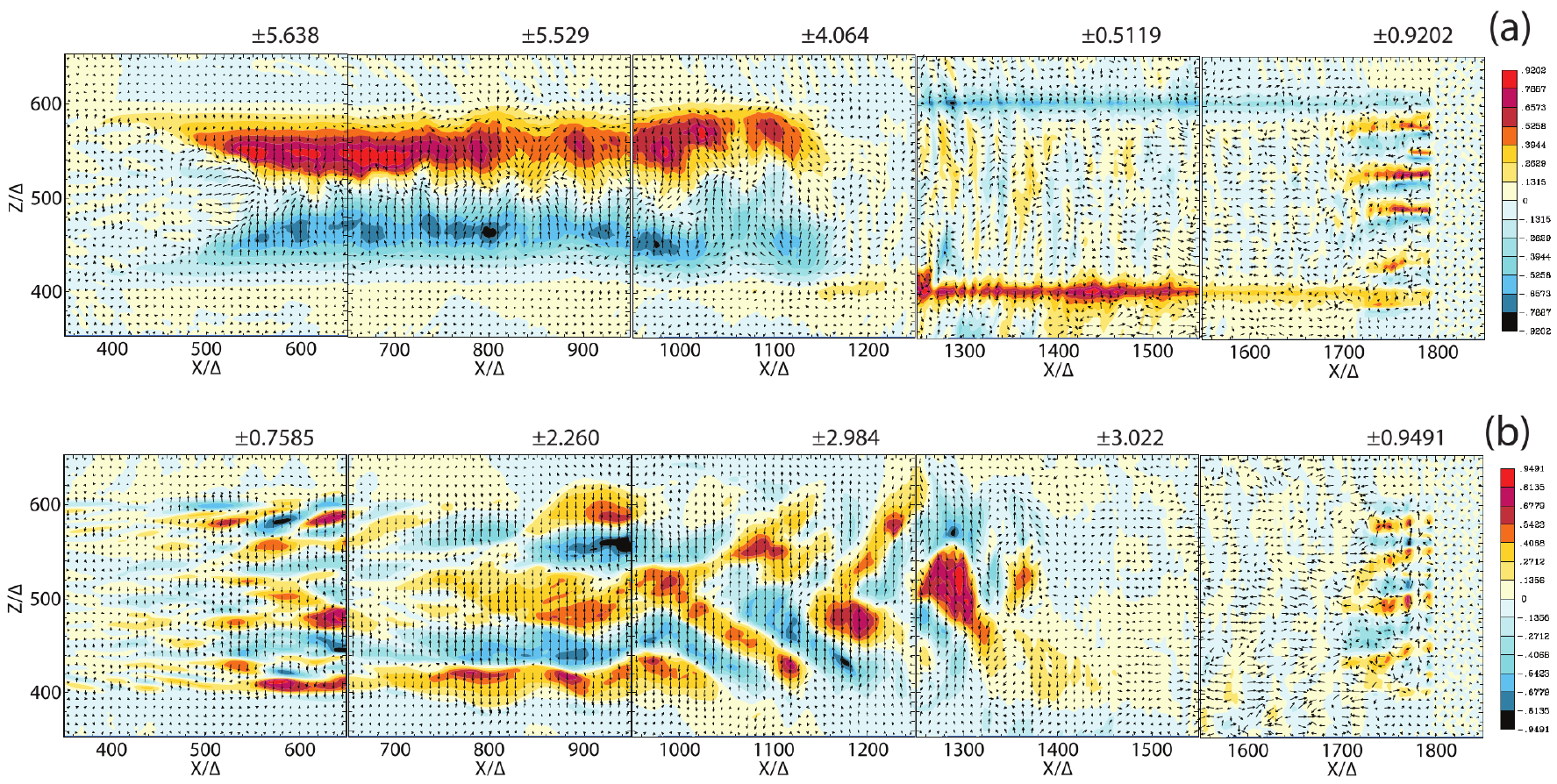}
\end{center}
\caption{2D slices of By for electron-ion plasma (a, upper panel) and (b, lower panel) pair plasma at time $t=1700\,{\ompe}^{-1}$. Arrows show ${E}_{x,z}$. The color scale at the right is for only the rightmost panel (region $1550 < x/\Delta < 1850$). The maximum and minimum values in each of the five jet regions are indicated at the upper right. Reproduced with permission from \citet{2016ApJ...820...94N} (\textcopyright  AAS). \label{Nish_f11}}
\end{figure}

The 3D simulations of \citet{2016ApJ...820...94N} involve the penetration of background plasma by a relativistic jet, and so they cover both the compression of plasma at the head of the jet and the shear at the sides, although the simulation box is too short to permit the leading shock to fully develop. For pair plasma they observe strong mixing of jet and background electrons, and the combined action of the kKH, Mushroom, and Weibel instabilities creates density fluctuations. Figure~\ref{Nish_f11} demonstrates that for electron-ion plasma the jet electrons are collimated by strong toroidal magnetic
fields generated by the Mushroom instability that change polarity at some distance along the jet.

Carefully modeling the radiation output of shear flows, \citet{2017PhRvE..96a3316P} demonstrated that the expected polarization signature of the linear phase of the kKH instability differs from that produced in later stages of the evolution, which may offer a means to experimentally deduce the presence of shear flows. In general, shear-flow scenarios such as the spine-sheath concept of AGN jets may have beaming patterns of radiation that vary between electron-ion jets and pair-plasma flows. Shear flows can in particular emit radiation in the forward direction that has a very hard spectrum compared to that seen at larger aspect angles, which may alleviate the need to postulate a very high minimum electron Lorentz factor in AGN radiation modeling \citep{2017ApJ...847...90L}. A~number of aspects in the evolution of the kKH instability at shear interfaces depend on the Lorentz factor of the flow and hence may be different for GRBs and AGN with typical Lorentz factors $\Gamma\approx 300$ and $\Gamma\approx 10$, respectively, and so one should not expect the same radiation signature for GRBs and AGN \citep{2018ApJ...854..129L}.

\subsection{TeV pair beams}
TeV-scale gamma rays are subject to absorption by pair production following collisions with extragalactic background radiation in the Optical or near-infrared \citep{Gould66}. Due to the inverse Compton scattering, these pairs will
emit secondary photons in the GeV band \citep{1994ApJ...423L...5A}. The observational analysis indicates, however, that the measured gamma-ray signal in the GeV energy band is smaller than the predicted cascade emission assuming that the pairs lose their energy only by inverse-Compton scattering \citep{Neronov10}. Thus, some other dissipation processes must be in play. 

One possible explanation relies on the existence of a large-scale hypothetical magnetic field in cosmic voids \citep{Elyiv09, Neronov10, Taylor11}. An alternative model posits that an electron-positron beam propagating through the IGM plasma drives the electrostatic (two-stream) instability resulting in energy loss to the plasma waves \citep{Broderick12,Puchwein12,RS12a,RS12,Miniati13}.
An observational analysis of cascade emission around radio galaxies supports the plasma-wave interpretation \citep{Broderick18}.

Although the electrostatic instability should in principle be easily accessible in PIC simulations, a substantial adjustment in parameter values is needed to render the growth rate large enough. Care must be exercised though to maintain the same physical regime. \citet{Shalaby17} noted the extreme narrowness of the velocity resonance in $\veck$ space, and they use 1D simulations to demonstrate the very high grid resolution that is needed to properly capture this mode in PIC simulations. The instability develops in the kinetic limit though, i.e. the warm-beam regime, for which dominant wave growth arises in oblique directions, mandating at least 2D simulations. The first 2D simulations were published by \citet{Sironi14} and showed saturation at moderate intensity levels, but substantial heating. \citet{Kempf16} argued that this heating arises from using a beam with higher energy density that is thermally carried by the background plasma. They found negligible energy transfer from the beam to the plasma, but their simulations were conducted in the reactive regime which doesn't apply to TeV pair beams. A list of requirements for scaling was formulated by \citet{Rafighi17} who also used 2D simulations to demonstrate the variation in the evolution of the system, if one or more of these requirements are violated. Even if one designs simulations in the appropriate regime, the code maintains energy conservation to a sufficient decree, and one mimics the non-Maxwellian distribution function of the pair beams, very little, if any energy loss is observed during the simulation \citep{Vafin18}. Significant energy loss of the pairs by inverse-Compton scattering of the cosmic microwave background happens over about a million growth times of the electrostatic instability, far too long for a PIC simulation the resolves the plasma frequency. One needs to scale the saturation mechanism and level from the simulation to real conditions which requires other techniques. Using analytical methods \citet{Vafin18} found that the energy loss by driving the electrostatic instability is likely dominant, suppressing the GeV-band cascade emission. Modeling the spectral evolution of the waves under nonlinear Landau damping instead suggests that there is no constant saturation level in the resonant band in $\veck$ space, which might invalidate the analytical extrapolation \citep{2019arXiv190109640V}.

\vfill
\newpage
\section{High-velocity non-relativistic systems: SNRs}
Supernova remnants (SNR) are observationally known to accelerate particles to very high energy, on account of the detection of TeV-scale gamma rays and X-ray synchrotron emission. The supernova explosion drives shocks into the ambient material with speeds of a few thousand kilometer per second. The shocks speed is for most of the lifetime of the SNR much higher than both the sound speed and the Alfv\'en speed in the ambient medium, implying large values of the sonic and the Alfv\'enic Mach number, $\ms\gg 1$ and $\ma\gg 1$. The compression at an MHD shock would depend only on the adiabatic constant, $\gamma$, and for the compression ratio one finds $\kappa=\simeq (\gamma+1)/(\gamma-1)$. \citet{1978MNRAS.182..147B} demonstrated that at such a fast shock diffusive shock acceleration can operate and provide a particle spectrum $N(p)\propto p^{-s}$, with $s=(\kappa+2)/(\kappa -1)\simeq 2$. 

PIC simulations have provided many insights into the structure of SNR shocks \citep{2016RPPh...79d6901M}. They also have demonstrated the acceleration of electrons and ions to moderately high energy, and in particular elucidated the processes through which particle acceleration arises. Pre-accelerated particles see the shock as a sharp discontinuity and can be further energized by diffusive shock acceleration, provided magnetic turbulence is continuously generated in the upstream region that scatters the particles. PIC simulations have been used to infer the saturation mechanism and level of cosmic-ray streaming instabilities that analytical estimates show to have a sufficiently high growth rate under SNR conditions for reaching the amplitudes that radiation modeling mandates. We shall discuss these three areas of inquiry in turn.

Nearly all PIC simulations, whose results we discuss on the following pages, were conducted in 2D3V geometry, and essentially all of them involve a shock speed higher than 10\% of the speed of light, a factor ten larger than those in real SNRs. Both modifications are mandated by a need to limit the computational expense, as is the choice of a reduced ion/electron mass ratio.

\subsection{Cosmic-ray streaming instabilities}
A charged particle propagating in a large-scale magnetic field follows a~helical trajectory. A circularly polarized electromagnetic wave, for example an Alfv\'en wave, typically imposes a rapidly oscillating $\mathbf{v}\times \mathbf{B}$ force at the location of the particle, unless a match is found between the Larmor frequency of the particle, $\oml$, its momentum, $\mathbf{p}$, and the frequency and wavevector of the wave, $\omega$ and $\mathbf{k}$. Such \emph{resonant} wave particle interactions with Alfv\'en waves have long been thought to be the main agent of cosmic-ray scattering \citep{1975MNRAS.172..557S,1975MNRAS.173..245S}, and the resonance condition is roughly a match between the wavenumber and the Larmor radius, $k\,r_\mathrm{L}\approx 1$. For a charged ion with Lorentz factor $\gamma$ the wavelength, which would have to be well resolved on the grid of a PIC simulation, is 
\be
\lambda\approx 2\pi\,\frac{c}{v_\mathrm{A}} \sqrt{\frac{\mi}{\me}}\,\gamma\,\lse .
\ee
As the skin depth, $\lse$, has to be resolved as well, it is obvious that a prohibitively large grid is required even for a strongly reduced ion/electron mass ratio and a typically high Alfv\'en speed, $v_\mathrm{A}$. In recent years techniques have been developed to merge the traditional PIC method with MHD for highly relativistic particles \citep[e.g.][]{2015ApJ...809...55B}. \citet{2018MNRAS.476.2779L} conducted 1D PIC-MHD simulations of the resonant instability driven by cosmic-ray pressure anisotropy with a view to infer the resulting parallel spatial diffusion of cosmic rays with Lorentz factor $\gamma=100$. They find agreement between the scattering mean free path deduced from the simulation with that derived analytically. \citet{2019ApJ...882....3H}
find that the late-time evolution of the resonant instabilities sensitively depend on the initial level of cosmic-ray anisotropy.

Plasma instabilities can also be driven non-resonantly by the current that streaming cosmic rays provide. The precursor of SNR shocks is composed of quasi-isotropic energetic cosmic rays that are predominantly ions, even if the same number of electrons and ions are injected into diffusive shock acceleration \citep{1993A&A...270...91P}, resulting in a current in the incoming upstream medium. \citet{2004MNRAS.353..550B} demonstrated analytically and with MHD simulations that under certain conditions modified Alfv\'en waves can grow to very large amplitudes, $\vert\delta B\vert \gg B_0$, and so Bell's nonresonant instability may provide the magnetic-field amplification that radiation modeling of SNRs appears to require. In the linear phase the real part of the phase speed is far less than the Alfv\'en speed, and so the waves are almost entirely magnetic. The wavelengths are small, $k\,r_\mathrm{L}\gg 1$, and the scattering efficiency of the waves has to be explored with simulations. The non-resonance strictly applies to the energetic ions only, as gyroresonance with the background ions appears to significantly shape the mode \cite{2019ApJ...872...48W}, that for fast beam drifts conforms with an electron-whistler mode excited by the gyro-motion of the plasma ions \cite{2019ApJ...873...57W}.

If the mode is supposed to amplify the magnetic field in the cosmic-ray precursor of an SNR shock, its growth rate must be much larger than the shock-capture rate, and the ratio of the two rates is an estimate of the number of e-folding cycles, $N_\mathrm{e-fold}$, the instability has for growth. For Bohm diffusion of particles with Lorentz factor $\gamma$, $D=\gamma\,c^2/(3\,\omci)$, one can expect
\be
N_\mathrm{e-fold}\approx \frac{1}{12}\,\frac{U_\mathrm{CR}}{U_\mathrm{bulk}}\,\ma\,,
\label{e-fold}
\ee
where the ratio between the energy density in cosmic rays, $U_\mathrm{CR}$, and that in the bulk flow of far-upstream plasma, $U_\mathrm{bulk}$, can reach values around 10\%. Remnants expanding into an environment of moderately high density should allow for about 10 e-folding cycles, and hence a large wave amplitude. The question arises at what amplitude saturation processes terminate the wave growth. 

The first PIC simulations were presented by \citet{2008ApJ...684.1174N} who considered slowly drifting cosmic rays of Lorentz factor $\gamma=50$ in 2-D geometry. The analytical condition $\omi \ll \omci$ was not exactly met, as the expected growth rate was commensurate with the ion gyrofrequency. Instead of Bell's mode an oblique variant was observed to grow quickly and to saturate at $\delta B/B_0\lesssim 10$, lower than in MHD simulations, on account of a reduction in the drift velocity between the cosmic rays and the cold plasma. The nonresonant mode was then observed in PIC \citep{2009ApJ...694..626R,2009ApJ...698..445O,2009ApJ...706...38S}
and hybrid simulations \citep{2010ApJ...711L.127G} as predicted analytically. The nonlinear behaviour was similar to that reported in \citep{2008ApJ...684.1174N}, despite the differences in the initial development: a saturation level $\delta B/B_0\lesssim 10$, deceleration of the bulk motion between cosmic rays and plasma, turbulent motion of wavelength-sized plasma cells, and an evolution toward larger wavelengths than those initially favored. Most secondary instabilities spawned by Bell's instability operate on scales too large for capture with PIC simulations \citep{2011MNRAS.410...39B}; in fact none of the published simulations had sufficient transverse extent to cover the cosmic-ray filamentation instability that is expected to develop in the nonlinear phase of Bell's instability \citep{2005MNRAS.358..181B,2012MNRAS.419.2433R}. \citet{2010ApJ...717.1054R} demonstrated that the perpendicular current of additional cosmic rays, that have a Larmor radii smaller than the scales of Bell's wave, can spawn a secondary instability, which in their simulation further amplified the magnetic field to a level $\delta B/B_0\lesssim 50$. 

The detailed processes of feedback are difficult to infer from the earlier PIC simulations, as none of them featured particle tracing. Some insight was provided by \citet{2010ApJ...709.1148N} who simulated Bell's instability driven by a relativistic proton beam. They noticed pitch-angle scattering at a rate commensurate with the Bohm limit in the amplified field, accompanied by a diffusive reduction in the momentum component parallel to $\veck$, approximately in line with the results of the Fokker-Planck analysis of \citet{1984JGR....89.2673W}.

All of the simulations described above employed periodic boundary conditions, which do not permit a self-consistent description of the bulk deceleration in the nonlinear phase. Real cosmic-ray precursors continuously run into unmodified plasma, and the bulk deceleration during the nonlinear phase of the instability will change the density ratio of cosmic rays and plasma, which determines the wavelength and rate of peak growth. 

\begin{SCfigure}[0.75]
\includegraphics[width=8.5cm]{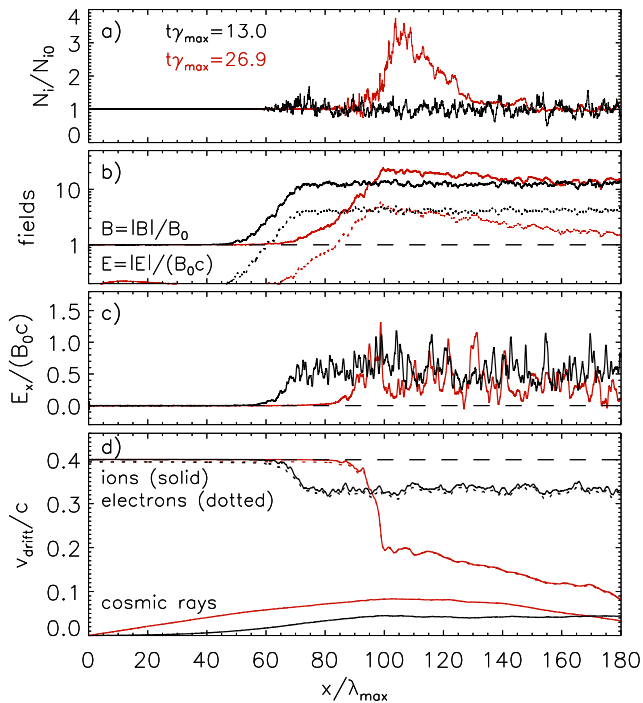}
\caption{Spatial profiles of the ion number density (a), magnetic and electric field amplitudes (b), $E_x$ component of the electric field (c) and bulk velocities (d). \emph{Black} lines refer to an intermediate time, whereas \emph{red} lines indicate the status at the end of the simulation. The compression zone at $x\approx 110 \lambda_\mathrm{max}$ is evident in panel (a). In panel (b), we see additional magnetic-field compression in this zone. In panel (d), \emph{dotted} lines are for plasma electrons, and \emph{solid} lines beginning at $v/c=0.4$ indicate the flow speed of plasma ions, which abruptly drops at the front of the compression zone. Taken with permission from \citet{2017MNRAS.469.4985K}.

}
\label{profiles}
\end{SCfigure} 
\citet{2017MNRAS.469.4985K} presented the first PIC simulation with open boundaries. They confirm magnetic-field amplification to $\delta B/B_0\lesssim 10$, strong turbulent motion, and substantial plasma heating. Cosmic-ray scattering is Bohm-like, if the Larmor radius of the particles is similar to the coherence length of the evolved turbulence. The most important new result is the development of a compression zone, shown in Figure~\ref{profiles}, where the instability turned nonlinear and the plasma is decelerated. Extrapolation to real conditions can be performed in the fluid picture by compensating the gradient of the CR pressure, $\delta\Pi_\mathrm{CR}\simeq U_\mathrm{CR}/3$, with the gradient of the plasma momentum flux, yielding for the total change in plasma flow velocity
\begin{equation}
\delta\Pi_\mathrm{CR} \simeq-\rho_\mathrm{up}\,V_\mathrm{up}\,\delta V \quad\Rightarrow\ \delta V\simeq\frac{V_\mathrm{up}}{6}\,\frac{U_\mathrm{CR}}{U_\mathrm{bulk}} .
\end{equation}
For very efficient cosmic-ray acceleration the bulk deceleration will be substantial \cite{2019ApJ...873...57W}, and the simulations of \citet{2017MNRAS.469.4985K} suggest that it is not continuous, but may steepen into an abrupt transition. 

To summarize, PIC simulations have demonstrated that cosmic rays can indeed amplify the magnetic field and be efficiently scattered in the precursors of SNR shocks, but that comes with significant heating and localized bulk deceleration of the upstream plasma. We do not yet know what impact that has on the sub-shock and its ability to accelerate particles.

\subsection{Quasi-parallel shocks}
\begin{SCfigure}[0.7]
\includegraphics[width=9cm]{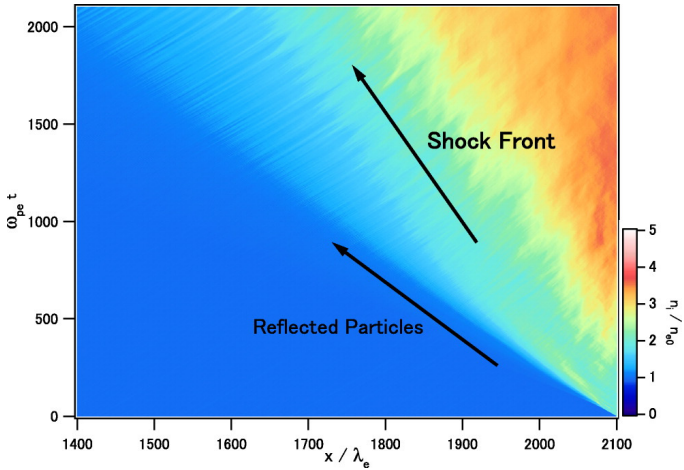}
\caption{Development of an unmagnetized shock in a PIC simulation. The color indicates the ion density that sharply increases at the so-called ramp. Reflected particles form the foreshock and drive turbulence there. At quasiperpendicular shocks the foreshock is small and usually referred to as the shock foot. Reproduced with permission from \citet{2008ApJ...681L..93K} (\textcopyright  AAS).

}
\label{kato-fg2}
\end{SCfigure}

All high-Mach-number shocks are shaped by reflection of upstream ions off the shock surface (see Fig.~\ref{kato-fg2}). The distinction between quasi-parallel and quasi-perpendicular shocks arises from the initial Larmor orbit of a reflected particle being completely detached from the shock in the former case, as opposed to returning the particle back to the shock in the latter situation.

At quasi-parallel shocks the reflected particles therefore escape to the far-upstream region, where they first need to excite turbulence before they can be captured again by the shock. A very extended foreshock develops that is neither homogeneous nor uniform, and in which localized but spatially separated beams of electrons and ions drive waves. Shocks in unmagnetized media can be treated as special case of quasi-parallel shocks with $\ma\rightarrow \infty$. 

\citet{2008ApJ...681L..93K} demonstrated that unmagnetized shocks are mediated by the Weibel instability, as are relativistic shocks \citep{2004ApJ...608L..13F,2008ApJ...673L..39S}. Weakly magnetized shocks with Alfv\'enic Mach numbers larger than 20 can be mediated by the Weibel instability as well, and the efficiency is in fact higher when a weak parallel magnetic field exists. \citet{2012ApJ...759.73N} report that the magnetic turbulence in the shock transition is up to an order of magnitude stronger in this case. Electron dynamics play an
important role in the development of the system, and \citet{2017PhPl...24f2105B,2018MNRAS.473..198D} demonstrate that the two-stream instability is of high relevance for the formation of mildly relativistic shocks. Ion–ion
or ion–electron streaming generally drives the turbulence,
which is mainly magnetic. 

It takes time for the shock to settle to a quasi-equilibrium though, following the initial development of the instabilities that shape it \cite{2015ApJ...803L..29S}. Fully kinetic PIC simulations typically cover only a few ion Larmor times, $\omci^{-1}$, and so the ion distribution downstream of the shock does not isotropize and little, if any, ion acceleration is observed. That is different for electrons. In the strictly parallel simulations of \citet{2012ApJ...759.73N} electron acceleration was not significant. As soon as the magnetic field is somewhat inclined to the shock normal, electron shock-surfing acceleration kicks in \citep{2009ApJ...690..244A}, that is driven by reflection of incoming electrons off the electrostatic structure of strong Buneman waves at the foot of the shock, and their subsequent acceleration in the motional electric field. We shall discuss this process in more detail in the context of quasi-perpendicular shock (see Section~\ref{quasiperp}). 

A process that can become increasingly important for larger magnetic-field obliquity, $\thbn$, is the shock-drift acceleration, essentially an adiabatic mirror reflection in the de Hoffman-Teller frame that leads to a speed $V_r \simeq 2 \vsh / \cos \thbn$ of the reflected particles in the upstream frame. The reflected electrons can excite waves in the upstream region that scatter them. The electrons would probably not excite Bell's instability \citep{2004MNRAS.353..550B}, because it is almost impossible to simultaneously meet the requirements of small growth rate, $\gamma_\mathrm{max}\ll\omci$, and large wavenumber, $k\gg r_\mathrm{L}$. They should be able to drive whistler waves instead, if the condition
\be
\ma \gtrsim \frac{\cos\thbn}{2}\,\frac{\mi}{\me}
\label{ah10}
\ee 
is met \citep{2010PhRvL.104r1102A}. For quasi-parallel SNR shocks that requires an Alfv\'en speed below a few km/s, or an upstream gas density of at least a few hydrogen atoms per cc for a 10-$\mu$G magnetic field. The whistler waves in the upstream region are not to be confused with the oblique whistler instability that operates in the foot of quasi-perpendicular shocks.

For a magnetic-field inclination as small as $\thbn=5.7^\circ$, and a shock speed of half the speed of light, \citet{2010A&A...509A..89D} observed in their 2D simulation the formation of short large-amplitude magnetic structures, so-called SLAMs, that efficiently scatter electrons, leading to their acceleration ahead of the shock. Similar structures have been detected at interplanetary shocks that accelerate electrons \citep{1997A&A...322..696C}. 

\citet{2015ApJ...802..115K} ran a very long PIC simulation of a quasi-parallel shock that demonstrated the driving of Alfv\'enic waves in the foreshock. \citet{2019HEDP...3300709O} confirmed that finding. The reflection of particles off quasi-parallel shocks, and their driving instabilities in the upstream region, can be simulated for a longer time, if one abandons multidimensionality. \citet{2015PhRvL.114h5003P} ran a 1D PIC simulation for about $310\,\omci^{-1}$ with $\thbn=30^\circ$. Reflected ions (and some electrons) were observed at a distance $L\gtrsim 13\ c/\omci$ ahead of the shock. As the peak growth rate of Bell's mode, $\gamma_\mathrm{max}$, can be related to the flux ratio of incoming (i) and reflected (r) ions,
\be
\gamma_\mathrm{max} = \omci\,\frac{\ma}{2}\,\frac{F_\mathrm{r}}{F_\mathrm{i}},
\ee
within the shock-capture time, $L/\vsh$, one expects for the parameters of their simulation
\be
N_\mathrm{e-fold} \approx 10\,\frac{100\,F_\mathrm{r}}{F_\mathrm{i}} \ll 100
\ee
exponential growth times before the plasma is captured by the shock and hence sufficient time for Bell's mode to operate. Here the last inequality derives from the requirement $\gamma_\mathrm{max} \ll \omci$ and their choice $\ma=20$. Indeed, \citet{2015PhRvL.114h5003P} observe magnetic fluctuations that appear consistent with those predicted by Bell. The interpretation is difficult though, because in the 1D simulation only waves with $\angle(\veck,\vecb)=30^\circ$ can exist, as opposed to the parallel waves for which Bell's treatment applies.

Longer simulations can be run with the hybrid technique which treats electrons as a massless fluid. Only ions are simulated as quasi-particles, and so the scale that needs to be resolved is the ion skin depth, about a factor $42$ larger than the electron skin depth that is the fundamental scale for full PIC simulations. Hybrid simulations can hence cover larger spatial and temporal scales at the expense of neglecting electron-induced instabilities that are evident in full PIC simulations \cite{2018ApJ...857...36Y}. \citet{2013ApJ...765L..20C} performed large-scale hybrid simulations of quasi-parallel shocks with $\ma=30$ and demonstrated that reflected ions fill a very extended foreshock, streaming at about three times the shock speed. There they drive parallel waves that on account of the $\vecj\times \delta\vecb$ force focus the energetic backstreaming ions and push the plasma away from the region of strong current. One of the consequences of this filamentation is a strong heating of the plasma before it passes through the shock. In later simulations of mildly relativistic shocks, \citet{2019MNRAS.485.5105C} find little upstream heating of electrons and the power-law high-energy tails in the spectra of both electrons and ions. 

The magnetic-field amplification upstream of quasi-parallel shocks appears to result from a competition between non-resonant (Bell-type) modes and resonant streaming instabilities. \citet{2014ApJ...794...46C} find that for $\ma\lesssim 30$ resonant wave-particle interactions dominate, whereas for stronger shocks non-resonant modes are driven far upstream that migrate to larger scales and eventually transition to resonant waves in the near precursor of the shock. The magnetic-field amplification factor appears to scale as $\sqrt{\ma}$.

\begin{SCfigure}[0.75]
\includegraphics[width=8.2cm]{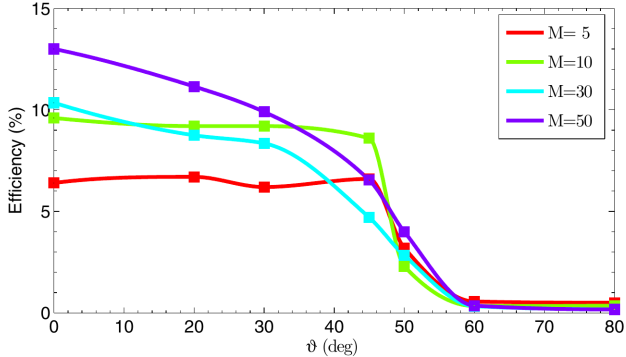}
\caption{Acceleration efficiency, defined as the relative post-shock
energy density carried by particles with $E\ge 5\,m\,v_\mathrm{sh}^2$ for several shock inclinations, $\thbn$, and Alfv\'enic Mach numbers, $M=\ma$. To be noted is the significant drop in the
acceleration efficiency for $\thbn\gtrsim 45^\circ$. Reproduced with permission from \citet{2014ApJ...783...91C} (\textcopyright  AAS).

}
\label{cs14}
\end{SCfigure}

Figure~\ref{cs14} demonstrates that approximately 10\% of the post-shock energy density appears to be carried by ions with more than ten times their initial flow energy, $E\ge 5\,m\,v_\mathrm{sh}^2$ \citep{2014ApJ...783...91C}. Following the trajectories of fast ions, \cite{2015ApJ...798L..28C} calculated the minimum energy needed for injection into diffusive shock acceleration as a function of $\thbn$. The energy spectrum of accelerated particles is that expected for diffusive shock acceleration of test particles, $N(p)\propto p^{-2}$. One would expect curved spectra on account of the high energy density of accelerated particles, but that is not observed in the hybrid simulations of \citet{2014ApJ...783...91C}, possibly because they use a strictly non-relativistic code, and so there is no difference in the adiabatic index of cold and energetic particles.

Hybrid simulations of ions with different mass number, $A$, and charge number, $Z$, indicate that the shock preaccelerates fully ionized metals to a similar spectrum as that of protons, whereas incompletely ionized metals would be preferentially accelerated \citep{2017PhRvL.119q1101C}. This type of source material should be  abundant in superbubbles and similar galactic environments, where short-lived cosmic rays like $^{60}$Fe must be produced whose recent detection suggests a very high injection efficiency \citep{2016Sci...352..677B}. Care must be exercized though, because material with different $A/Z$ will trigger two-stream modes in the upstream regions that broadens the ion distribution functions and increases the acceleration efficiency of the shock \cite{2019PhRvS..22d3401K}.

\subsection{Quasi-perpendicular shocks}\label{quasiperp}

Figure~\ref{cs14} indicates that in the hybrid simulations of \citet{2014ApJ...783...91C} the acceleration efficiency drops to the per-cent level and below, once the shock obliquity $\thbn \gtrsim 50^\circ$. For those angles only a few, if any, ions are reflected to the far upstream and produce magnetic turbulence there, although the shock is still clearly subluminal (cf. Eq.~\ref{eq:dHT}).  Apparently the reflected ions are not fast enough to evade the shock. \mpo{Recent MHD-PIC simulations do not quite agree with this picture, in which both particle acceleration and magnetic-field amplification was observed for large inclination angles, albeit on large time scale and involving shock corrugation on a scale that requires a large simulation box to be captured \cite{2018MNRAS.473.3394V}. There is also an apparent conflict with the findings of \cite{2013MNRAS.430.2873R} who used a spherical-harmonics expansion of the Vlasov-Fokker-Planck equation to find a quasi-universal behaviour of shocks, because the magnetic field in the immediate upstream region was completed disordered, whatever the orientation of the far-upstream field. \citet{2013JGRA..118.1132S} conducted 2D simulations to find that in quasi-perpendicular shocks one observes two populations of reflected ions, one field-aligned and the other one gyro-phase bunched. In any case,} electrons are typically faster and may be able to fill the foreshock and drive turbulence there. Fully kinetic PIC simulations of oblique shocks are therefore needed. 
\begin{SCfigure}[0.7]
\includegraphics[width=8.4cm]{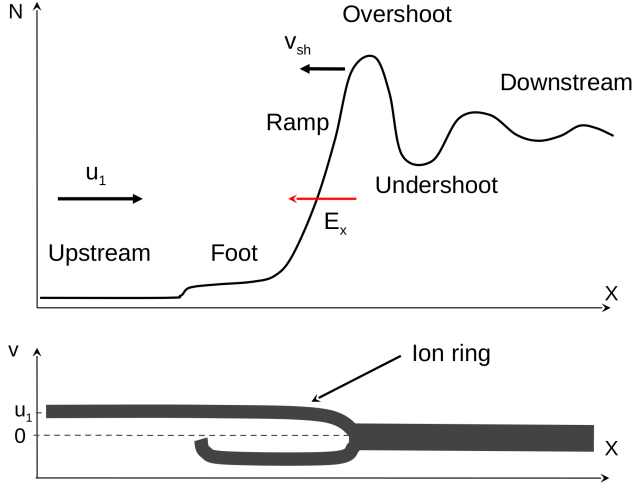}
\caption{Structure of a perpendicular shock. Reflected ions are directed back to the shock after a half-gyration around the magnetic field (bottom panel). Electron-ion counterstreaming in the foot region drives Buneman waves that heat electrons. Superposition of incoming, reflected, and returning ions causes density spikes such as the overshoot (upper panel). The figure is reproduced by courtesy of Dr. Artem Bohdan.

}
\label{perpshock}
\end{SCfigure} 

The shock transition itself is partially shaped by electron dynamics, and electrons might be pre-accelerated, even if they do not produce a foreshock. To be noted is that this acceleration first arises from processes operating at the shock itself; it is not the conventional diffusive shock acceleration that builds on repeated crossing of the shock and hence a mean free path that is larger than the thickness of the shock. A back-of-the-envelope estimate equates the mean free path with the Larmor radius of the electrons which then must be larger than the ion Larmor radius that is a proxy of the shock thickness,
\be
\lambda_\mathrm{mfp}
\simeq \frac{p_\mathrm{e}}{\me\omce}
\gg \frac{V_\mathrm{sh}}{\omci}
\qquad\Rightarrow\ p_\mathrm{e}\gg \mi V_\mathrm{sh} \ .
\label{elecdsa}
\ee
We can conclude that for diffusive shock acceleration to operate the electrons must be accelerated to a momentum far in excess of that of the incoming ions. For supernova remnants with $V_\mathrm{sh}\approx 0.01\,c$ that would be about $10$~MeV/c, and so the electrons have to be pre-accelerated to $\gamma_\mathrm{e} \gg 20$.

Figure~\ref{perpshock} provides a sketch of the structure of a perpendicular shock that with small modifications is applicable to quasi-perpendicular shocks as well. Typically, a fraction $\epsilon\approx 0.2$ of ions is reflected and counterstreams to the incoming plasma. At small Mach numbers the counterstreaming ions may drive oblique whistler waves \citep{2011ApJ...733...63R}. It is in competition with the electron cyclotron drift instability that, however, becomes weaker in 2D simulations with large box, possibly on account of shock rippling that also somewhat affects the growth rate of the oblique whistler waves \citep{2014PhPl...21b2102U}.

At larger Mach numbers they provide fertile ground for the Buneman instability. Assuming the reflected ions retain their speed in the shock frame, $v_\mathrm{i}=v_\mathrm{sh}$, charge and current neutrally mandate that the incoming electrons stream with speed $v_\mathrm{e}=v_\mathrm{sh} (1-\epsilon)/(1+\epsilon)$ \citep{Matsumoto2012}. For Buneman waves to grow, the thermal speed of the electrons should be smaller than the velocity difference between electrons and reflected ions, leading to \citep{2016ApJ...820...62W}
\be
\ms \gtrsim \frac{1+\epsilon}{2}\sqrt\frac{\mi}{\me}\sqrt\frac{T_\mathrm{e}}{T_\mathrm{i}}\ .
\label{buneman1}
\ee
Simple estimates of the energy transferred to Buneman waves and their ability to trap electrons in competition with Larmor deflection permit deriving a trapping condition, 
\be
\ma \gtrsim (1+\epsilon)\left(\frac{\mi}{\me}\right)^{2/3}\ ,
\label{buneman2}
\ee
above which electron shock-surfing acceleration can significantly energize electrons. The simulations of \citet{Matsumoto2012} demonstrated electron acceleration, or its absence, in line with these conditions. 

Care must be exercised in applying the driving and trapping formula to 2D3V PIC simulations though, because the 2D spatial grid restricts the $\veck$ vectors of any wave to the simulation plane \citep{2007ApJ...661L.171O}. Only if the large-scale magnetic field is oriented strictly out of the simulation plane, the Larmor orbit of reflected ions, and hence the streaming velocity relative to the incoming electrons, $\delta \vecv$, is contained in the simulation plane, and the Buneman waves with $\veck \parallel \delta\vecv$ are fully captured. \citet{2017ApJ...847...71B} present modified versions of equation~\ref{buneman1} and ~\ref{buneman2} that are supposed to account for the projection effect but do not work perfectly, possibly on account of a modified alignment of the motional electric field and the equipotential surfaces of the Buneman waves in simulations with homogeneous magnetic field in the simulation plane.

It was noted early that turbulence in the shock transition modifies the reflection of ions and leads to a patchy distribution of Buneman wave intensity \citep{2008ApJ...681L..85U,2009JGRA..114.3217L}. Filamentary structures are produced by the ion-Weibel instability that also provides a considerable amplification of the magnetic field \citep{2010PhPl...17c2114K,2010ApJ...721..828K}. The simulations of \citet{2013PhRvL.111u5003M} indicate that the ion-acoustic instability is driven as well. The current sheets resulting from the ion-Weibel modes are subject to magnetic reconnection which may be a secondary way of accelerating electrons besides the shock surfing acceleration operating in the Buneman wave zone \citep{2015Sci...347..974M}. That may be needed, as the 2D3V simulations of \citet{2019ApJ...885...10B} for various electron-ion mass ratios suggest that the contribution of SSA to the downstream high-energy-electron population is small for the real mass ratio. 

The complex interplay of the various instabilities that shape the shock is compromised in 2D PIC simulation. We already noted that the Buneman modes are fully captured only, if the large-scale magnetic field is oriented along the normal of the simulation plane \cite{2019ApJ...878....5B}. Similar, but not identical restrictions apply to other instabilities, and so only 3D simulations can provide a complete picture of electron pre-acceleration. 
\citep{2017PhRvL.119u5101M} presented the first 3D full-PIC simulation of a quasi-perpendicular shock, showing that the electron acceleration rate is higher than that in 2D PIC simulations both with in-plane and with out-of-plane orientation of the magnetic field. Electrons that were initially accelerated by the surfing mechanism are further energized by pitch-angle diffusion in the magnetic turbulence in the shock transition region. Hence, the ion-Weibel modes are a central agent in the efficient acceleration of electrons. \citet{2019MNRAS.482.1154T} also find that shock rippling leads to efficient electron energization at superluminal shocks.

\subsection{SNR shocks in partially ionized media}
Near many SNR shocks hydrogen lines can be observed, which suggests that the neutral fraction of the ambient material can be up to unity \citep{2000ApJ...535..266G}. Neutral particles pass through the shock and form a cold population with high bulk speed that for shocks at the projected periphery  of the remnant essentially lies in the plane of the sky, thus producing narrow Balmer lines. The neutrals charge-exchange in the downstream region, producing a slightly suprathermal ion and a hot neutral that gives rise to a broad Balmer line. 

Of particular interest is the possibility of a hot neutral to return to the upstream region, where it may charge-exchange again, producing a hot pick-up ion that may trigger instabilities. Very much like the feedback imposed by a cosmic-ray precursor, the returning neutrals would modify the flow profile and hence soften the spectrum of accelerated particles  \citep{2012ApJ...755..121B}. For a shock speed below $3000\ $km/s, about 10\% of the upstream neutral particles leak back into the upstream region \citep{2012ApJ...758...97O}. \citet{2016ApJ...827...36O} conducted 3D hybrid simulations and demonstrated that the pick-up ions drive fast-mode turbulence that upon contact with the shock induce vorticity and drive further magnetic-field amplification akin that described by \citet{2007ApJ...663L..41G}. Particles can be accelerated to high energy by shock-drift acceleration in the motional electric field in the upstream region. The simulations suggest that the pick-up process of returning neutral provides a means of pre-acceleration and hence injection into diffusive shock acceleration.

\vfill
\newpage
\section{Low-velocity high-beta systems: Clusters shocks}
Clusters of galaxies and the filaments that connect them are the largest structures in the Universe. Current structure formation theory assumes that clusters form through a hierarchical sequence of mergers and accretion of smaller systems. Galaxy mergers dissipate huge amounts of gravitational energy ($10^{63}-10^{64}$ ergs) mainly at shocks that form, and this energy is channeled into heating of the gas in the intra-cluster-medium (ICM), large-scale ICM motion, but also into non-thermal particles and magnetic fields.

Merging galaxy clusters show radio synchrotron emission from relativistic electrons \citep[e.g.,][]{Lindner-14,van-Weeren-10}. This emission is seen primarily in the form of so-called giant radio halos in the center of galaxy clusters and giant radio relics located at cluster peripheries. Radio relics have linear sizes at Mpc scales and are associated with large-scale shocks that propagate in the ICM during mergers and accelerate electrons. They are also thought to be possible sources of UHECRs above $\sim10^{18}$ eV, though the $\gamma$-ray emission from galaxy clusters, which would be a unique signature of cosmic-ray protons, has not been detected so far. On the other hand, radio halo emission is usually assumed to originate from electrons accelerated through scattering on MHD turbulence and/or secondary electrons resulting from cosmic-ray proton interactions with ICM \cite[for review see, e.g.,][]{2014IJMPD..2330007B}.

Merger shocks have been detected through X-ray observations \citep[e.g.,][]{2017A&A...600A.100A, Markevitch-02}. They indicate that these shocks are weak, i.e., their sonic Mach numbers are very low, $\ms\lesssim 4$, though probably still supercritical. Their velocities are similar to those of middle-age SNRs ($\vsh\sim 10^3$ km/s), but they propagate in the high-temperature ($T\sim 1-10$ keV) and magnetized ($\sigma\sim 0.1$, magnetic field strength $B\sim 1 \mu{\rm G}$, Alfv\'en Mach number $\ma\lesssim 10$) dilute (plasma density $n\sim 10^{-4}-10^{-5}$ cm$^{-3}$) medium, in which the plasma beta is high ($\beta\gg 1$). The connection of radio relics with shocks suggests that relativistic electrons are produced through diffusive shock acceleration. Shocks may accelerate locally injected electrons or re-accelerate pre-existing energetic electrons. However, particle acceleration, and especially electron injection mechanisms are poorly understood in this regime. In particular, at low Mach numbers the Buneman instability cannot be triggered \citep{Matsumoto2012}, and therefore electron injection via SSA does not work. In this respect, the electron injection seems harder to obtain, as the available free energy for microinstabilities to operate in the shock transition is much smaller than at high Mach number shocks.  
Note that the Mach numbers of cluster shocks are of the same order as those of shocks observed in the heliosphere, i.e, the Earth's bow shock and interplanetary shocks. However, the plasma beta is much smaller in the heliosphere. Similar conditions of low Mach number and high $\beta$ can be met at fast mode shocks that occur in the magnetic reconnection outflows in solar flares \cite[see, e.g.,][]{Park-12,Park-13}, though other physical parameters differ considerably ($n\sim 5\cdot 10^{9}$ cm$^{-3}$, $B\sim 6 {\rm G}$, $T\sim 0.8$ keV). Observations by {\it Yohkoh} and {\it RHESSI} in hard X-rays reveal electron energization above flare loop tops and footpoints \cite{2003ApJ...595L..69L}.  

Electron acceleration at low-Mach-number high-$\beta$ shocks has only recently been studied with 1D and 2D PIC simulations by a handful of authors. First 1D simulations \cite{Matsukiyo-11} and later 2D simulations \cite{Park-12,Park-13} showed that in such shocks electrons can be efficiently energized via SDA. 
In high-$\beta$ conditions and at {\it subluminal} shocks some SDA-accelerated electrons are reflected off the shock and form a non-equilibrium velocity distribution in the foreshock region that leads to instabilities which generate waves, provided the sonic Mach number is $\gtrsim 2.25$ \cite{2018ApJ...864..105H}. It was proposed that electrons can be scattered on these waves back towards the shock to experience further acceleration \cite{Matsukiyo-11}. This scenario has been confirmed in 2D simulations \cite{Guo-14a, Guo-14b} that demonstrate that upstream scattering allows for repeated SDA cycles, similar to a sustained first-order Fermi process. The upstream waves have been identified with the oblique mode of the \jn{electron} firehose instability \jn{(EFI)}, driven by the electron temperature anisotropy caused by electrons reflected off the shock and streaming along the mean magnetic field ($T_{e\parallel}/T_{e\perp}> 1$). Systematic investigations indicated that these mechanisms of wave generation and electron scattering work at low-Mach-number shocks for temperatures relevant for galaxy clusters in a wide range of magnetic field obliquity,
in particular in high-$\beta$ plasma, $\beta\gtrsim 20$ \cite{Guo-14b}.
For shock obliquities that allow large fractions of reflected electrons and thus strong temperature anisotropy, a non-thermal power-law tail with slope $p=2.4$ in the energy spectra ($dn/d\gamma\propto\gamma^{-p}$) was found, giving a radio synchrotron index of $-0.7$, compatible with observations. However such spectra were shown to be formed only in the {\it upstream} region of the shock, and downstream spectra were still close to thermal distributions.
Electrons can undergo irreversible non-adiabatic heating in low-Mach-number shocks \cite{Guo-17,Guo-18}.
\jn{The most recent studies show that electron pre-acceleration occurs only in shocks above an ``EFI critical Mach number'' $M^*_{\rm ef}\approx 2.3$, that is higher than typical critical Mach number $M_{\rm crit}\approx 1.26$ derived from MHD jump conditions in low $\beta$ shocks \cite{2019ApJ...876...79K}. This means that shocks with $\ms\lesssim 2.3$ cannot accelerate electrons. On the other hand, at supercritical shocks with $\ms\gtrsim M^*_{\rm ef}$ electrons may not reach high enough energies to be injected to DSA, because EFI was observed to saturate and did not generate long-wavelength modes.
In this case additional electron pre-accelerations are required for DSA in ICM shocks, such as the re-acceleration of fossil electrons with flat power-law spectra \cite[e.g.,][]{2016JKAS...49...83K}.}

A different view has recently been reported in 2D simulations of shocks with $\beta=3$ \cite{Matsukiyo-15}, that were earlier studied with 1D simulations \cite{Matsukiyo-11} and showed efficient SDA. A simulation with a large transverse size of the computational box demonstrated the importance of the multi-scale shock structure that includes ion-scale fluctuations in the form of shock surface corrugations. The origin of rippling wave modes in the shock overshoot is considered to be due to a downstream temperature anisotropy of ions provided by ions that were reflected off the shock potential and advected back to the shock \citep[e.g.,][]{Lowe03,2009ApJ...695..574U,Winske88}. In this case the Alfv\'{e}n ion cyclotron (AIC) instability is triggered. The rippling forms local regions of weaker magnetic field along the shock surface, and electrons that approach the overshoot are transmitted downstream and are not reflected in the SDA process. The effect is significant, since all electrons encounter the weak field region during their SDA interaction with the shock. However, some non-thermal electrons can still be found at the shock, but they result from local wave-particle interactions in the shock transition.      
With growing plasma beta, the temperature anisotropy becomes smaller, the growth rate of the AIC instability decreases, and the rippling modes have larger wavelengths. It was estimated that the wavelength of the ripples in simulations by Guo et al. for $\beta=20$ \cite{Guo-14a, Guo-14b} is much larger than the transverse system size they used, so that the modes could not be captured. On the other hand, it was shown that at magnetizations $\sigma\gtrsim 0.1$ close to that used in \cite{Matsukiyo-15} ($\sigma=1/9$), although electron reflection via SDA is efficient, the Fermi-like acceleration is inhibited because the strong magnetic pressure suppresses the growth of the firehose instability. The shock rippling effects in conditions in which sustained Fermi-like process can operate must therefore be studied. This is a topic of active current research.

\vfill
\newpage
\section{Low-velocity low-beta systems: The heliosphere}
After the discovery of the Earth's bow shock in front of the Earth's magnetosphere by in-situ spacecraft observations \citep{Sonnet63, Ness64}, the physics of collisionless shocks attracted the attention of many researchers.  The mean free path of binary collisions is by many orders of magnitude larger than the typical thickness of the Earth's bow shock, and the physical processes of energy dissipation at the shock front are attributed to particle scattering by collective plasma waves self-consistently generated by non-Maxwellian velocity distribution functions.  When ions and electrons are crossing the shock front, they are quickly heated by waves, and some of them are further accelerated to much higher energies to form a non-thermal high-energy tail.

In observational and computational studies of the Earth's bow shock in 1970s, it is found that the shock structure for a quasi-perpendicular shock varies according to the shock Mach number, $M_\mathrm{A}$.  For a subcritical Mach number, $M_\mathrm{A} < M_\mathrm{cr}$, anomalous resistivity generated by a non-Maxwellian distribution function at the shock ramp can provide enough dissipation to maintain a stationary collisionless shock structure. The critical Mach number depends on both the plasma $\beta$ and the shock normal angle, $\thbn$, defined as the angle between the shock normal vector and the magnetic field upstream, and the critical Mach number $M_\mathrm{cr}$ for a perpendicular shock is known to be $M_\mathrm{cr} \simeq 2.76$ \cite{Edmiston84}.  

In a supercritical shock with $M_\mathrm{A} > M_\mathrm{cr}$, it is known that a stationary shock structure cannot be realized only by energy dissipation through anomalous resistivity, and viscous dissipation as additional process are needed to obtain the shock structure in a magneto-hydrodynamic (MHD) framework (see Appendix~A).  In the Vlasov/kinetic plasma framework, however, additional dissipation is provided by the reflection of the incoming ions at the shock ramp due to the shock electrostatic potential and the magnetic mirror. During the reflection, the gyrating reflected ions can gain energy from the motional electric field, and then those accelerated ions gyrate into the downstream region. The multi-component nature of the incoming and reflected ions in front of the shock (i.e., the so-called foot) triggers various plasma instabilities which lead to plasma thermalization. 

In this section, we will review recent progress on the microphysics of collisionless shocks, concerning plasma instabilities and plasma heating in and around the shock front region for planetary and interplanetary shocks.  We will also touch on the heliospheric termination shock, which is believed to be responsible for the generation of Anomalous Cosmic Rays in the heliosphere.  We are aware that we will not able to include all important works, but can focus only on several studies linked to micro-plasma physics.  

\subsection{Ion reflection off supercritical shocks}
A combination of theoretical works and satellite observations has shown that a part of upstream cold ions flowing into the shock front is reflected back at a supercritical shock with $M_\mathrm{A} > 3$.  The number density of the reflected particles is about $20 \% - 30 \%$, but those reflected ions carry most of energy dissipated in the shock.  The reflection mechanisms at the shock front can be provided by the compressed magnetic field and/or the ambipolar electric field induced by the relative motion between ions and electrons. The reflected ions also excite various types of plasma waves that heat the plasma upstream of the shock \cite{Wu84}.

\begin{figure}[tb]
\begin{center}
\includegraphics[width=13.5cm]{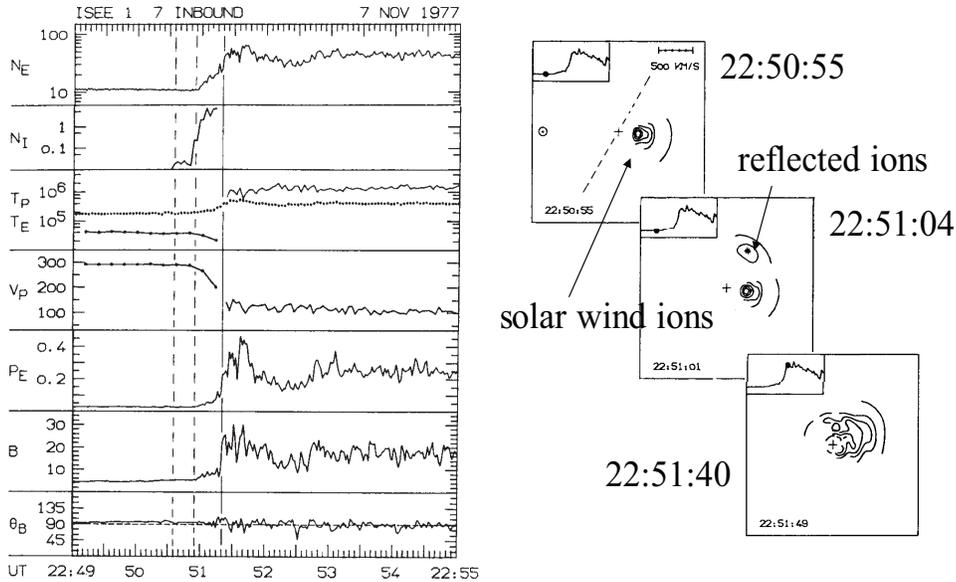}
\end{center}
\caption{Structure of a supercritical shock observed at the Earth's bow shock with $M_\mathrm{A} =8.6$ and $\theta_{BN}=85^\circ$. The reflected ion component appears in the foot region. Reproduced with permission from \citet{Sckopke83}. \label{fig3}}
\end{figure}

Figure \ref{fig3} shows an example of a supercritical shock observed at the Earth's bow shock by the ISEE satellite \cite{Sckopke83}.  From the top in the left-hand panel, the figure displays the plasma (electron) density, the density of the reflected ions, the ion and electron temperatures, the plasma flow speed, the electron gas pressure, the magnitude of the magnetic field, and the angle $\thbn$. 
$\thbn \approx 90^\circ$ suggests a quasi-perpendicular shock.  The horizontal axis is the universal time.  The reflected ions (second panel) can be seen at some distance from the shock, and their number density increases toward the shock front.  

The relationship between the incoming solar wind and the reflected ions can be clearly seen in the velocity distribution functions in the right-hand panels in Figure \ref{fig3}.  Three different time stages of ion velocity distribution functions in two-dimensional space are shown.  The top panel is taken when the satellite is situated in far upstream, and only the solar wind population can be observed.  In the foot region of the middle panel, we can see the reflected ions in addition to the solar wind flowing into the shock front.  In the bottom panel, due to the rapid heating, the downstream distribution function relaxes to a quasi-thermalized but complicated distribution.  During the ion reflection, those ions are accelerated in the motional electric field, and then the accelerated ions may overcome the shock potential induced by the ambipolar electric field and can pass through the shock.

\begin{figure}[tb]
\begin{center}
\includegraphics[width=12cm]{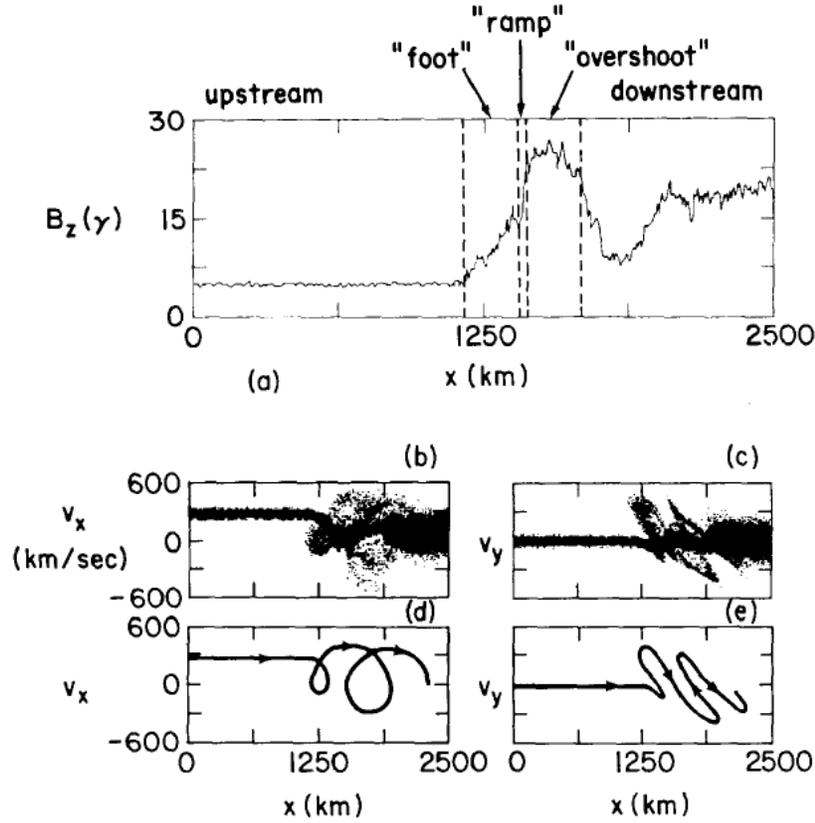}
\end{center}
\caption{Supercritical shock structure obtained with a hybrid simulation. (a) compressed magnetic field with foot and overshoot. (b) Ion phase-space distribution with reflected component. Reproduced with permission from \citet{Wu84}. \label{fig4}}
\end{figure}

The structure of super-critical shocks has been also investigated with hybrid simulations \cite{Leroy81,Leroy82,Wu84}.  The top magnetic-field plot in Figure~\ref{fig4} clearly shows the three distinct structures of the foot, the ramp, and the overshoot. The upper panels in the bottom show the ion phase space distribution, and the lower panels display typical particle trajectories. One can see the reflected ions in association with the foot structure, which is surprisingly consistent with the ISEE observation shown in Figure~\ref{fig4}.  The hybrid model can capture most of the ion dynamics in supercritical shocks.  Note that the phenomenological anomalous resistivity to dissipate the electric current energy is included in the hybrid model, and the thickness of the ramp is controlled by the magnitude of the anomalous resistivity.

\subsection{Shock reformation and shock surface rippling}
Iit has been shown that quasi-perpendicular shocks are intrinsically nonstationary, and so is the ion reflection process. Much attention has been paid to the self-reformation of the shock front, or shock reformation \cite{Biskamp72,Lembege03}, because it would influence the energy dissipation in shocks \cite{Scholer04, Shimada05}. While additional dissipation is provided by plasma instabilities in the foot region, where both incoming and reflected ions coexist, the strength of the plasma instabilities may be related to the magnitude of the additional dissipation to maintain the steady-state shock structure.  The strength of the plasma instabilities in the foot region could depend on the number density of the reflected ions, the relative velocities between the incoming and the reflected ions, and other properties that control the reflection process such as the shock potential, etc.  

It has been shown that shock reformation can be clearly observed, if the ion plasma $\beta_i$ in the upstream region is not necessarily large, i.e., $\beta_i < 0.5$ for mildly high-Mach-number shocks with $M_\mathrm{A} \approx 5$ \cite{Scholer03}. This tendency has been also confirmed for high-Mach-number shocks with $M_\mathrm{A} > 20$ \cite{Shimada05}.  Roughly speaking, the fraction of reflected ions may increase with increasing upstream ion temperature, because the portion of the high-energy population that exceeds the shock potential energy becomes large in the incoming Maxwellian plasma. It may suggest that a large fraction of the reflected ions for high ion plasma $\beta_i$ could provide enough energy dissipation for a stationary shock structure.  By increasing the Alfv\'enic Mach number, $M_\mathrm{A}$, but keeping the upstream ion plasma $\beta_i$ the same, the relative velocity between the incoming and the reflected ions becomes large, and then the plasma instabilities in the foot region lead to a very dynamical and turbulent state. As a result, the shock reformation becomes weak \cite{Shimada05}.

In addition to the shock reformation, another important non-stationarity would be the shock surface rippling appearing as a multi-dimensional effect \cite{Lowe03,Burgess07,Johlander16}.  We will make some remarks on shock rippling in kinetic and collisionless shocks here, but it should be noted that the term of corrugation instability is commonly used in hydrodynamic and magnetohydrodynamic approaches.  The corrugation instability in the fluid framework has been discussed by a number of authors \cite{Dyakov54,Gardner64,Stone95}, and it is known that parallel and perpendicular fast magnetosonic shock waves are stable against the corrugation instability for polytropic gases with $\gamma < 3$ \cite{Gardner64}.

The rippling that occurs along the shock front in kinetic collisionless shocks plays different important roles depending on the scale of the rippling.  On a scale smaller than the ion inertial length, lower-hybrid drift waves are often generated in the ramp due to the strong gradient of the compressed magnetic field, which in turn heats electrons to maintain the energy dissipation.  On the MHD scale, Alfv\'enic wave rippling wass demonstrated with hybrid simulations \cite{Lowe03}.  It is controversial how the MHD scale rippling is generated, but the ion temperature anisotropy \cite{Winske88} and the Kelvin-Helmholtz instability driven by the velocity shear introduced by reflected ions flowing along the shock surface \cite{Balogh13} are suggested as possible mechanisms to excite the rippling surface waves.  It is interesting to note that the MMS satellite, that has a capability to measure the temporal resolution with an order of magnitude better than before, observed the shock surface ripples in the Earth's bow shock \cite{Johlander16}.  

\subsection{Termination shock in the Heliosphere}
\begin{figure}[tb]
\begin{center}
\includegraphics[width=13cm]{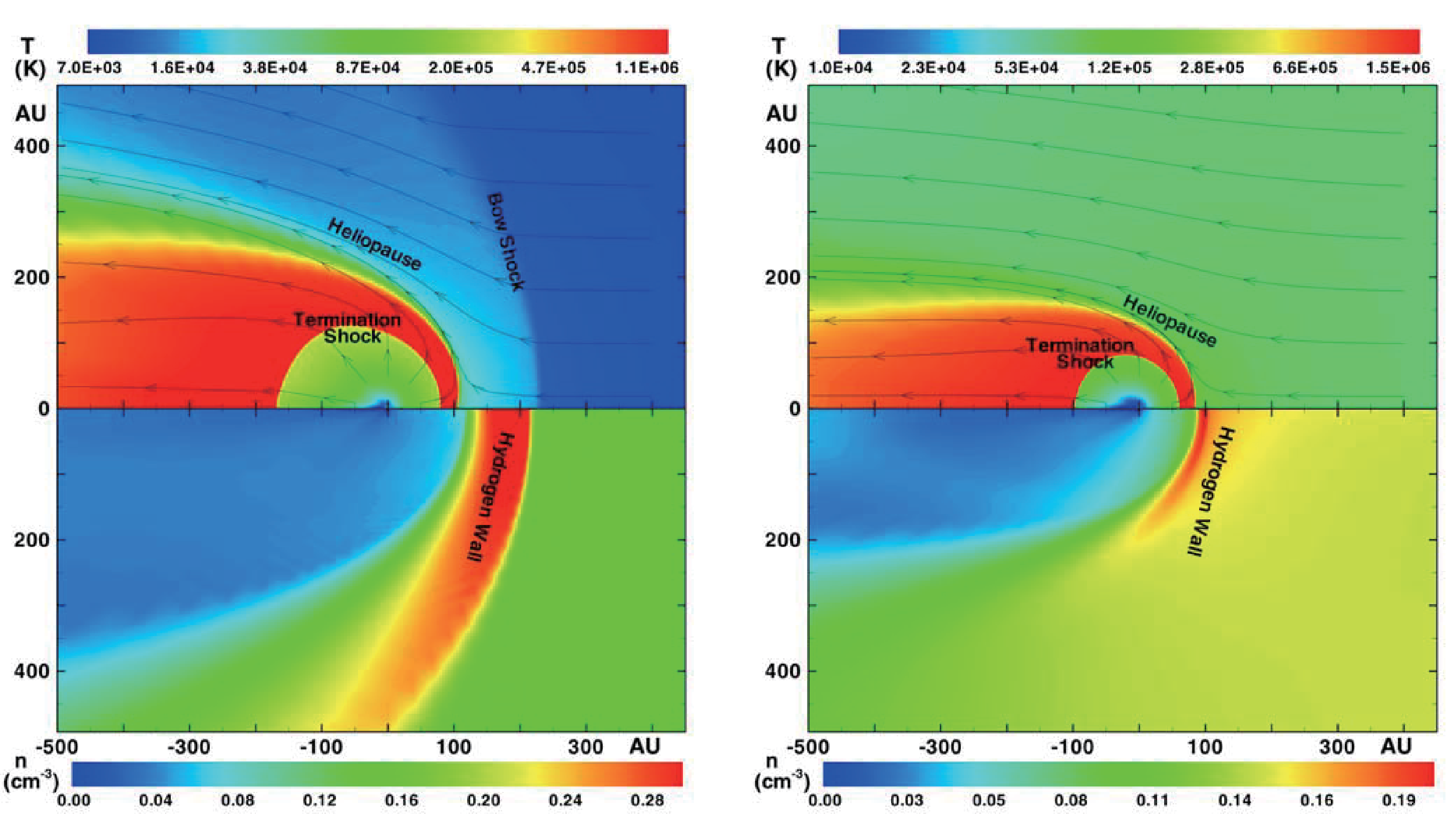}
\end{center}
\caption{Two different models for the global heliosphere: (left) The LISM is assumed to be a supersonic flow. Top and bottom panels show the temperature distribution of the solar wind and interstellar plasma and the density of neutral hydrogen, respectively. (right) The LISM is assumed to be a subsonic flow. The same format as the left panel. Reproduced with permission from \citet{Zank03}. \label{fig_zank}}
\end{figure}

Our heliosphere, surrounded by the interstellar gas, is an interesting astrophysical object that involves the dynamical interaction of two gases of different origin: the super-sonic solar wind plasma emanating from the sun and the interstellar gas. Pressure balance between the solar wind dynamic pressure and the gas pressure of the interstellar medium roughly gives the size of the heliosphere as $100 $~AU.  By taking into account the interstellar neutral hydrogen, namely, the collisional interaction including charge exchange under the multi-component nature of the solar wind and interstellar medium, hydrodynamic global simulations have been carried out by \citet{Zank03}.  Figure \ref{fig_zank} shows the global structures of the heliosphere for two different models with supersonic (left) and subsonic (right) flow in the local interstellar medium (LISM). We can see several distinct boundary structures of the termination shock at about $80 $~AU, the heliopause at about $110$~AU, the bow shock at about $230$~AU for the supersonic case in the left-hand panel.  In addition to these boundary structures, we can also see a density enhancement at the nose of the heliosphere due to the accumulation and compression of the local interstellar medium (LISM).  For the sub-sonic case in the right the structures are basically the same as in the supersonic case, but obviously there is no bow shock formation.

Among many intriguing phenomena observed by Voyager 1 and 2 in the outer heliosphere, we will quickly touch on anomalous cosmic rays (ACRs), which is an enhancement of particles at energies between $10^7$ to $10^8$~eV, between the energetic solar wind population and Galactic cosmic rays. ACRs are thought to originate from interstellar neutral atoms that have been ionized either by solar ultraviolet photons or through charge exchange with solar wind ions.  The ionized charged particles are picked up by the solar wind, and then the pick-up ions (PUIs) are transported to the heliospheric termination shock together with the solar wind \cite{Fisk74}.  PUIs are theoretically believed to be accelerated at the termination shock (TS) by diffusive shock acceleration \cite{Pesses81}.
However, the energy spectrum observed by the Voyager 1 spacecraft was not consistent with theoretical predictions based on DSA, and the intensity of ACRs increased as Voyager 1 went deeper into the downstream region of the TS, i.e., the heliosheath, which is the region between the TS and the heliopause (HP) \cite{Stone05,Burlaga05,Decker08}.

The contamination of PUIs in the termination shock (TS) is expected to modify the heating and structure of the shock from that of a standard magnetosonic shock. \citet{Matsukiyo14} studied the microstructure of the TS on ion and electron scales using one-dimensional PIC simulations.  When the relative pickup ion density is about 20\%-30\%, the reflection of pickup ions results in the formation of a cross-shock potential extended over the whole foot region, and most of the downstream thermal energy is gained by the pickup ions, while some heating of the solar wind ions and electrons occurs.  Due to the reflection of pickup ions, an increase of the cross-shock potential can be observed as well.  However, the shock ion surfing acceleration \cite{Sagdeev66}, which is proposed to be an injection process of ACRs \cite{Lee96, Zank96}, did not occur in their simulations, because the thickness of the electrostatic potential responsible to the shock surfing acceleration appears to be much larger than the electron inertial length.

This kinetic simulation result does not contradict the Voyager observations of a continuous increase of ACRs flux when entering the heliosheath through the termination shock \cite{Stone05,Burlaga05,Decker08}, because the notion of injection by surfing acceleration does not necessarily work under the PUIs population.  The Voyager observations may suggest that ACRs are generated much deeper in the heliosheath, and other mechanisms such as turbulent reconnection under the striped solar wind may operate \cite{Drake10}.  However, it seems to be premature to rule out the generation of ACRs at the termination shock by diffusive shock acceleration.


\vfill
\newpage

\section{Summary}
The minor role of collisions in space plasma mandates a kinetic description of many processes that involve energy transfer, the most obvious of which is the acceleration of particles to very high energies. The distribution function of particles is generally not known. It is shaped by interactions with collective electromagnetic fields that in many situations can be described as an ensemble of plasma waves. Asymmetries in the distribution function of the particles lead to the driving of waves, which in its earliest linear phase can be analytically followed. The later phase, when saturation is possibly achieved and the waves modify the distribution of particles, is inherently non-linear. The particle-in-cell technique has been developed to study such non-linear plasma interactions in computer experiments. By following individual quasi-particles that stand for many electrons or ions, we can investigate the evolution of collective electromagnetic fields and the distribution function of the particles. Current computer experiments involve billions of quasi-particles, and continuous algorithm development provides a high level of veracity.

This review describes the current state of research. It summarizes recent results and puts them in perspective. It also provides an overview over current questions. We decided to sort the vast multitude of results by application, magnetized or not, fast shocks or slow shocks, the various acceleration mechanisms, and so forth. To a fair degree this kind of sorting coincides with a classification by source class. Each section is written so that understanding it should not require perusal of the other sections, with the exception of the introduction and some material that we moved to the appendix.

We conclude by stating that the spectrum of results achieved with PIC simulations is very impressive. We owe a large fraction of our understanding of shocks and particle acceleration to PIC simulations, and we are confident that there is more to come.

\vfill
\newpage
\appendix

\section{Critical Mach number}
Let us review the physics of the critical Mach number.  The standard Rankine-Hugoniot relations in the shock rest frame for a perpendicular shock can be given as follows,
\begin{eqnarray}
\rho_1 v_1 &=& \rho_2 v_2, \\
\rho_1v_1^2+P_1+\frac{B_1^2}{8 \pi} &=& \rho_2v_2^2+P_2+\frac{B_2^2}{8 \pi}, \\
\rho_1 v_1 \left(\frac{v_1^2}{2}+\frac{\gamma}{\gamma-1}\frac{P_1}{\rho_1}
+\frac{B_1^2}{4 \pi \rho_1} \right) &=&
\rho_2 v_2 \left(\frac{v_2^2}{2}+\frac{\gamma}{\gamma-1}\frac{P_2}{\rho_2}
+\frac{B_2^2}{4 \pi \rho_2} \right), \\
v_1 B_1 &=& v_2 B_2,
\end{eqnarray}
where $\rho_J$, $v_j$, $P_j$, and $B_j$ are the plasma density, the velocity parallel to the shock normal, the plasma pressure, and the magnetic field perpendicular to the shock normal; the suffix $1$ and $2$ denotes the physical quantities upstream and downstream of the shock, respectively.  For simplicity, we assume the upstream plasma pressure is zero.  We also introduce the following normalizations: $n=\rho_2/\rho_1$, $u=v_2/v_1$, $b=B_1/B_2$, $p=P_2/(B_1^2/8 \pi)$, and $M_A=v_1/V_{A1}$.  The ratio of the specific heat is set to be $\gamma=5/3$.  Then we obtain,
\begin{eqnarray}
n u &=& 1, \\
1 + \frac{1}{2 M_A^2} &=& u + \frac{1}{2 M_A^2} (p + b^2), \\
1+\frac{2}{M_A^2} &=& u^2 + \frac{1}{M_A^2} \left( \frac{5}{2} pu + 2 b^2 u \right), \\
b u &=& 1.
\end{eqnarray}
By solving the above Rankine-Hugoniot relations, we have,
\begin{equation}
(u-1) \left( 8 M_A^2 u^2 - (5 + 2 M_A^2)u -1 \right) = 0,
\end{equation}
and we get the compression ratio of $n = 1/u =4$ for $M_A \rightarrow \infty$.

Note that the downstream plasma flow velocity $v_2$ is not necessarily slower than the downstream sound speed, $c_{s,2} = \sqrt{\gamma P_2/\rho_2}$. Below the so-called critical Mach number, $M_A = M_{cr} \simeq 2.76$, the sound speed $c_{s,2}$ is slower than the downstream plasma speed $v_2$, but it is easy to confirm that the downstream plasma speed is subsonic for the fast mode speed of $v_f = \sqrt{c_{s,2}^2 + v_{A,2}^2}$.

In the standard Rankine-Hugoniot relations, the thickness of the shock front is neglected, and it is implicitly assumed that some kind of energy dissipation at the shock front maintains the steady-state shock structure.  In the MHD framework, the energy dissipation may be provided by Ohmic heating and/or fluid viscosity.  Let us first study the shock structure provided by only the Ohmic dissipation.

The modified Rankine-Hugoniot relations by taking into account the Ohmic dissipation can be written as the function of the distance $x$ from the shock front position as follows,
\begin{align}
\rho_1 v_1 &= \rho_2(x) v_2(x), \\
\rho_1v_1^2 + \frac{B_1^2}{8 \pi} &= \rho_2(x)v_2(x)^2+P_2(x)+\frac{B_2(x)^2}{8 \pi}, \\
\frac{v_1^2}{2}+\frac{B_1^2}{4 \pi \rho_1}  &=
\frac{v_2(x)^2}{2}+\frac{5}{2}\frac{P_2(x)}{\rho_2(x)}
+\frac{B_2(x)^2}{4 \pi \rho_2(x)} 
-\frac{c^2}{32 \pi^2 \sigma} \frac{\partial B_2(x)^2}{\partial x} , \\
v_1 B_1 &= v_2(x) B_2(x) - \frac{c^2}{4 \pi \sigma} \frac{\partial B_2(x)}{\partial x}.
\end{align}

By normalizing in the same way as the standard Rankine-Hugoniot relations, we obtain,
\begin{align}
n(x) u(x) &= 1, \\
1 + \frac{1}{2 M_A^2} &= u(x) + \frac{1}{2 M_A^2} (p(x) + b^2(x)), \\
1+\frac{2}{M_A^2} &= u(x)^2 + \frac{u(x)}{M_A^2}
\left( \frac{5}{2} p(x) + 2 b(x)^2 \right)
- \frac{1}{R_{M}\,M_A^2} \frac{\partial b(x)^2}{\partial \chi}, \\
b(x) u(x) &= 1 + R_{M}^{-1} \frac{\partial b(x)}{\partial \chi} ,
\end{align}
where the normalized spatial scale, $\chi$, is defined as $\chi = x (\omci/v_1)$.  We have used the ion gyro frequency $\omci$ and the magnetic Reynolds number $R_\mathrm{M}= 4 \pi \sigma v_1^2/(c^2 \omci)$.

After some cumbersome calculation, one can obtain,
\begin{equation}
  \frac{\partial}{\partial \chi} u(x) = \frac{1}{M_A^2}
  \frac{u(x)}{u_s^2 - u(x)^2}
  \left( 1 + \frac{5}{3} R_\mathrm{M}^{-1} \frac{\partial}{\partial \chi} b(x) \right)
  \left( \frac{\partial}{\partial \chi} b(x) \right),
\end{equation}
where $u_s$ is the normalized downstream sound velocity defined by $u_s = c_{s,2}/v_1$ at $x \rightarrow \infty$.  As $u(x)$ ($b(x)$) is a monotonically decreasing (increasing) function of $x$, the above differential equation has a singularity at $u_s = u(x)$, if $M_A > M_{cr}$.  Therefore, the steady-state, magnetosonic shock cannot be maintained only by Ohmic dissipation in the MHD framework.  However, instead of Ohmic heating term, if we include the viscous heating term and perform a similar mathematical calculation, we can easily find that there is no singularity in the presence of viscous dissipation.

\section{Plasma instabilities at {nonrelativistic} shock front}\label{appendixB}
We now discuss some basic instabilities associated with the dynamical nature of particle reflection off {a nonrelativistic} shock front, because these reflected particles play an important role in the generation of various electromagnetic and electrostatic waves, which are responsible for plasma heating and particle acceleration \cite{2009ApJ...699..990B}. 
Among the many plasma instabilities expected in the shock region \cite{Wu84}, we first review the ion cyclotron beam instability that occurs if the relative velocity between the incoming and the reflected ions exceeds the Alfven speed.  

The dispersion relation for transverse electromagnetic waves propagating parallel to the ambient magnetic field, i.e., $\veck \| \vecb_0$, is given as,
\begin{equation}
  1 - \frac{k^2c^2}{\omega^2} -
     \sum_{j=i,e,b} \frac{\ompj^2}{(\omega - k v_{0j} \pm \omcj) \omega} \left( \frac{\omega - k v_{0j}}{\omega} \right) = 0,
\end{equation}
where $\omega$ and $k$ are the wave frequency and the wave vector for the transverse electromagnetic wave, and $\ompj=\sqrt{4 \pi n_j e_j^2/m_j}$ and $\omcj=e_j B_0/m_j c$ are the plasma frequency and the cyclotron frequency for the species $j$, respectively.  $v_{0j}$ is the bulk plasma flow for the species $j$ moving parallel to the ambient magnetic field.  We have assumed that the plasma temperature is cold.  Note that for ion and electron plasma at rest the dispersion relation in the low frequency limit, $\omega \ll \omce$, can be approximated by
\begin{equation}
  \omega^2 = k^2 V_A^2 \left(1 \pm \frac{\omega}{\omci} \right),
\end{equation}
where $V_A=B_0/\sqrt{4 \pi (m_i+m_e)n}$ is the Alfven speed, and $n=n_i=n_e$ under the charge neutrality. One can easily find two branches of the whistler wave and the ion cyclotron wave.

When we have a beam population, i.e., the reflected ions, in addition to the background ions and electrons, the above dispersion relation can be written as,
\begin{equation}
  \omega^2 = k^2 V_A^2 + V_A^2 \left(\frac{\delta \ompi^2}{c^2}\right)
     \left(\frac{\omega - kv_{0b}}{\omega - k v_{0b} + \omci}\right),
\end{equation}
where $v_{0b}$ is the bulk speed of the beam component in the shock upstream frame.  We have assumed the charge neutrality of $n_i + n_b = n_e$ and $n_i/n_b = (1-\delta)/\delta$.  The second term of the right-hand side represents the cyclotron resonance between the whistler wave and the beam ions, and the instability occurs around a point of intersection of $\omega = k V_A$ and $\omega - k v_{0b} + \omci=0$, if the beam velocity, $v_{0b}$, exceeds the Alfven speed.  For $\delta \ll 1$, we obtain,
\begin{eqnarray}
  \frac{\mathrm{Re}(\omega)}{\omci} \sim \frac{V_A}{v_{0b}-V_A} \sim \frac{1}{2 M_A}, \\
  \frac{I\mathrm{m}(\omega)}{\omci} \sim \left(\frac{\delta}{2}\right)^{1/3}.
\end{eqnarray}

The nonlinear evolution of the ion-cyclotron beam instability was studied using PIC simulations \citep{1984JGR....89.2673W,Hoshino85}.  For the linear stage of the instability they confirmed that the right-hand polarized Alfv\'en wave is excited on a time scale commensurate with the linear growth rate.  In the nonlinear stage they observed that the amplitude of the Alfv\'en wave reaches a value several times larger than the ambient magnetic field, $\delta B/B_0\gg 1$, and that the ion beam population relaxed into a diffuse ion population due to pitch angle scattering by the large-amplitude waves.

Not only the interaction between the incoming ions and the reflected ions, but also the interaction between ions and electrons can generate many waves.  The relative velocity between the reflected ions and incoming electrons provides free energy to generate a relatively high-frequency wave emission.  For electrostatic waves with unmagnetized ions and electrons, we have the dispersion relation of Buneman instability as,
\begin{equation}
  1-\frac{\ompe^2}{\omega^2}-(1-\delta)\frac{\ompi^2}{\omega^2}
   - \frac{\delta \ompi^2}{(\omega-k v_{0b})^2}=0,
\end{equation}
where $\delta$ is the number-density ratio of reflected and incoming ions.  For $\delta \ll 1$, we obtain
\begin{eqnarray}
   \frac{\mathrm{Re}(\omega)}{\ompe} \sim \delta^{1/3} \left(\frac{m_e}{16 m_i} \right)^{1/3}, \\
   \frac{\mathrm{Im}(\omega)}{\ompe} \sim \delta^{1/3} \left(\frac{3 m_e}{16 m_i} \right)^{1/3}, 
\end{eqnarray}
and the unstable wave number is $k \sim \ompe/v_{0b}$.  Note that if no ion beam population exists, we would find the normal Langmuir wave with $\omega^2 \simeq \ompe^2$ in cold-plasma approximation.  
The Buneman instability at high-Mach-number shocks was proposed by \citet{Papadopoulos88}, and PIC simulations were used to study electron heating and acceleration by nonlinear interactions {with Buneman waves} \citep{Shimada00,Schmitz02}.  They found that a large-amplitude, bipolar electric field in association with electron phase space holes can be generated during the nonlinear stage of the Buneman instability.  For high-Mach-number shocks such as planetary bow shocks in the outer heliosphere \cite{Masters16}, the electric field can become large enough compared to the ambient magnetic field to allow strong electron heating and acceleration \citep{Hoshino02}. Several authors discussed that the electron can be trapped in the large-amplitude electric field, and the trapped electron can subsequently gain energy from the shock motional electric field, leading to the generation of nonthermal electrons at high-Mach-number shocks such as supernova remnant shocks \cite{Hoshino02,Amano07,2013PhRvL.111u5003M,2015Sci...347..974M}. 

While the Buneman instability/two-stream instability driven by relative motion between ions and electrons is known to play an important role in the pre-heating of electrons at high-Mach-number shocks, it may not be necessarily excited at the Earth's bow shock where the Mach number is less than about 10.  Instead of the Buneman instability, the modified two-stream instability can be excited for shocks with moderate Mach numbers. The modified two-stream instability occurs under the coupling between the Buneman mode and the whistler wave for oblique propagation.  The dispersion relation is,
\begin{eqnarray}
  \left(1 - \frac{\ompe^2}{\omega^2}
     - (1-\delta) \frac{\ompi^2}{\omega^2}
      - \frac{\delta \ompi^2}{(\omega-k v_{0b})^2} \right)
 \left(1 - \frac{\omce^2 \cos^2 \theta}{\omega^2(1+H_e^2)^2} \right) \nonumber \\
 -\frac{\omce^2 \sin^2 \theta}{\omega^2(1+H_e)}
 \left(1- (1-\delta) \frac{\ompi^2}{\omega^2}
  - \frac{\delta \ompi^2}{(\omega-k v_{0b}^2)} \right) =0,
\end{eqnarray}
where $H_e=\omce^2/k^2c^2$.  We have assumed that ions are unmagnetized and $\ompe^2 \gg \omce > \omega \gg \omci$ and $H_e < 1$.
The angle between the ambient magnetic field and the wave vector, $k$, is $\theta$, and for simplicity we have dealt with three components, the incoming solar wind ions and electrons and the reflected ions from the shock front, where $\delta$ is the number-density ratio of reflected to incoming ions.  It is clear that the first two terms in the left-hand side represent the coupling between the Buneman mode and the obliquely propagating whistler wave.  The oscillation frequency, $\mathrm{Re}({\omega})$, and the growth rate, $\mathrm{Im}({\omega})$, can be given as,
\begin{eqnarray}
  \frac{\mathrm{Re}(\omega)}{\omci} \sim \frac{m_i}{m_e}
  \frac{\cos \theta}{\sqrt{1+H_e}} G^{-1/2}, \\
  \frac{\mathrm{Im}(\omega)}{\omci} \sim \frac{\sqrt{3}}{2} \delta^{1/3}
    \left(\frac{m_i}{m_e}\right)^{2/3}
    \left(\frac{\cos \theta \sin^2 \theta}
         {2\sqrt{1+H_e}} \right)^{1/3} G^{-1/2},
\end{eqnarray}
where $G= (1+H_e)+(\omce^2/\ompe^2)\sin^2 \theta$. PIC simulations for a finite temperature plasma demonstrated that the modified two-stream instability can be generated in the Earth's bow shock and plays an important role in the electron heating parallel to the magnetic field \citep{Matsukiyo03,Scholer03}.  

\mh{
\section{Plasma instabilities at a relativistic shock front}\label{appendixC}
}
Here we introduce two examples of plasma instabilities that operate at relativistic shocks: one is the Weibel instability \cite{Weibel59} which can be excited in an unmagnetized or a weakly magnetized shock, and the other is the synchrotron maser instability \cite{Schneider59} which generates a large-amplitude precursor wave in a magnetized shock.

As we are interested in the plasma instabilities at the shock front, we may focus on the velocity distribution function that consists of both incoming and reflected components.  For a (weakly) magnetized perpendicular shock, the incoming particles are reflected back to the upstream region by the Lorentz force/mirror force During the gyro-motion of the reflected particles we may assume that the velocity distribution function can be approximated by the ring distribution, namely,
\begin{equation}
  F_j(u_{\|},u_{\perp}) = \frac{1}{2 \pi u_{0j}} \delta(u_\|) \delta(u_\perp -u_{0j}),
\end{equation}
where $\|$ and $\perp$ denote the directions parallel and the perpendicular to the ambient magnetic field, and $u_{0j}$ may be almost equal to the incoming flow speed/shock upstream flow speed in the shock rest frame.  The distribution function is normalized as follows,
\begin{equation}
  \int_{-\infty}^{\infty} \int_{0}^{\infty} 2 \pi F_j(u_{\|},u_{\perp}) u_{\perp} d u_{\perp} d u_{\|} =1.
\end{equation}

We assume a uniform plasma for the ring-distribution region, and the ambient magnetic field is parallel to $z$ axis.  Then the plasma dispersion relation for a longitudinal wave propagating parallel to the ambient magnetic field can be given by \cite{Yoon87,2018ApJ...858...93I},
\begin{align}
  1-\frac{c^2 k^2}{\omega^2} &  \nonumber \\
  &=\sum_{j=i,e} \frac{\ompj^2}{\omega^2} \int_{-\infty}^{\infty} \int_{0}^{\infty}
  \frac{(\omega - k_{\|} v_{\|}) \partial F_s/\partial u_{\|} +
    k_{\|} v_{\perp} \partial F_s /\partial u_{\|}} {\omega - k_{\|} v_{\|} \pm \omcj}
  \pi u_{\perp}^2 d u_{\perp} d u_{\|} \nonumber \\
  &=\sum_{j=i,e} \frac{\ompj^2}{\omega ( \omega \pm \omcj)} +
  \frac{1}{2} \left(1 - \frac{c^2 k_{\|}^2}{\omega^2} \right) \sum_{j=i,e} \beta_{0j}^2
  \frac{\ompj^2}{(\omega \pm \omcj)^2} ,
\end{align}
which includes the relativistic plasma frequency, $\ompj = \sqrt{4 \pi n_j e^2/ m_j \gamma}$, the relativistic cyclotron frequency, $\omcj = e B/m_j c \gamma$, the Lorentz factor, $\gamma= \sqrt{1 + u_{\|}^2 + u_{\perp}^2}$, the parallel and perpendicular velocities, $v_{\|}/c = u_{\|}/\gamma$, $v_{\perp}/c=u_{\perp}/\gamma$, and the normalized speed, $\beta_{0j} = v_{0j}/c$.

For a pair plasma in the unmagnetized limit, $\Omega_{c\pm} \to 0$, the dispersion relation above can be simply written as,
\begin{equation}
 (\omega^2 - c^2 k_\|^2)(\omega^2 + \beta_0^2 \omega_{p\pm}^2) - 2 \omega^2 \omega_{p\pm}^2 = 0.
\end{equation}
This quadratic equation for $\omega^2$ has two solutions of $\omega^2 > 0$ and $\omega^2 < 0$ for $\beta_0 >0$.  The mode with $\omega^2<0$ shows the Weibel instability, and we find the growth rate,
\begin{equation}
  \frac{ \rm{Im}(\omega)}{\omega_{p\pm}} \simeq \beta_0,
\end{equation}
for $c k_{\|}/\omega_{c\pm} > 1$.

For the magnetized case with finite $\Omega_{c\pm}$ and $\Omega_{c\pm}^2/\omega_{p\pm}^2 \gg 1$, the growth rate of the magnetized Weibel instability is given by $\rm{Im}(\omega) \sim \beta_0 \omega_{p\pm}/\sqrt{2}$ and the oscillation frequency $\rm{Re}(\omega) \sim \Omega_{c\pm}$.  The basic behavior of the wave growth is similar to the Weibel instability in an unmagnetized plasma.

Next we show the synchrotron maser instability propagating perpendicular to the ambient magnetic field with $\vec{k}_{\perp} \| \vec{e}_x$.  The dispersion relation of the wave with electric field polarized in the $x-y$ plane is given by \cite{Baldwin69},
\begin{equation}
  \frac{c^2 k_{\perp}^2}{\omega^2} = \epsilon_{yy} - \frac{\epsilon_{xy} \epsilon_{yx}}{\epsilon_{xx}},
\end{equation}
where the components of the dielectric tensor are,
\begin{equation}
  \epsilon_{ij}= \delta_{ij} + 2 \pi \sum_{s=i,e} \sum_{n=-\infty}^{\infty} \int_{-\infty}^{\infty} \int_{0}^{\infty}
  \frac{\partial F_s}{\partial u_{\perp}} \frac{\omega_{ps}^2}{\omega(\omega- n \Omega_{cs})} \Psi_{snij}
   u_{\perp}^2 d u_{\perp} d u_{\|},  
\end{equation}
where
\begin{equation}
  \Psi_{snij}= \left(  \begin{array}{cc}
    (n/z_s)J_n^2(z_s) & i (n/z_s) J_n(z_s) J_n^{\prime}(z_s) \\
    -i (n/z_s) J_n(z_s) J_n^{\prime}(z_s)  & (J_n^{\prime}(z_s))^2
                       \end{array} \right).
\end{equation}
$J_n(z_s)$ is the ordinary Bessel function of the first kind with index $n$, and the argument is $z_s = k_{\perp} v_{\perp s}/\Omega_{cs}$ for a particle of species $s$.

By substituting the cold ring distribution function and integrating by parts, we obtain the elements of the dielectric tensor as \cite{Hoshino91},
\begin{eqnarray}
  \epsilon_{xx} &=& 1 -\sum_{s=i,e} \sum_{n=-\infty}^{\infty} \frac{n^2 \omega_{ps}^2}{\omega (\omega - n \Omega_{cs})}
  \left( 2\frac{J_n}{z_s}J_n^{\prime}-\beta_s^2 \frac{\omega}{\omega-n\Omega_{cs}}J_n^2 \right), \\
  \epsilon_{xy} &=& -\epsilon_{yx}=i\sum_{s=i,e} \sum_{n=-\infty}^{\infty} \frac{n \omega_{ps}^2}{\omega (\omega-n \Omega_{cs})} \nonumber \\
  &\times& \left( \frac{J_n}{z_s}J_n^{\prime}+(J_n^{\prime})^2+J_n J_n^{``} 
  -\beta_s^2 \frac{\omega}{\omega-n\Omega_{cs}} \frac{J_n}{z_s}J_n^{'} \right), \\
  \epsilon_{yy} &=& 1 -\sum_{s=i,e} \sum_{n=-\infty}^{\infty} \frac{\omega_{ps}^2}{\omega (\omega - n \Omega_{cs})} \nonumber \\
  &\times& \left( 2 (J_n^{\prime})^2 + 2 z_s J_n^{'} J_n^{``} - \beta_s^2 \frac{\omega}{\omega-n\Omega_{cs}} (J_n^{'})^2 \right). 
\end{eqnarray}

For a positron-electron plasma, we can easily show that $\epsilon_{xy}=0$ because of the charge and mass symmetry.  In this case, the transverse extraordinary mode is decoupled from the longitudinal upper-hybrid mode, and the dispersion relations of the transverse extraordinary mode and the longitudinal upper-hybrid mode are respectively given by
\begin{equation}
  \epsilon_{yy} =  \frac{c^2 k_{\perp}^2}{\omega^2},  \nonumber
\end{equation}
and
\begin{equation}
  \epsilon_{xx} = 0.  \nonumber
\end{equation}
For a pair plasma with zero ring velocity of $\beta_{\pm}=0$ and $\gamma_{\pm}=1$, we can easily obtain the component of the dielectric component as,
\begin{equation}
  \epsilon_{xx}=\epsilon_{yy}=1 - \frac{2 \omega_{p\pm}^2}{\omega^2-\Omega_{c\pm}^2}.
\end{equation}
Therefore, the dispersion relation for the transverse extraordinary mode becomes
\begin{equation}
  \frac{c^2 k_{\perp}^2}{\omega^2} = 1 - \frac{2 \omega_{p\pm}^2}{\omega^2-\Omega_{c\pm}^2},
\end{equation}
while the dispersion relation for the longitudinal upper-hybrid mode is,
\begin{equation}
 \omega^2 = 2 \omega_{p\pm}^2+\Omega_{c\pm}^2.
\end{equation}
These dispersion relations are consistent with those derived from the cold-fluid equations.

For a relativistic cold ring and pair plasma with $\gamma_{\pm} \gg 1$, the dispersion relations for the transverse mode and the longitudinal mode become, respectively,
\begin{equation}
  \frac{k_{\perp}^2 c^2}{\omega^2} \cong 1 + \sum_{s=e^{\pm}} \frac{\Omega_{ps}^2}{\omega^2} \frac{v_{0s}^2}{c^2} (J_n^{\prime})^2
  \left(1 - \frac{n \Omega_{cs}}{\omega} \right)^{-2},
\end{equation}
and
\begin{equation}
  1 + \sum_{s=e^{\pm}} \frac{\Omega_{ps}^2}{\omega^2} \frac{n^2 v_{0s}^2}{z_s^2 c^2} J_n^2
  \left(1 - \frac{n \Omega_{cs}}{\omega} \right)^{-2} \cong 0.
\end{equation}

We may assume that unstable modes appear around the frequency satisfying the cyclotron resonance condition, $\omega \simeq n \Omega_{c\pm}$.  By assuming that $\omega= k_{\perp}c = n \Omega_{c\pm} + \delta$ and $\delta \ll n \Omega_{c\pm}$, we obtain the increment of $\delta$ for the transverse extraordinary mode as,
\begin{equation}
  \frac{\delta}{\Omega_{c\pm}} \cong \left\{ \begin{array}{@{\,}lll}
      \left(\frac{3}{2^{11}} \right)^{1/9} \left(\frac{\Gamma(2/3)}{\pi} \right)^{2/3}
      \frac{1 + i \sqrt{3}}{\sigma_{\pm}^{1/3}} \left( \frac{\Omega_{c\pm}}{\omega} \right)^{1/9} &
      \rm{for}~ n = \frac{\omega}{\Omega_{c\pm}} \ll \gamma_{\pm}^3 \\
         \\
      \left(\frac{1}{2^4 \pi \gamma_{\pm}} \right)^{1/3}
      \frac{1 + i \sqrt{3}}{\sigma_{\pm}^{1/3}} \exp \left( - \frac{2 n}{9 \gamma_{\pm}} \right) &
      \rm{for}~ n = \frac{\omega}{\Omega_{c\pm}} \gg \gamma_{\pm}^3  \\
                                           \end{array} \right.
\end{equation}
where $\sigma_{\pm}=\Omega_{c\pm}^2/\omega_{p\pm}^2$.  We have used the asymptotic forms of the Bessel functions for large index, $n \gg 1$, and the argument $z_s= n \varepsilon$ with $n \gg \varepsilon > 0$.  From the solution above, we find that the high-harmonic modes up to $n \sim \gamma_{\pm}^3$ have significantly large growth rates, while the growth rates quickly decay in $n \gg \gamma_{\pm}^3$.  It should be noted that the growth rates for high harmonic modes may be subject to a finite temperature of the ring distribution function, and that the growth rates decrease with increasing temperature \cite[e.g.][]{Amato06}

On the other hand, the increment of $\delta$ for the longitudinal mode is,
\begin{equation}
  \frac{\delta}{\Omega_{c\pm}} \cong \left\{ \begin{array}{@{\,}lll}
      \frac{i}{\sigma_{\pm}^{1/2}} 
      \left(\frac{\Gamma(1/3)}{2^{2/3} 3^{1/6} \pi} \right)
      \left( \frac{\Omega_{c\pm}}{\omega} \right)^{1/3} &
      \rm{for}~ n = \frac{\omega}{\Omega_{c\pm}} \ll \gamma_{\pm}^3 \\
         \\
      \frac{i}{\sigma_{\pm}^{1/2}} 
      \left(\frac{\gamma_{\pm}}{2 \pi} \right)
      \left( \frac{\Omega_{c\pm}}{\omega} \right)^{1/2}
      \exp \left(- \frac{2 n}{9 \gamma_{\pm}^3} \right) &
      \rm{for}~ n = \frac{\omega}{\Omega_{c\pm}} \gg \gamma_{\pm}^3  \\
                                           \end{array} \right.
\end{equation}
By comparing the growth rate of the longitudinal mode to that of the transverse mode, we find a faster reduction with increasing $n$ for the longitudinal mode than for the transverse mode.


\section*{Acknowledgements}
J.N. and M.H. acknowledge support of the Polish-Japanese joint research project for the years 2017-2019 under the agreement on scientific cooperation between the Polish Academy of Sciences and the Japan Society for the Promotion of Science (JSPS). J.N. also has been partially supported by Narodowe Centrum Nauki through research project DEC-2013/10/E/ST9/00662.

\section*{References}
\bibliographystyle{elsarticle-num}

\begin{thebibliography}{10}
\expandafter\ifx\csname url\endcsname\relax
  \def\url#1{\texttt{#1}}\fi
\expandafter\ifx\csname urlprefix\endcsname\relax\def\urlprefix{URL }\fi
\expandafter\ifx\csname href\endcsname\relax
  \def\href#1#2{#2} \def\path#1{#1}\fi

\bibitem[Achterberg et al.(2001)]{2001MNRAS.328..393A} Achterberg, A., Gallant, Y.~A., Kirk, J.~G., \& Guthmann, A.~W.\ 2001, \mnras, 328, 393
\bibitem[Aharonian et al.(1994)]{1994ApJ...423L...5A} Aharonian, F.~A., Coppi, P.~S., \& Voelk, H.~J.\ 1994, \apjl, 423, L5 \bibitem[Akamatsu et al.(2017)]{2017A&A...600A.100A} Akamatsu, H., Mizuno, M., Ota, N., et al.\ 2017, \aap, 600, A100 
\bibitem[Alsop \& Arons(1988)]{1988PhFl...31..839A} Alsop, D., \& Arons, J.\ 1988, Physics of Fluids, 31, 839
\bibitem[Alves et al.(2012)]{2012ApJ...746L..14A} Alves, E.~P., Grismayer, T., Martins, S.~F., et al.\ 2012, \apjl, 746, L14 
\bibitem[Alves et al.(2015)]{2015PhRvE..92b1101A} Alves, E.~P., Grismayer, T., Fonseca, R.~A., \& Silva, L.~O.\ 2015, \pre, 92, 021101 
\bibitem[Alves et al.(2016)]{2016PPCF...58a4025A} Alves, E.~P., Grismayer, T., Silveirinha, M.~G., Fonseca, R.~A., \& Silva, L.~O.\ 2016, Plasma Physics and Controlled Fusion, 58, 014025 
\bibitem[Amano \& Hoshino(2007)]{Amano07} Amano, T., \& Hoshino, M.\ 2007, \apj, 661, 190
\bibitem[Amano \& Hoshino(2009)]{2009ApJ...690..244A} Amano, T., \& Hoshino, M.\ 2009, \apj, 690, 244 
\bibitem[Amano \& Hoshino(2010)]{2010PhRvL.104r1102A} Amano, T., \& Hoshino, M.\ 2010, Physical Review Letters, 104, 181102 
\bibitem[Amato \& Arons(2006)]{Amato06} Amato, E., \& Arons, J.\ 2006, \apj, 653, 325
\bibitem{Arons79} Arons, J.\ 1979, \ssr, 24, 437
\bibitem[Bai et al.(2015)]{2015ApJ...809...55B} Bai, X.-N., Caprioli, D., Sironi, L., \& Spitkovsky, A.\ 2015, \apj, 809, 55 
\bibitem[Ball \& Melrose(2001)]{2001PASA...18..361B} Ball, L., \& Melrose, D.~B.\ 2001, PASA, 18, 361 
\bibitem[Ball et al.(2018)]{2018ApJ...862...80B} Ball, D., Sironi, L., \& {\"O}zel, F.\ 2018, \apj, 862, 80
\bibitem[Balogh \& Treumann(2013)]{Balogh13} Balogh, A., \& Treumann, R.~A.\ 2013, Physics of Collisionless Shocks: Space Plasma Shock Waves, ISSI Scientific Report Series, Volume 12.~ISBN 978-1-4614-6098-5.~Springer Science+Business Media New York, 2013
\bibitem[Baldwin et al.(1969)]{Baldwin69} Baldwin, D.~E., Bernstein, I.~B., \& Weenink, M.~P.~H.\ 1969, Advances in Plasma Physics, 3, 1
\bibitem[Baumann et al.(2013)]{2013ApJ...771...93B} Baumann, G., Haugb{\o}lle, T., \& Nordlund, {\AA}.\ 2013, \apj, 771, 93
\bibitem[Beckwith \& Stone(2011)]{2011ApJS..193....6B} Beckwith, K., \& Stone, J.~M.\ 2011, \apjs, 193, 6 
\bibitem[Bednarz \& Ostrowski(1998)]{1998PhRvL..80.3911B} Bednarz, J., \& Ostrowski, M.\ 1998, Physical Review Letters, 80, 3911 
\bibitem[Bell(1978)]{1978MNRAS.182..147B} Bell, A.~R.\ 1978, MNRAS, 182, 147 
\bibitem[Bell(2004)]{2004MNRAS.353..550B} Bell, A.~R.\ 2004, \mnras, 353, 550
\bibitem[Bell(2005)]{2005MNRAS.358..181B} Bell, A.~R.\ 2005, \mnras, 358, 181 
\bibitem[Belyaev(2015)]{Belyaev15} Belyaev, M.~A.\ 2015, \mnras, 449, 2759
\bibitem[Bernstein \& Ahearne (1968)]{Bernstein68} Bernstein, I.B., Ahearne, J.F. 1968, Ann. Phys. 49, 1
\bibitem[Binns et al.(2016)]{2016Sci...352..677B} Binns, W.~R., Israel, M.~H., Christian, E.~R., et al.\ 2016, Science, 352, 677 
\bibitem[Birdsall \& Langdon(2005)]{bl05}  Birdsall, C.K., Langdon, A.B.\ 2005, {\it Plasma physics via computer simulations}, IoP
\bibitem[Biskamp(1993)]{1993noma.book.....B} Biskamp, D.\ 1993, Cambridge Monographs on Plasma Physics, Cambridge [England]; New York, NY: Cambridge University Press
\bibitem[Biskamp \& Welter(1972)]{Biskamp72} Biskamp, D., \& Welter, H.\ 1972, Physical Review Letters, 28, 410
\bibitem[Blandford \& Eichler(1987)]{1987PhR...154....1B} Blandford, R., \& Eichler, D.\ 1987, \physrep, 154, 1 
\bibitem[Blandford et al.(2017)]{Blandford17} Blandford, R., Yuan, Y., Hoshino, M., \& Sironi, L.\ 2017, \ssr, 207, 291
\bibitem[Blasi et al.(2012)]{2012ApJ...755..121B} Blasi, P., Morlino, G., Bandiera, R., Amato, E., \& Caprioli, D.\ 2012, \apj, 755, 121
\bibitem[Bohdan(2017)]{Bohdan2017} Bohdan, A., PhD thesis\ 2017
\bibitem[Bohdan et al.(2017)]{2017ApJ...847...71B} Bohdan, A., Niemiec, J., Kobzar, O., \& Pohl, M.\ 2017, \apj, 847, 71 
\bibitem[Bohdan et al.(2019)]{2019ApJ...878....5B} Bohdan, A., Niemiec, J., Pohl, M., et al.\ 2019, \apj, 878, 5
\bibitem[Bohdan et al.(2019)]{2019ApJ...885...10B} Bohdan, A., Niemiec, J., Pohl, M., et al.\ 2019, \apj, 885, 10
\bibitem[Boozer(2018)]{2018JPlPh..84a7102B} Boozer, A.~H.\ 2018, Journal of Plasma Physics, 84, 715840102
\bibitem[Boris(1970)]{Boris1970} Boris, J.~P., ``Relativistic plasma simulation-optimization of a hybrid code,'' in Proceedings of the Fourth Conference on Numerical Simulation Plasmas (Naval Research Laboratory, Washington, D.C., 1970) pp. 3–67
\bibitem[Brackbill \& Forslund(1982)]{1982JCoPh..46..271B} Brackbill, J.~U., \& Forslund, D.~W.\ 1982, Journal of Computational Physics, 46, 271 
\bibitem[Brackbill \& Cohen(1985)]{1985mts..conf.....B} Brackbill, J.~U., \& Cohen, B.~I.\ 2014, Multiple time scales, Vol.~3., Academic Press
\bibitem[Bret et al.(2005)]{2005PhRvE..72a6403B} Bret, A., Firpo, M.-C., \& Deutsch, C.\ 2005, \pre, 72, 016403
\bibitem[Bret(2009)]{2009ApJ...699..990B} Bret, A.\ 2009, \apj, 699, 990
\bibitem[Bret et al.(2014)]{2014PhPl...21g2301B} Bret, A., Stockem, A., Narayan, R., et al.\ 2014, Physics of Plasmas, 21, 072301
\bibitem[Bret, \& Dieckmann(2017)]{2017PhPl...24f2105B} Bret, A., \& Dieckmann, M.~E.\ 2017, Physics of Plasmas, 24, 062105
\bibitem[Bret, \& Pe'er(2018)]{2018JPlPh..84c9011B} Bret, A., \& Pe'er, A.\ 2018, Journal of Plasma Physics, 84, 905840311
\bibitem[Bret, \& Narayan(2018)]{2018JPlPh..84f9004B} Bret, A., \& Narayan, R.\ 2018, Journal of Plasma Physics, 84, 905840604
\bibitem[Broderick et al.(2012)]{Broderick12}Broderick, A. E., Chang, P., \& Pfrommer, C.\ 2012, ApJ, 752, 22
\bibitem[Broderick et al.(2018)]{Broderick18} Broderick, A.~E., Tiede, P., Chang, P., et al.\ 2018, \apj, 868, 87 
\bibitem[Brunetti \& Jones(2014)]{2014IJMPD..2330007B} Brunetti, G., \& Jones, T.~W.\ 2014, Int. J. Mod. Phys. D, 23, 1430007-98
\bibitem[Burgess {et~al.}(1989)]{burgess} Burgess, D., Wilkinson, W.~P., \& Schwartz, S.~J. 1989, Journal of Geophysical
  Research: Space Physics, 94, 8783
\bibitem{Burgess07} Burgess, D., \& Scholer, M.\ 2007, Phys. Pl., 14, 012108
\bibitem[Burlaga et al.(2005)]{Burlaga05} Burlaga, L.~F., Ness, N.~F., Acu{\~n}a, M.~H., et al.\ 2005, Science, 309, 2027 
\bibitem[Bykov et al.(2014)]{2014ApJ...789..137B} Bykov, A.~M., Ellison, D.~C., Osipov, S.~M., \& Vladimirov, A.~E.\ 2014, \apj, 789, 137 
\bibitem[Bykov et al.(2012)]{2012SSRv..173..309B} Bykov, A., Gehrels, N., Krawczynski, H., et al.\ 2012, \ssr, 173, 309 
\bibitem[Bykov et al.(2011)]{2011MNRAS.410...39B} Bykov, A.~M., Osipov, S.~M., \& Ellison, D.~C.\ 2011, \mnras, 410, 39
\bibitem[Capdessus et al.(2012)]{2012PhRvE..86c6401C} Capdessus, R., d'Humi{\`e}res, E., \& Tikhonchuk, V.~T.\ 2012, \pre, 86, 036401
\bibitem[Caprioli \& Spitkovsky(2013)]{2013ApJ...765L..20C} Caprioli, D., \& Spitkovsky, A.\ 2013, \apjl, 765, L20 
\bibitem[Caprioli \& Spitkovsky(2014a)]{2014ApJ...783...91C} Caprioli, D., \& Spitkovsky, A.\ 2014, \apj, 783, 91 
\bibitem[Caprioli \& Spitkovsky(2014b)]{2014ApJ...794...46C} Caprioli, D., \& Spitkovsky, A.\ 2014, \apj, 794, 46 
\bibitem[Caprioli et al.(2015)]{2015ApJ...798L..28C} Caprioli, D., Pop, A.-R., \& Spitkovsky, A.\ 2015, \apjl, 798, L28
\bibitem[Caprioli et al.(2017)]{2017PhRvL.119q1101C} Caprioli, D., Yi, D.~T., \& Spitkovsky, A.\ 2017, Physical Review Letters, 119, 171101 
\bibitem[Cerutti \& Philippov(2017)]{Cerutti17} Cerutti, B., \& Philippov, A.~A.\ 2017, \aap, 607, A134
\bibitem[Cerutti et al.(2013)]{2013ApJ...770..147C} Cerutti, B., Werner, G.~R., Uzdensky, D.~A., \& Begelman, M.~C.\ 2013, \apj, 770, 147
\bibitem[Chang et al.(2008)]{2008ApJ...674..378C} Chang, P., Spitkovsky, A., \& Arons, J.\ 2008, \apj, 674, 378 
\bibitem[Chen et al.(2002)]{2002PhRvL..89p1101C} Chen, P., Tajima, T., \& Takahashi, Y.\ 2002, Physical Review Letters, 89, 161101
\bibitem[Chen et al.(2013)]{2013JCoPh.236..220C} Chen, M., Cormier-Michel, E., Geddes, C.~G.~R., et al.\ 2013, Journal of Computational Physics, 236, 220
\bibitem[Chen, \& Beloborodov(2014)]{Chen14} Chen, A.~Y., \& Beloborodov, A.~M.\ 2014, \apjl, 795, L22
\bibitem[Chen et al.(2017)]{2017JGRA..12210318C} Chen, Y., T{\'o}th, G., Cassak, P., et al.\ 2017, Journal of Geophysical Research (Space Physics), 122, 10 
\bibitem[Classen \& Mann(1997)]{1997A&A...322..696C} Classen, H.-T., \& Mann, G.\ 1997, \aap, 322, 696 
\bibitem[Comisso, \& Sironi(2018)]{2018PhRvL.121y5101C} Comisso, L., \& Sironi, L.\ 2018, \prl, 121, 255101
\bibitem[Courant et al.(1928)]{cfl}Courant, R., Friedrichs, K., Lewy, H.\ 1928, Mathematische Annalen (in German), 100, 32–74
\bibitem[Crumley et al.(2019)]{2019MNRAS.485.5105C} Crumley, P., Caprioli, D., Markoff, S., et al.\ 2019, \mnras, 485, 5105
\bibitem[Daldorff et al.(2014)]{2014JCoPh.268..236D} Daldorff, L.~K.~S., T{\'o}th, G., Gombosi, T.~I., et al.\ 2014, Journal of Computational Physics, 268, 236
\bibitem[Dawson(1962)]{da62} Dawson, J.M.\ 1962, Phys. Fl. 5, 445 
\bibitem[Dawson(1983)]{da83} Dawson, J.M.\ 1983, Rev. Mod. Phys. 55-2, 403 
\bibitem{Decker08} Decker, R.~B., Krimigis, S.~M., Roelof, E.~C., et al.\ 2008, \nat, 454, 67
\bibitem[Dieckmann et al.(2010)]{2010A&A...509A..89D} Dieckmann, M.~E., Murphy, G.~C., Meli, A., \& Drury, L.~O.~C.\ 2010, \aap, 509, A89
\bibitem[Dieckmann, \& Bret(2018)]{2018MNRAS.473..198D} Dieckmann, M.~E., \& Bret, A.\ 2018, \mnras, 473, 198
\bibitem[Doma{\'n}ski \& Badziak(2018)]{2018PhLA..382.3412D} Doma{\'n}ski, J., \& Badziak, J.\ 2018, Physics Letters A, 382, 3412 
\bibitem[Drake et al.(2006)]{2006Natur.443..553D} Drake, J.~F., Swisdak, M., Che, H., \& Shay, M.~A.\ 2006, \nat, 443, 553 
\bibitem[Drake et al.(2010)]{Drake10} Drake, J.~F., Opher, M., Swisdak, M., \& Chamoun, J.~N.\ 2010, \apj, 709, 963 
\bibitem[Drouin et al.(2010)]{2010JCoPh.229.4781D} Drouin, M., Gremillet, L., Adam, J.-C., \& H{\'e}ron, A.\ 2010, Journal of Computational Physics, 229, 4781
\bibitem[Drury \& Strong(2017)]{2017A&A...597A.117D} Drury, L.~O.~'., \& Strong, A.~W.\ 2017, \aap, 597, A117 
\bibitem[Dyakov(1954)]{Dyakov54} D'yakov, S.~P.\ 1954,  Zh. Eksp. Teor. Fiz. 27, 288
\bibitem{Edmiston84} Edmiston, J.~P., \& Kennel, C.~F.\ 1984, J. Pl. Ph., 32, 429
\bibitem[Ellison \& Double(2004)]{2004APh....22..323E} Ellison, D.~C., \& Double, G.~P.\ 2004, Astroparticle Physics, 22, 323 
\bibitem[Elyiv et al.(2009)]{Elyiv09}Elyiv, A., Neronov, A., \& Semikoz, D. V.\ 2009, Phys. Rev. D, 80, 023010
\bibitem[Esirkepov(2001)]{2001CoPhC.135..144E} Esirkepov, T.~Z.\ 2001, Computer Physics Communications, 135, 144
\bibitem[Fermi(1949)]{1949PhRv...75.1169F} Fermi, E.\ 1949, Physical Review, 75, 1169 
\bibitem{Fisk74} Fisk, L.~A., Kozlovsky, B., \& Ramaty, R.\ 1974, \apjl, 190, L35
\bibitem[Fleishman(2006)]{2006ApJ...638..348F} Fleishman, G.~D.\ 2006, \apj, 638, 348
\bibitem[Fonseca et al.(2013)]{2013PPCF...55l4011F} Fonseca, R.~A., Vieira, J., Fiuza, F., et al.\ 2013, Plasma Physics and Controlled Fusion, 55, 124011
\bibitem[Frederiksen et al.(2004)]{2004ApJ...608L..13F} Frederiksen, J.~T., Hededal, C.~B., Haugb{\o}lle, T., \& Nordlund, {\AA}.\ 2004, \apjl, 608, L13
\bibitem[Frederiksen et al.(2010)]{2010ApJ...722L.114F} Frederiksen, J.~T., Haugb{\o}lle, T., Medvedev, M.~V., \& Nordlund, {\AA}.\ 2010, \apjl, 722, L114 
\bibitem[Gallant \& Achterberg(1999)]{1999MNRAS.305L...6G} Gallant, Y.~A., \& Achterberg, A.\ 1999, \mnras, 305, L6
\bibitem[Gardner \& Kruskal(1964)]{Gardner64} Gardner, C.~S., \& Kruskal, M.~D.\ 1964, Physics of Fluids, 7, 700 
\bibitem[Gargat{\'e} et al.(2010)]{2010ApJ...711L.127G} Gargat{\'e}, L., Fonseca, R.~A., Niemiec, J., et al.\ 2010, \apjl, 711, L127 
\bibitem[Germaschewski et al.(2016)]{2016JCoPh.318..305G} Germaschewski, K., Fox, W., Abbott, S., et al.\ 2016, Journal of Computational Physics, 318, 305 
\bibitem[Ghavamian et al.(2000)]{2000ApJ...535..266G} Ghavamian, P., Raymond, J., Hartigan, P., \& Blair, W.~P.\ 2000, \apj, 535, 266 
\bibitem[Giacalone \& Jokipii(2007)]{2007ApJ...663L..41G} Giacalone, J., \& Jokipii, J.~R.\ 2007, \apjl, 663, L41 
\bibitem[Godfrey \& Vay(2013)]{2013JCoPh.248...33G} Godfrey, B.~B., \& Vay, J.-L.\ 2013, Journal of Computational Physics, 248, 33
\bibitem[Gould \& Schr\'eder(1966)]{Gould66} Gould, R. J., \& Schr\'eder, S.\ 1966, PhRvL, 16, 748
\bibitem[Greenwood et al.(2004)]{2004JCoPh.201..665G} Greenwood, A.~D., Cartwright, K.~L., Luginsland, J.~W., \& Baca, E.~A.\ 2004, Journal of Computational Physics, 201, 665 
\bibitem[Grismayer et al.(2013)]{2013PhRvL.111a5005G} Grismayer, T., Alves, E.~P., Fonseca, R.~A., \& Silva, L.~O.\ 2013, Physical Review Letters, 111, 015005 
\bibitem[Grismayer et al.(2017)]{2017PhRvE..95b3210G} Grismayer, T., Vranic, M., Martins, J.~L., Fonseca, R.~A., \& Silva, L.~O.\ 2017, \pre, 95, 023210 
\bibitem[Guo et al.(2014a)]{Guo-14a} Guo,~X., Sironi,~L., \& Narayan,~R.\ 2014, \apj, 794, 153 
\bibitem[Guo et al.(2014b)]{Guo-14b} Guo,~X., Sironi,~L., \& Narayan,~R.\ 2014, \apj, 797, 47 
\bibitem[Guo et al.(2014c)]{Guo14} Guo, F., Li, H., Daughton, W., \& Liu, Y.-H.\ 2014, Physical Review Letters, 113, 155005
\bibitem[Guo et al.(2017)]{Guo-17} Guo,~X., Sironi,~L., \& Narayan,~R.\ 2017, \apj, 851, 134 
\bibitem[Guo et al.(2018)]{Guo-18} Guo,~X., Sironi,~L., \& Narayan,~R.\ 2018, \apj, 858, 95
\bibitem[Ha et al.(2018)]{2018ApJ...864..105H} Ha, J.-H., Ryu, D., Kang, H., et al.\ 2018, \apj, 864, 105
\bibitem[Hakobyan et al.(2019)]{2019ApJ...877...53H} Hakobyan, H., Philippov, A., \& Spitkovsky, A.\ 2019, \apj, 877, 53
\bibitem[Haugb{\o}lle(2011)]{2011ApJ...739L..42H} Haugb{\o}lle, T.\ 2011, \apjl, 739, L42
\bibitem[Haugb{\o}lle et al.(2013)]{2013PhPl...20f2904H} Haugb{\o}lle, T., Frederiksen, J.~T., \& Nordlund, \AA.\ 2013, Physics of Plasmas, 20, 062904
\bibitem[Heavens \& Drury(1988)]{1988MNRAS.235..997H} Heavens, A.~F., \& Drury, L.~O.\ 1988, \mnras, 235, 997
\bibitem[Hededal(2005)]{2005PhDT.........2H} Hededal, C.\ 2005, Ph.D.~Thesis, arXiv:astro-ph/0506559
\bibitem[Hededal et al.(2004)]{2004ApJ...617L.107H} Hededal, C.~B., Haugb{\o}lle, T., Frederiksen, J.~T., \& Nordlund, {\AA}.\ 2004, \apjl, 617, L107
\bibitem[Hededal \& Nishikawa(2005)]{2005ApJ...623L..89H} Hededal, C.~B., \& Nishikawa, K.-I.\ 2005, \apjl, 623, L89 
\bibitem[Hockney \& Eastwood(1981)]{he81} Hockney, R.W., Eastwood, J.W.\ 1981, {\it Computer simulation using particles}, McGraw-Hill 
\bibitem[Holcomb, \& Spitkovsky(2019)]{2019ApJ...882....3H} Holcomb, C., \& Spitkovsky, A.\ 2019, \apj, 882, 3
\bibitem[Hoshino(2008)]{2008ApJ...672..940H} Hoshino, M.\ 2008, \apj, 672, 940 
\bibitem[Hoshino \& Terasawa(1985)]{Hoshino85} Hoshino, M., \& Terasawa, T.\ 1985, \jgr, 90, 57
\bibitem[Hoshino \& Arons(1991)]{Hoshino91} Hoshino, M., \& Arons, J.\ 1991, Physics of Fluids B, 3, 818
\bibitem[Hoshino et al.(1992)]{Hoshino92} Hoshino, M., Arons, J., Gallant, Y.~A., \& Langdon, A.~B.\ 1992, \apj, 390, 454
\bibitem[Hoshino \& Shimada(2002)]{Hoshino02} Hoshino, M., \& Shimada, N.\ 2002, \apj, 572, 880
\bibitem[Hoshino(2012)]{Hoshino12} Hoshino, M.\ 2012, Physical Review Letters, 108, 135003
\bibitem[Hoshino \& Lyubarsky(2012)]{Hoshino12b} Hoshino, M., \& Lyubarsky, Y.\ 2012, \ssr, 173, 521
\bibitem[Ikeya \& Matsumoto(2015)]{2015PASJ...67...64I} Ikeya, N., \& Matsumoto, Y.\ 2015, \pasj, 67, 64
\bibitem[Iwamoto et al.(2017)]{2017ApJ...840...52I} Iwamoto, M., Amano, T., Hoshino, M., \& Matsumoto, Y.\ 2017, \apj, 840, 52 
\bibitem[Iwamoto et al.(2018)]{2018ApJ...858...93I} Iwamoto, M., Amano, T., Hoshino, M., \& Matsumoto, Y.\ 2018, \apj, 858, 93
\bibitem[Iwamoto et al.(2019)]{Iwamoto19} Iwamoto, M., Amano, T., Hoshino, M., et al.\ 2019, \apjl, 883, L35
\bibitem[Jaroschek \& Hoshino(2009)]{2009PhRvL.103g5002J} Jaroschek, C.~H., \& Hoshino, M.\ 2009, Physical Review Letters, 103, 075002
\bibitem[Jaroschek et al.(2005)]{2005ApJ...618..822J} Jaroschek, C.~H., Lesch, H., \& Treumann, R.~A.\ 2005, \apj, 618, 822 
\bibitem[Johlander et al.(2016)]{Johlander16} Johlander, A., Schwartz, S.~J., Vaivads, A., et al.\ 2016, Physical Review Letters, 117, 165101
\bibitem[Kalapotharakos et al.(2018)]{Kalapotharakos18} Kalapotharakos, C., Brambilla, G., Timokhin, A., et al.\ 2018, \apj, 857, 44
\bibitem[Kang(2016)]{2016JKAS...49...83K} Kang, H.\ 2016, Journal of Korean Astronomical Society, 49, 83
\bibitem[Kang et al.(2019)]{2019ApJ...876...79K} Kang, H., Ryu, D., \& Ha, J.-H.\ 2019, \apj, 876, 79
\bibitem[Kato(2007)]{2007ApJ...668..974K} Kato, T.~N.\ 2007, \apj, 668, 974
\bibitem[Kato \& Takabe(2008)]{2008ApJ...681L..93K} Kato, T.~N., \& Takabe, H.\ 2008, \apjl, 681, L93 
\bibitem[Kato \& Takabe(2010a)]{2010PhPl...17c2114K} Kato, T.~N., \& Takabe, H.\ 2010, Physics of Plasmas, 17, 032114 
\bibitem[Kato \& Takabe(2010b)]{2010ApJ...721..828K} Kato, T.~N., \& Takabe, H.\ 2010, \apj, 721, 828 
\bibitem[Kato(2015)]{2015ApJ...802..115K} Kato, T.~N.\ 2015, \apj, 802, 115 
\bibitem[Kelley(2003)]{Kelley2003} Kelley, C.\ 2003, Solving Nonlinear Equations with Newton’s Method, Fundamentals of Algorithms, SIAM, Philadelphia
\bibitem[Kempf et al.(2016)]{Kempf16}Kempf, A., Kilian, P., \& Spanier, F.\ 2016, A\&A, 585, A132
\bibitem{Kennel84} Kennel, C.~F., \& Coroniti, F.~V.\ 1984, \apj, 283, 710 
\bibitem[Keshet et al.(2009)]{2009ApJ...693L.127K} Keshet, U., Katz, B., Spitkovsky, A., \& Waxman, E.\ 2009, \apjl, 693, L127 
\bibitem[Keshet \& Waxman(2005)]{2005PhRvL..94k1102K} Keshet, U., \& Waxman, E.\ 2005, Physical Review Letters, 94, 111102
\bibitem[Kirk et al.(2000)]{2000ApJ...542..235K} Kirk, J.~G., Guthmann, A.~W., Gallant, Y.~A., \& Achterberg, A.\ 2000, \apj, 542, 235
\bibitem[Kirk \& Schneider(1987)]{1987ApJ...315..425K} Kirk, J.~G., \& Schneider, P.\ 1987, \apj, 315, 425
\bibitem{Kirk03} Kirk, J.~G., \& Skj{\ae}raasen, O.\ 2003, \apj, 591, 366
\bibitem[Knoll \& Keyes(2004)]{2004JCoPh.193..357K} Knoll, D.~A., \& Keyes, D.~E.\ 2004, Journal of Computational Physics, 193, 357
\bibitem[Kobzar et al.(2017)]{2017MNRAS.469.4985K} Kobzar, O., Niemiec, J., Pohl, M., \& Bohdan, A.\ 2017, \mnras, 469, 4985 
\bibitem[Kolberg, \& Schlickeiser(2018)]{2018PhR...783....1K} Kolberg, U., \& Schlickeiser, R.\ 2018, \physrep, 783, 1
\bibitem[Krall, \& Trivelpiece(1973)]{1973ppp..book.....K} Krall, N.~A., \& Trivelpiece, A.~W.\ 1973, 
Principles of Plasma Physics, McGraw-Hill Book Company, New York
\bibitem[Krauss-Varban \& Wu(1989)]{1989JGR....9415367K} Krauss-Varban, D., \& Wu, C.~S.\ 1989, JGR, 94, 15367 
\bibitem[Kucharek \& Scholer(1995)]{Kucharek95} Kucharek, H., \& Scholer, M.\ 1995, \jgr, 100, 1745
\bibitem[Kumar et al.(2019)]{2019PhRvS..22d3401K} Kumar, R., Sakawa, Y., D{\"o}hl, L.~N.~K., et al.\ 2019, Physical Review Accelerators and Beams, 22, 043401
\bibitem[Kuramitsu et al.(2008)]{2008ApJ...682L.113K} Kuramitsu, Y., Sakawa, Y., Kato, T., Takabe, H., \& Hoshino, M.\ 2008, \apjl, 682, L113
\bibitem[Langdon et al.(1983)]{1983JCoPh..51..107L} Langdon, A.~B., Cohen, B.~I., \& Friedman, A.\ 1983, Journal of Computational Physics, 51, 107
\bibitem[Lapenta(2008)]{2008PhRvL.100w5001L} Lapenta, G.\ 2008, Physical Review Letters, 100, 235001 
\bibitem[Lapenta(2017)]{2017JCoPh.334..349L} Lapenta, G.\ 2017, Journal of Computational Physics, 334, 349
\bibitem[Lapenta \& Bettarini(2011)]{2011EL.....9365001L} Lapenta, G., \& Bettarini, L.\ 2011, EPL (Europhysics Letters), 93, 65001 
\bibitem[Lapenta et al.(2017)]{2017JPlPh..83b7005L} Lapenta, G., Gonzalez-Herrero, D., \& Boella, E.\ 2017, Journal of Plasma Physics, 83, 705830205
\bibitem[Lapenta \& Lazarian(2012)]{2012NPGeo..19..251L} Lapenta, G., \& Lazarian, A.\ 2012, Nonlinear Processes in Geophysics, 19, 251
\bibitem[Lazarian \& Vishniac(1999)]{1999ApJ...517..700L} Lazarian, A., \& Vishniac, E.~T.\ 1999, \apj, 517, 700 
\bibitem[Lebiga et al.(2018)]{2018MNRAS.476.2779L} Lebiga, O., Santos-Lima, R., \& Yan, H.\ 2018, \mnras, 476, 2779 
\bibitem{Lee96} Lee, M.~A., Shapiro, V.~D., \& Sagdeev, R.~Z.\ 1996, \jgr, 101, 4777
\bibitem[Lee et al.(2019)]{2019ApJ...871...74L} Lee, S.-Y., Ziebell, L.~F., Yoon, P.~H., et al.\ 2019, \apj, 871, 74
\bibitem[Lehe et al.(2016)]{2016PhRvE..94e3305L} Lehe, R., Kirchen, M., Godfrey, B.~B., Maier, A.~R., \& Vay, J.-L.\ 2016, \pre, 94, 053305
\bibitem[{{Lembege}(2003)}]{2003LNP...615...54L}
{Lembege}. 2003, in Lecture Notes in Physics, Berlin Springer Verlag, Vol. 615, Space Plasma Simulation, ed. J.~{B{\"u}chner}, C.~{Dum}, \& M.~{Scholer}, 54--78
\bibitem[{{Lembege} \& {Savoini}(1992)}]{1992PhFlB...4.3533L}
{Lembege}, B., \& {Savoini}, P. 1992, Physics of Fluids B, 4, 3533
\bibitem[Lemb{\`e}ge et al.(2003)]{Lembege03} Lemb{\`e}ge, B., Savoini, P., Balikhin, M., Walker, S., \& Krasnoselskikh, V.\ 2003, Journal of Geophysical Research (Space Physics), 108, 1256 
\bibitem[{Lemb{\`e}ge} {et~al.}(2009)]{2009JGRA..114.3217L}
{Lemb{\`e}ge}, B., {Savoini}, P., {Hellinger}, P., \& {Tr{\'a}vn{\'{\i}}{\v
  c}ek}, P.~M. 2009, \jgr, 114, 3217
\bibitem[Lemoine \& Pelletier(2003)]{2003ApJ...589L..73L} Lemoine, M., \& Pelletier, G.\ 2003, \apjl, 589, L73 
\bibitem[Lemoine et al.(2006)]{2006ApJ...645L.129L} Lemoine, M., Pelletier, G., \& Revenu, B.\ 2006, \apjl, 645, L129
\bibitem[Lemoine(2019)]{2019PhRvD..99h3006L} Lemoine, M.\ 2019, \prd, 99, 083006
\bibitem[Lemoine et al.(2019)]{2019PhRvL.123c5101L} Lemoine, M., Gremillet, L., Pelletier, G., et al.\ 2019, \prl, 123, 035101
\bibitem[Lemoine et al.(2019)]{2019PhRvE.100c3209L} Lemoine, M., Vanthieghem, A., Pelletier, G., et al.\ 2019, \pre, 100, 033209
\bibitem[Lemoine et al.(2019)]{2019PhRvE.100c3210L} Lemoine, M., Pelletier, G., Vanthieghem, A., et al.\ 2019, \pre, 100, 033210
\bibitem[Leroy {et~al.}(1981)]{leroy1981}
Leroy, M.~M., Goodrich, C.~C., Winske, D., Wu, C.-C.~S., \& Papadopoulos, K.~D. 1981, Geophysical Research Letters, 8, 1269
\bibitem[Leroy et al.(1981)]{Leroy81} Leroy, M.~M., Goodrich, C.~C., Winske, D., Wu, C.~S., \& Papadopoulos, K.\ 1981, \grl, 8, 1269
\bibitem[Leroy {et~al.}(1982)]{leroy1982}
Leroy, M.~M., Winske, D., Goodrich, C.~C., Wu, C.-C.~S., \& Papadopoulos, K.~D. 1982, Journal of Geophysical Research: Space Physics, 87, 5081
\bibitem[Leroy et al.(1982)]{Leroy82} Leroy, M.~M., Winske, D., Goodrich, C.~C., Wu, C.~S., \& Papadopoulos, K.\ 1982, \jgr, 87, 5081
\bibitem[Leroy (1983)]{Leroy83} Leroy, M.~M.\ 1983, Phys. Fl., 26, 2742
\bibitem[Li et al.(2017)]{2017CoPhC.214....6L} Li, F., Yu, P., Xu, X., et al.\ 2017, Computer Physics Communications, 214, 6 
\bibitem[Liang et al.(2013a)]{2013ApJ...766L..19L} Liang, E., Boettcher, M., \& Smith, I.\ 2013, \apjl, 766, L19\bibitem[Liewer et al.(1993)]{Liewer93} Liewer, P.~C., Goldstein, B.~E., \& Omidi, N.\ 1993, \jgr, 98, 15
\bibitem[Liang et al.(2013b)]{2013ApJ...779L..27L} Liang, E., Fu, W., Boettcher, M., Smith, I., \& Roustazadeh, P.\ 2013, \apjl, 779, L27 
\bibitem[Liang et al.(2017)]{2017ApJ...847...90L} Liang, E., Fu, W., \& B{\"o}ttcher, M.\ 2017, \apj, 847, 90 
\bibitem[Liang et al.(2018)]{2018ApJ...854..129L} Liang, E., Fu, W., B{\"o}ttcher, M., \& Roustazadeh, P.\ 2018, \apj, 854, 129 
\bibitem[Lin et al.(2003)]{2003ApJ...595L..69L} Lin, R.~P., Krucker, S., Hurford, G.~J., et al.\ 2003, \apjl, 595, L69
\bibitem[Lindner et al.(2014)]{Lindner-14} Lindner,~R.~R., Baker,~A.~J., Hughes,~J.~P., et al.\ 2014, \apj, 786, 49
\bibitem[Lipatov(2002)]{Lipatov2002} Lipatov, A. S.\ 2002, The Hybrid Multiscale Simulation Technology (Berlin: Springer-Verlag)
\bibitem[Lowe \& Burgess(2000)]{Lowe00} Lowe, R.~E., \& Burgess, D.\ 2000, \grl, 27, 3249
\bibitem[Lowe \& Burgess(2003)]{Lowe03} Lowe, R.~E., \& Burgess, D.\ 2003, Annales Geophysicae, 21, 671
\bibitem[Lucek \& Bell(2000)]{2000MNRAS.314...65L} Lucek, S.~G., \& Bell, A.~R.\ 2000, \mnras, 314, 65 
\bibitem[Lyubarsky(2006)]{2006ApJ...652.1297L} Lyubarsky, Y.\ 2006, \apj, 652, 1297 
\bibitem{Lyubarsky01} Lyubarsky, Y., \& Kirk, J.~G.\ 2001, \apj, 547, 437
\bibitem[Ma et al.(2018)]{2018JGRA..123.3742M} Ma, Y., Russell, C.~T., Toth, G., et al.\ 2018, Journal of Geophysical Research (Space Physics), 123, 3742
\bibitem[Makwana et al.(2018)]{2018PhPl...25h2904M} Makwana, K.~D., Keppens, R., \& Lapenta, G.\ 2018, Physics of Plasmas, 25, 082904
\bibitem[Markidis et al.(2018)]{2018FrP.....6..100M} Markidis, S., Olshevsky, V., Sishtla, C.~P., et al.\ 2018, Frontiers in Physics, 6, 100
\bibitem[Marcowith et al.(2016)]{2016RPPh...79d6901M} Marcowith, A., Bret, A., Bykov, A., et al.\ 2016, Reports on Progress in Physics, 79, 046901 
\bibitem[Markevitch et al.(2002)]{Markevitch-02} Markevitch,~M., Gonzalez,~A.~H., David,~L., et al.\ 2002, \apjl, 567, L27
\bibitem[Markidis \& Lapenta(2011)]{2011JCoPh.230.7037M} Markidis, S., \& Lapenta, G.\ 2011, Journal of Computational Physics, 230, 7037
\bibitem[Martins et al.(2009)]{2009ApJ...695L.189M} Martins, S.~F., Fonseca, R.~A., Silva, L.~O., \& Mori, W.~B.\ 2009, \apjl, 695, L189
\bibitem[Masters et al.(2016)]{Masters16} Masters, A., Sulaiman, A.~H., Sergis, N., et al.\ 2016, \apj, 826, 4
\bibitem[Matsukiyo \& Scholer(2003)]{Matsukiyo03} Matsukiyo, S., \& Scholer, M.\ 2003, Journal of Geophysical Research (Space Physics), 108, 1459
\bibitem[Matsukiyo et al.(2011)]{Matsukiyo-11} Matsukiyo,~S., Ohira,~Y.,Yamazaki,~R., \& Umeda,~T.\ 2011, \apj, 742, 47 
\bibitem[Matsukiyo \& Scholer(2014)]{Matsukiyo14} Matsukiyo, S., \& Scholer, M.\ 2014, Journal of Geophysical Research (Space Physics), 119, 2388  
\bibitem[Matsukiyo \& Matsumoto(2015)]{Matsukiyo-15} Matsukiyo,~S., \& Matsumoto,~Y.\ 2015, Journal of Physics Conference Series, 642, 012017
\bibitem[Matsumoto et al.(2012)]{Matsumoto2012} Matsumoto, Y., Amano, T., \& Hoshino, M.\ 2012, \apj, 755, 109
\bibitem[Matsumoto et al.(2013)]{2013PhRvL.111u5003M} Matsumoto, Y., Amano, T., \& Hoshino, M.\ 2013, Physical Review Letters, 111, 215003 
\bibitem[Matsumoto et al.(2015)]{2015Sci...347..974M} Matsumoto, Y., Amano, T., Kato, T.~N., \& Hoshino, M.\ 2015, Science, 347, 974
\bibitem[Matsumoto et al.(2017)]{2017PhRvL.119u5101M} Matsumoto, Y., Amano, T., Kato, T.~N., \& Hoshino, M.\ 2017, Physical Review Letters, 119, 105101
\bibitem[Medvedev(2000)]{2000ApJ...540..704M} Medvedev, M.~V.\ 2000, \apj, 540, 704 
\bibitem[Medvedev(2006)]{2006ApJ...637..869M} Medvedev, M.~V.\ 2006, \apj, 637, 869 
\bibitem[Medvedev et al.(2011)]{2011ApJ...737...55M} Medvedev, M.~V., Frederiksen, J.~T., Haugb{\o}lle, T., \& Nordlund, {\AA}.\ 2011, \apj, 737, 55
\bibitem[Medvedev \& Loeb(1999)]{1999ApJ...526..697M} Medvedev, M.~V., \& Loeb, A.\ 1999, \apj, 526, 697
\bibitem{Michel71} Michel, F.~C.\ 1971, \planss, 19, 1580
\bibitem[Mignone et al.(2009)]{2009MNRAS.393.1141M} Mignone, A., Ugliano, M., \& Bodo, G.\ 2009, \mnras, 393, 1141\bibitem[Miniati \& Elyiv(2013)]{Miniati13} Miniati, F., \& Elyiv, A.\ 2013, ApJ, 770, 54
\bibitem[Mignone et al.(2018)]{2018ApJ...859...13M} Mignone, A., Bodo, G., Vaidya, B., et al.\ 2018, \apj, 859, 13
\bibitem[Nagata et al.(2008)]{Nagata08} Nagata, K., Hoshino, M., Jaroschek, C.~H., \& Takabe, H.\ 2008, \apj, 680, 627 
\bibitem[Neronov \& Vovk(2010)]{Neronov10}Neronov, A., \& Vovk, I.\ 2010, Science, 328, 73
\bibitem[Nerush et al.(2011)]{2011PhRvL.106c5001N} Nerush, E.~N., Kostyukov, I.~Y., Fedotov, A.~M., et al.\ 2011, Physical Review Letters, 106, 035001 
\bibitem[Ness et al.(1964)]{Ness64} Ness, N.~F., Scearce, C.~S., \& Seek, J.~B.\ 1964, \jgr, 69, 3531
\bibitem[Niemiec \& Ostrowski(2004)]{2004ApJ...610..851N} Niemiec, J., \& Ostrowski, M.\ 2004, \apj, 610, 851 
\bibitem[Niemiec \& Ostrowski(2006)]{2006ApJ...641..984N} Niemiec, J., \& Ostrowski, M.\ 2006, \apj, 641, 984
\bibitem[Niemiec et al.(2006)]{2006ApJ...650.1020N} Niemiec, J., Ostrowski, M., \& Pohl, M.\ 2006, \apj, 650, 1020
\bibitem[Niemiec et al.(2008)]{2008ApJ...684.1174N} Niemiec, J., Pohl, M., Stroman, T., \& Nishikawa, K.-I.\ 2008, \apj, 684, 1174 
\bibitem[Niemiec et al.(2010)]{2010ApJ...709.1148N} Niemiec, J., Pohl, M., Bret, A., \& Stroman, T.\ 2010, \apj, 709, 1148 
\bibitem[Niemiec et al.(2012)]{2012ApJ...759.73N} Niemiec, J., Pohl, M., Bret, A., \& Wieland, V.\ 2012, \apj, 759, 73 
\bibitem[Nishikawa et al.(2003)]{2003ApJ...595..555N} Nishikawa, K.-I., Hardee, P., Richardson, G., et al.\ 2003, \apj, 595, 555 
\bibitem[Nishikawa et al.(2005)]{2005ApJ...622..927N} Nishikawa, K.-I., Hardee, P., Richardson, G., et al.\ 2005, \apj, 622, 927
\bibitem[Nishikawa et al.(2009)]{2009ApJ...698L..10N} Nishikawa, K.-I., Niemiec, J., Hardee, P.~E., et al.\ 2009, \apjl, 698, L10
\bibitem[Nishikawa et al.(2014)]{2014ApJ...793...60N} Nishikawa, K.-I., Hardee, P.~E., Du{\c t}an, I., et al.\ 2014, \apj, 793, 60 
\bibitem[Nishikawa et al.(2016)]{2016ApJ...820...94N} Nishikawa, K.-I., Frederiksen, J.~T., Nordlund, {\AA}., et al.\ 2016, \apj, 820, 94 
\bibitem[Ohira \& Takahara(2007)]{2007ApJ...661L.171O} Ohira, Y., \& Takahara, F.\ 2007, \apjl, 661, L171 
\bibitem[Ohira et al.(2009)]{2009ApJ...698..445O} Ohira, Y., Reville, B., Kirk, J.~G., \& Takahara, F.\ 2009, \apj, 698, 445 
\bibitem[Ohira(2012)]{2012ApJ...758...97O} Ohira, Y.\ 2012, \apj, 758, 97 
\bibitem[Ohira(2016)]{2016ApJ...827...36O} Ohira, Y.\ 2016, \apj, 827, 36 
\bibitem[Olshevsky et al.(2019)]{2019CoPhC.235...16O} Olshevsky, V., Bacchini, F., Poedts, S., et al.\ 2019, Computer Physics Communications, 235, 16
\bibitem[{Omidi} \& {Winske}(1992)]{1992JGR....9714801O}
{Omidi}, N., \& {Winske}, D. 1992, \jgr, 97, 14801
\bibitem[Ostrowski \& Bednarz(2002)]{2002A&A...394.1141O} Ostrowski, M., \& Bednarz, J.\ 2002, \aap, 394, 1141
\bibitem[Otsuka et al.(2019)]{2019HEDP...3300709O} Otsuka, F., Matsukiyo, S., \& Hada, T.\ 2019, High Energy Density Physics, 33, 100709
\bibitem[Panaitescu \& Kumar(2002)]{2002ApJ...571..779P} Panaitescu, A., \& Kumar, P.\ 2002, \apj, 571, 779\bibitem[Park et al.(2012)]{Park-12} Park, J., Workman, J.~C., Blackman, E.~G., Ren, C., \& Siller, R.\ 2012, Phys. Plasmas, 19, 062904
\bibitem[Papadopoulos(1988)]{Papadopoulos88} Papadopoulos, K.\ 1988, \apss, 144, 535
\bibitem[Park et al.(2013)]{Park-13} Park,~J., Ren,~C., Workman,~J.~C., Blackman,~E.~G.\ 2013, \apj, 765, 147
\bibitem[Park et al.(2015)]{2015PhRvL.114h5003P} Park, J., Caprioli, D., \& Spitkovsky, A.\ 2015, Physical Review Letters, 114, 085003 
\bibitem[Pausch et al.(2017)]{2017PhRvE..96a3316P} Pausch, R., Bussmann, M., Huebl, A., et al.\ 2017, \pre, 96, 013316 
\bibitem{Pesses81} Pesses, M.~E., Eichler, D., \& Jokipii, J.~R.\ 1981, \apjl, 246, L85
\bibitem[Petropoulou et al.(2016)]{2016MNRAS.462.3325P} Petropoulou, M., Giannios, D., \& Sironi, L.\ 2016, \mnras, 462, 3325
\bibitem[Petropoulou, \& Sironi(2018)]{2018MNRAS.481.5687P} Petropoulou, M., \& Sironi, L.\ 2018, \mnras, 481, 5687
\bibitem[Petropoulou et al.(2019)]{2019ApJ...880...37P} Petropoulou, M., Sironi, L., Spitkovsky, A., et al.\ 2019, \apj, 880, 37
\bibitem[Piran(2004)]{2004RvMP...76.1143P} Piran, T.\ 2004, Reviews of Modern Physics, 76, 1143
\bibitem[Philippov, \& Spitkovsky(2018)]{Philippov18} Philippov, A.~A., \& Spitkovsky, A.\ 2018, \apj, 855, 94
\bibitem[Pohl(1993)]{1993A&A...270...91P} Pohl, M.\ 1993, \aap, 270, 91 
\bibitem[Pohl et al.(2015)]{2015A&A...574A..43P} Pohl, M., Wilhelm, A., \& Telezhinsky, I.\ 2015, \aap, 574, A43 
\bibitem[Puchwein et al.(2012)]{Puchwein12}Puchwein, E., Pfrommer, C., Springel, V., Broderick, A. E., \& Chang, P.\ 2012, MNRAS, 423, 149
\bibitem[Rafighi et al.(2017)]{Rafighi17}Rafighi, I., Vafin, S., Pohl, M., \& Niemiec, J.\ 2017, A\&A, 607, A112
\bibitem[Reville \& Bell(2012)]{2012MNRAS.419.2433R} Reville, B., \& Bell, A.~R.\ 2012, \mnras, 419, 2433 
\bibitem[Reville, \& Bell(2013)]{2013MNRAS.430.2873R} Reville, B., \& Bell, A.~R.\ 2013, \mnras, 430, 2873
\bibitem[Reville \& Kirk(2010)]{2010ApJ...724.1283R} Reville, B., \& Kirk, J.~G.\ 2010, \apj, 724, 1283
\bibitem[Rieger \& Duffy(2006)]{2006ApJ...652.1044R} Rieger, F.~M., \& Duffy, P.\ 2006, \apj, 652, 1044 
\bibitem[Ripperda et al.(2018)]{2018ApJS..235...21R} Ripperda, B., Bacchini, F., Teunissen, J., et al.\ 2018, \apjs, 235, 21
\bibitem[Riquelme \& Spitkovsky(2009)]{2009ApJ...694..626R} Riquelme, M.~A., \& Spitkovsky, A.\ 2009, \apj, 694, 626 
\bibitem[Riquelme \& Spitkovsky(2010)]{2010ApJ...717.1054R} Riquelme, M.~A., \& Spitkovsky, A.\ 2010, \apj, 717, 1054
\bibitem[Riquelme \& Spitkovsky(2011)]{2011ApJ...733...63R} Riquelme, M.~A., \& Spitkovsky, A.\ 2011, \apj, 733, 63 
\bibitem{Sagdeev66} Sagdeev, R.~Z.\ 1966, Reviews of Plasma Physics, 4, 23 
\bibitem[Savoini et al.(2013)]{2013JGRA..118.1132S} Savoini, P., Lembege, B., \& Stienlet, J.\ 2013, Journal of Geophysical Research (Space Physics), 118, 1132\bibitem[Schlickeiser et al.(2012a)]{RS12a}Schlickeiser, R., Elyiv, A., Ibscher, D., \& Miniati, F.\ 2012a, ApJ, 758, 101
\bibitem[Schlickeiser et al.(2012b)]{RS12}Schlickeiser, R., Ibscher, D., \& Supsar, M.\ 2012b, ApJ, 758, 102
\bibitem[Schmitz et al.(2002)]{Schmitz02} Schmitz, H., Chapman, S.~C., \& Dendy, R.~O.\ 2002, \apj, 579, 327 
\bibitem[Schneider(1959)]{Schneider59} Schneider, J.\ 1959, \prl, 2, 504
\bibitem[Scholer et al.(2003)]{Scholer03} Scholer, M., Shinohara, I., \& Matsukiyo, S.\ 2003, \jgr, 108, 1014 
\bibitem[Scholer \& Matsukiyo(2004)]{Scholer04} Scholer, M., \& Matsukiyo, S.\ 2004, Annales Geophysicae, 22, 2345
\bibitem[Sckopke et al.(1983)]{Sckopke83} Sckopke, N., Paschmann, G., Bame, S.~J., Gosling, J.~T., \& Russell, C.~T.\ 1983, \jgr, 88, 6121
\bibitem[Shalaby et al.(2017)]{Shalaby17} Shalaby, M., Broderick, A.~E., Chang, P., et al.\ 2017, \apj, 848, 81 
\bibitem[Shalaby et al.(2018)]{Shalaby18}Shalaby, M., Broderick, A. E., Chang, P., et al.\ 2018, ApJ, 859, 45
\bibitem[Shalchi(2012)]{2012PhPl...19j2901S} Shalchi, A.\ 2012, Physics of Plasmas, 19, 102901 
\bibitem[Shimada \& Hoshino(2000)]{Shimada00} Shimada, N., \& Hoshino, M.\ 2000, \apjl, 543, L67
\bibitem[Shimada \& Hoshino(2004)]{Shimada04} Shimada, N., \& Hoshino, M.\ 2004, Physics of Plasmas, 11, 1840
\bibitem[Shimada \& Hoshino(2005)]{Shimada05} Shimada, N., \& Hoshino, M.\ 2005, Journal of Geophysical Research (Space Physics), 110, A02105
\bibitem[Silva et al.(2003)]{2003ApJ...596L.121S} Silva, L.~O., Fonseca, R.~A., Tonge, J.~W., et al.\ 2003, \apjl, 596, L121
\bibitem[Sironi \& Giannios(2014)]{SironiG14} Sironi, L., \& Giannios, D.\ 2014, ApJ, 787, 49
\bibitem[Sironi \& Spitkovsky(2009)]{2009ApJ...698.1523S} Sironi, L., \& Spitkovsky, A.\ 2009, \apj, 698, 1523
\bibitem[Sironi \& Spitkovsky(2009)]{2009ApJ...707L..92S} Sironi, L., \& Spitkovsky, A.\ 2009, \apjl, 707, L92 
\bibitem[Sironi \& Spitkovsky(2011)]{2011ApJ...726...75S} Sironi, L., \& Spitkovsky, A.\ 2011, \apj, 726, 75
\bibitem[Sironi \& Spitkovsky(2011)]{Sironi11} Sironi, L., \& Spitkovsky, A.\ 2011, \apj, 741, 39
\bibitem[Sironi et al.(2013)]{2013ApJ...771...54S} Sironi, L., Spitkovsky, A., \& Arons, J.\ 2013, \apj, 771, 54
\bibitem[Sironi \& Spitkovsky(2014)]{Sironi14} Sironi, L., \& Spitkovsky, A.\ 2014, \apjl, 783, L21
\bibitem[Skilling(1975a)]{1975MNRAS.172..557S} Skilling, J.\ 1975, \mnras, 172, 557\bibitem[Skilling(1975b)]{1975MNRAS.173..245S} Skilling, J.\ 1975, \mnras, 173, 245
\bibitem[Sonnet (1963)]{Sonnet63} Sonnet, C.~P.\ 1963, \jgr, 4, 1265
\bibitem[Spitkovsky(2005)]{2005AIPC..801..345S} Spitkovsky, A.\ 2005, Astrophysical Sources of High Energy Particles and Radiation, 801, 345
\bibitem[Spitkovsky(2008)]{2008ApJ...673L..39S} Spitkovsky, A.\ 2008, \apjl, 673, L39 
\bibitem[Spitkovsky(2008)]{2008ApJ...682L...5S} Spitkovsky, A.\ 2008, \apjl, 682, L5
\bibitem[Stockem Novo et al.(2015)]{2015ApJ...803L..29S} Stockem Novo, A., Bret, A., Fonseca, R.~A., et al.\ 2015, \apjl, 803, L29
\bibitem[Stockem Novo et al.(2016)]{2016NJPh...18j5002S} Stockem Novo, A., Bret, A., \& Sinha, U.\ 2016, New Journal of Physics, 18, 105002
\bibitem[Stone et al.(2017)]{Stone05} Stone, E.~C., Cummings, A.~C., McDonald, F.~B., et al.\ 2005, Science, 309, 2017 
\bibitem[Stone \& Edelman(1995)]{Stone95} Stone, J.~M., \& Edelman, M.\ 1995, \apj, 454, 182
\bibitem[Stroman et al.(2009)]{2009ApJ...706...38S} Stroman, T., Pohl, M., \& Niemiec, J.\ 2009, \apj, 706, 38 
\bibitem[Taylor et al.(2011)]{Taylor11}Taylor, A. M., Vovk, I., \& Neronov, A.\ 2011, A\&A, 529, A144
\bibitem[Timokhin(2010)]{2010MNRAS.408.2092T} Timokhin, A.~N.\ 2010, \mnras, 408, 2092
\bibitem{Timokhin13} Timokhin, A.~N., \& Arons, J.\ 2013, \mnras, 429, 20 
\bibitem[Trotta, \& Burgess(2019)]{2019MNRAS.482.1154T} Trotta, D., \& Burgess, D.\ 2019, \mnras, 482, 1154
\bibitem[Tzoufras et al.(2006)]{2006PhRvL..96j5002T} Tzoufras, M., Ren, C., Tsung, F.~S., et al.\ 2006, Physical Review Letters, 96, 105002
\bibitem[Umeda et al.(2003)]{2003CoPhC.156...73U} Umeda, T., Omura, Y., Tominaga, T., \& Matsumoto, H.\ 2003, Computer Physics Communications, 156, 73
\bibitem[{{Umeda} \& {Yamazaki}(2006)}]{2006EP&S...58E..41U}
{Umeda}, T., \& {Yamazaki}, R. 2006, Earth, Planets, and Space, 58, 41
\bibitem[Umeda et al.(2008)]{2008ApJ...681L..85U} Umeda, T., Yamao, M., \& Yamazaki, R.\ 2008, \apjl, 681, L85 
\bibitem[Umeda et al.(2009)]{2009ApJ...695..574U} Umeda, T., Yamao, M., \& Yamazaki, R.\ 2009, \apj, 695, 574
\bibitem[Umeda et al.(2014)]{2014PhPl...21b2102U} Umeda, T., Kidani, Y., Matsukiyo, S., \& Yamazaki, R.\ 2014, Physics of Plasmas, 21, 022102 
\bibitem{Usov94} Usov, V.~V.\ 1994, \mnras, 267, 1035
\bibitem[van~Weeren et al.(2010)]{van-Weeren-10} van~Weeren,~R.~J., R\"{o}ttgering,~H.~J.~A., Br\"{u}uggen,~M., \& Hoeft,~M.\ 2010, Science, 330, 347
\bibitem[Vafin et al.(2018)]{Vafin18} Vafin, S., Rafighi, I., Pohl, M., \& Niemiec, J.\ 2018, \apj, 857, 43
\bibitem[Vafin et al.(2019)]{2019arXiv190109640V} Vafin, S., Deka, P.~J., Pohl, M., et al.\ 2019, \apj, 873, 10
\bibitem[van Marle et al.(2018)]{2018MNRAS.473.3394V} van Marle, A.~J., Casse, F., \& Marcowith, A.\ 2018, \mnras, 473, 3394
\bibitem[Vay(2007)]{2007PhRvL..98m0405V} Vay, J.-L.\ 2007, Physical Review Letters, 98, 130405 
\bibitem[Vay(2008)]{2008PhPl...15e6701V} Vay, J.-L.\ 2008, Physics of Plasmas, 15, 056701 
\bibitem[Vay et al.(2011)]{2011JCoPh.230.5908V} Vay, J.-L., Geddes, C.~G.~R., Cormier-Michel, E., \& Grote, D.~P.\ 2011, Journal of Computational Physics, 230, 5908 
\bibitem[Villasenor \& Buneman(1992)]{1992CoPhC..69..306V} Villasenor, J., \& Buneman, O.\ 1992, Computer Physics Communications, 69, 306
\bibitem[Vranic et al.(2015)]{2015CoPhC.191...65V} Vranic, M., Grismayer, T., Martins, J.~L., Fonseca, R.~A., \& Silva, L.~O.\ 2015, Computer Physics Communications, 191, 65
\bibitem[Weibel(1959)]{Weibel59} Weibel, E.~S.\ 1959, \prl, 2, 83
\bibitem[Weidl et al.(2019)]{2019ApJ...872...48W} Weidl, M.~S., Winske, D., \& Niemann, C.\ 2019, \apj, 872, 48
\bibitem[Weidl et al.(2019)]{2019ApJ...873...57W} Weidl, M.~S., Winske, D., \& Niemann, C.\ 2019, \apj, 873, 57
\bibitem[Wieland et al.(2016)]{2016ApJ...820...62W} Wieland, V., Pohl, M., Niemiec, J., Rafighi, I., \& Nishikawa, K.-I.\ 2016, \apj, 820, 62
\bibitem[Winske \& Leroy(1984)]{1984JGR....89.2673W} Winske, D., \& Leroy, M.~M.\ 1984, \jgr, 89, 2673 
\bibitem[Winske \& Quest(1988)]{Winske88} Winske, D., \& Quest, K.~B.\ 1988, \jgr, 93, 9681 
\bibitem[Wu et al.(1984)]{Wu84} Wu, C.~S., Winske, D., Zhou, Y.~M., et al.\ 1984, \ssr, 37, 63
\bibitem[Yan \& Lazarian(2008)]{2008ApJ...673..942Y} Yan, H., \& Lazarian, A.\ 2008, \apj, 673, 942-953 
\bibitem[Yang et al.(2018)]{2018ApJ...857...36Y} Yang, Z., Lu, Q., Liu, Y.~D., et al.\ 2018, \apj, 857, 36
\bibitem[Yoon, \& Davidson(1987)]{Yoon87} Yoon, P.~H., \& Davidson, R.~C.\ 1987, \pra, 35, 2718
\bibitem[Yoon et al.(2016)]{2016PhRvE..93c3203Y} Yoon, P.~H., Ziebell, L.~F., Kontar, E.~P., et al.\ 2016, \pre, 93, 033203
\bibitem[Yu et al.(2014)]{2014JCoPh.266..124Y} Yu, P., Xu, X., Decyk, V.~K., et al.\ 2014, Journal of Computational Physics, 266, 124
\bibitem[Yu et al.(2015)]{2015CoPhC.192...32Y} Yu, P., Xu, X., Decyk, V.~K., et al.\ 2015, Computer Physics Communications, 192, 32
\bibitem{Zank96} Zank, G.~P., Pauls, H.~L., Cairns, I.~H., \& Webb, G.~M.\ 1996, \jgr, 101, 457
\bibitem[Zank \& M{\"u}ller(2003)]{Zank03} Zank, G.~P., \& M{\"u}ller, H.-R.\ 2003, Journal of Geophysical Research (Space Physics), 108, 1240
\bibitem[Zank et al.(2015)]{2015ApJ...814..137Z} Zank, G.~P., Hunana, P., Mostafavi, P., et al.\ 2015, \apj, 814, 137 
\bibitem[Zekovi{\'c}(2019)]{2019PhPl...26c2106Z} Zekovi{\'c}, V.\ 2019, Physics of Plasmas, 26, 032106\bibitem[Zenitani \& Hoshino(2001)]{Zenitani01} Zenitani, S., \& Hoshino, M.\ 2001, \apjl, 562, L63
\bibitem[Zenitani \& Hoshino(2005)]{Zenitani05} Zenitani, S., \& Hoshino, M.\ 2005, Physical Review Letters, 95, 095001
\bibitem[Zenitani \& Hoshino(2007)]{Zenitani07} Zenitani, S., \& Hoshino, M.\ 2007, \apj, 670, 702
\bibitem[Zheleznyakov \& Suvorov(1972)]{Zheleznyakov72} Zheleznyakov, V.~V., \& Suvorov, E.~V.\ 1972, \apss, 15, 24

\end{thebibliography}

\end{document}